\documentclass[aps,prb,twocolumn,amssymb,amsmath,amsfonts,floatfix]{revtex4-2}
\usepackage{CJK}
\pdfoutput=1
\usepackage{subfig}
\usepackage{bm}
\usepackage{hyperref}
\usepackage{color}
\usepackage{psfrag,euscript,mathrsfs}
\usepackage{graphicx}
\usepackage{epsfig}
\usepackage{amssymb}

\def\XXint#1#2#3{{\setbox0=\hbox{$#1{#2#3}{\int}$}
    \vcenter{\hbox{$#2#3$}}\kern-.5\wd0}}

\definecolor{dgreen}{rgb}{0.2 ,0.54, 0.2}

\begin{document}
\begin{CJK*}{GB}{} 
\title{Confinement of spinons in  the XXZ spin-1/2 chain in presence of a transverse magnetic field}
\author{{Sergei~B.~Rutkevich}}
\affiliation{Fakult\"at f\"ur Mathematik und Naturwissenschaften, Bergische Universit\"at Wuppertal, 42097 Wuppertal, Germany.}
\maketitle
\end{CJK*}
\date{January 9, 2024}
\begin{abstract}
We study the tuning effect of a transverse magnetic field on the confinement of spinons  
 in the infinite XXZ spin-1/2 chain. The spinon confinement in this model takes place  in the gapped
antiferromagnetic phase upon application of a staggered  longitudinal magnetic field.  The tuning transverse
magnetic field has mutually orthogonal uniform  and  staggered components. The energy spectra  of the 
two-spinon bound states  (the `mesons')
in the  confinement regime are analytically calculated in this model using two different perturbative schemes.
The first one applies in the extreme anisotropic (Ising) limit and employs the inverse anisotropy constant 
 as a small parameter. The second perturbative scheme, which applies at any 
anisotropy in the gapped antiferromagnetic domain, exploits the
integrability of the XXZ spin chain at zero magnetic field. The small parameters in the  second technique are 
the components of the transverse, and staggered longitudinal magnetic fields. 
It is shown, that the weak transverse magnetic field mixes the transverse and longitudinal  meson modes, and leads to an avoided crossing of their energies upon increase of its strength. The explicit formulas for the two-spinon contribution to the dynamical structure factors of local spin operators  are obtained as well in this model in the weak confinement regime for 
 wave-vectors 
 close
to the points $\mathrm{k}=0$ and $\mathrm{k}=\pi$. 
 \end{abstract}
\maketitle
\section{Introduction}
The phenomenon of confinement is commonly associated with  high energy physics \cite{Nar04}. 
The problem of a consistent description  of the quark  confinement
 in hadrons in the frame of  quantum chromodynamics (QCD) in 
  four-dimensional space-time  remains to be  one of the most long-standing  challenging problems in  theoretical physics. The difficulty of this problem stems from the intrinsic 
non-perturbative nature of the quark confinement, and from the lack of clear understanding of its 
physical mechanism  in  QCD in four dimensions. 
 
Confinement of elementary excitations takes place also in certain two-dimensional quantum field theories (QFT)
and spin-chain models. It is worth to note, that the confinement in such models is provided by 
a rather simple and well understood mechanism, which cannot be responsible for the 
color confinement 
in the four-dimensional QCD. Nevertheless, studying the confinement in the two-dimensional space-time can give a
 useful insight into some aspect of the quark confinement in the high energy physics, see e.g. \cite{Wilson74,Hooft74}.
 Note also, that the  two-particle bound states of the two-dimensional QFT
and spin-chain models in the confinement regime are often
referred to  as ``mesons",  due to the analogy with QCD. 

It was shown by {}'t Hooft in 1974 \cite{Hooft74}, that 
the quarks in two-dimensional  QCD with an infinite number of colors are confined 
forming the  mesons, whose masses
 are exactly determined by the Bethe-Salpeter equation.  
 
  In 1978, McCoy and Wu \cite{McCoy78}  
studied the effect of  uniform magnetic field $h$ on the particle content of the 
 Ising field theory  (IFT) in the two-dimensional space-time.  At zero magnetic field $h=0$,
this relativistic field theory has two degenerate vacua in the ferromagnetic phase due to  
spontaneous breaking of the $\mathbb{Z}_2$-symmetry, and kinks interpolating between these vacua as elementary excitations.
McCoy and Wu showed in \cite{McCoy78}, that the 
magnetic field $h>0$ explicitly breaking the $\mathbb{Z}_2$-symmetry of the  model Hamiltonian, induces a 
linear attractive  potential acting between  neighboring kinks, which were initially free at $h=0$.  
The strength of the linear attractive potential   $U(x)=f|x|$  is characterized by the positive ``string tension" $f\sim h$.
Treating the two kinks as non-relativistic fermions with  dispersion law 
\begin{equation}\label{dl}
\omega_0(p)=m_0+\frac{p^2}{2 m_0},
\end{equation}
McCoy and Wu obtained \cite{McCoy78}  for the masses of the ``lightest mesons" (the energies  of the  two-kink bound states
with zero total momentum) at small $h>0$
 the simple formula: 
\begin{equation}\label{MW}
E_n=E_0+ \alpha z_n, \quad \text{with  } n=1,2,3\ldots,
\end{equation}
where  $E_0=2 m_0$, $\alpha =f^{2/3} m_0^{-1/3}$, and the numbers $-z_n$ are the zeros of the Airy function,
$\mathrm{Ai}(-z_n)=0$.

To our knowledge, the first study of the confinement of magnetic excitations in the spin-chain models was
reported by Shiba \cite{Shiba80} in 1980. He considered the antiferromagnetic XXZ spin-1/2 chain model in the presence of a staggered longitudinal magnetic field  in the limit of the strong uniaxial anisotropy, and  calculated the energy spectra of the two-kink bound states (``the Zeeman ladder")  by means of a 
strong-coupling expansion.  Shiba also suggested, that the effective staggered 
 field, that induces the kink confinement, can arise  at low temperatures  in the quasi-one-dimensional (1D)  uniaxial  antiferromagnetic crystals due to a weak interchain interaction in the three-dimensionally ordered antiferromagnetic phase. 
Note, that since the kinks in the antiferromagnetic XXZ spin chain carry   spin $\pm1/2$, they are also often called
`spinons'. We shall use both terms as synonyms.

The spinon confinement provided by the   outlined above  physical mechanism was 
observed later in  inelastic neutron scattering \cite{Gr15,Bera17} and terahertz 
spectroscopy \cite{Wang15,Wang19} experiments in the
quasi-1D spin-1/2 antiferromagnetic compounds  
$\mathrm{BaCo}_{2}\mathrm{V}_{2}\mathrm{O}_{8}$   and $\mathrm{SrCo}_{2}\mathrm{V}_{2}\mathrm{O}_{8}$ with 
 Heisenberg-Ising (XXZ) anisotropy.  However, though  Shiba's theory
  based on a strong-coupling expansion gives a useful starting point for the understanding 
of some qualitative features of the spinon confinement  in such magnetic crystals, it could not provide a quantitative description of 
the observed energy  spectra of the spinon bound states, since  the uniaxial anisotropy in the studied compounds is rather moderate. 
This is why the experimentally observed energy spectra of the spinon bound states in the confinement regime were usually interpreted 
in terms of the simple phenomenological McCoy-Wu formula \eqref{MW} with fitting parameters $E_0$ and $\alpha$, or  compared 
with the results of direct calculations of the energy spectra in the appropriate spin-chain model by means of different 
numerical techniques \cite{Gr15,Bera17,Wang15,Faur17}.

In the recent paper \cite{Rut22}, we described analytic perturbative calculations of the meson energy spectra 
in the gapped antiferromagnetic XXZ spin-chain model  at any value of the easy-axis anisotropy in the {\it weak confinement regime}, which
is realized in this model in the presence of a weak staggered  magnetic field $h_z$ parallel to the magnetic easy axis $z$. 
Preliminary results of this work were published in \cite{Rut18}. Two different perturbative techniques have been used in \cite{Rut22}.
Both  techniques exploit  the integrability of the XXZ spin-1/2 chain model in the deconfined phase at $h_z=0$ and use the staggered magnetic field $h_z>$0 in the confinement regime as a small parameter. 

The first more rigorous and systematic technique is based on a 
perturbative analysis of the Bethe-Salpeter equation, which was derived for the XXZ spin-chain model in \cite{Rut22}. For the 
IFT, the analogous Bethe-Salpeter equation was obtained and studied previously by Fonseca and Zamolodchikov 
\cite{FonZam2003,FZ06}. 

The second so-called {\it semiclassical  technique} is not  rigorous, but rather heuristic and intuitive. 
 It can be viewed as a generalization of  McCoy and Wu's scenario of  confinement to  systems, in which the 
kinks in the deconfined phase (i) have a non-quadratic dispersion law, and (ii) are not free, but can interact at short  distances.
Initially this technique was introduced in  \cite{FZ06,Rut05} in order to interpret  the mass spectrum of heavy mesons 
in the  Ising field theory.   Later the semiclassical technique was successfully applied to the calculation of the meson energy spectra in different two-dimensional  QFT and spin-chain models exhibiting confinement \cite{RutP09,Mus22,Rut08a,Rut18,Lagnese_2022,Ram20,rutkevich2023soliton}.
As in the McCoy and Wu picture, the two kinks forming a 
meson are treated in this approach as classical particles, that move  along the line and attract one another with a linear potential. 
However, the kinetic energy of these classical  particles is now not quadratic in their momenta, but is given by the
kink  dispersion law in the deconfined phase. The meson energy spectrum in this approach is
determined by means of the semiclassical quantization from the Bohr-Sommerfeld quantization rule. If the kinks interact at short distances already in the deconfined phase, their pair interaction is  accounted for their non-trivial two-particle scattering phase, which is added in this approach to the left-hand side of the Bohr-Sommerfeld  quantization condition.

It was shown in \cite{Rut22}, that to  leading order in the 
staggered magnetic field $h_z$,  the two perturbative techniques outlined above lead to the same result for the 
meson energy spectra in  the antiferromagnetic XXZ spin-chain model. 

In this paper, we continue to study  kink confinement in the gapped antiferromagnetic XXZ spin chain
induced by the  longitudinal staggered magnetic field. Here we address the problem of the effect of a 
weak external transverse magnetic field on the energy spectrum and spin polarization of the meson states.
Our interest in this subject is motivated by recent  experimental  studies  \cite{Kimura13,Kimura08,Wang15,Faur17,Faure_21,Wang16,Wang19,Muss21,Wang_22,Takay23}
of the influence of  external magnetic fields on the magnetic properties of the quasi-1D antiferromagnetic compounds  
$\mathrm{BaCo}_{2}\mathrm{V}_{2}\mathrm{O}_{8}$   and $\mathrm{SrCo}_{2}\mathrm{V}_{2}\mathrm{O}_{8}$.
Besides a number of phase transitions triggered by strong enough magnetic fields, a substantial modification
of the magnetic excitations induced by the weak external magnetic fields was observed. It turns out, that the weak longitudinal  
and  transverse uniform
external magnetic fields act in a very different way  on the meson energy spectra  in the confinement regime.
 It was reported in \cite{Wang15,Wang19,Faur17}, that the weak  longitudinal (parallel to the 
Ising axis) magnetic field leads to a simple Zeeman splitting of the meson energies, which is linear in the  
 applied  field. In contrast, the variation of the meson energy spectra with the applied  transverse magnetic field 
measured in  $\mathrm{BaCo}_{2}\mathrm{V}_{2}\mathrm{O}_{8}$  in the inelastic neutron scattering  \cite{Faur17} 
 and terahertz spectroscopic experiments \cite{Wang_22}  displays a rather peculiar non-linear dependence. In particular, it was observed 
 in  \cite{Faur17},  that the  transverse and longitudinal meson modes, which are characterized at zero transverse magnetic field by the $z$-projection of the spin $s=\pm 1$, and $s=0$, respectively, hybridize upon increase of the 
 applied transverse  field. Avoided crossing of the energy curves of 
 different meson modes with increasing transverse field was detected as well.

While the impact of a strong transverse magnetic field on the properties of quasi-1D antiferromagnetic crystals has been thoroughly  
studied in literature \cite{Dmitr02,Tako18,Muss21,Kimura_21,Wang_22,Wang23a}, the observed effect of a weak transverse field 
on the magnetic excitations in such crystals in the confinement regime is much less understood and requires theoretical explanations.
 The goal of  this work is to show, that the  unusual features of the   spin dynamics  
 observed in 
 \cite{Faur17}  in $\mathrm{BaCo}_{2}\mathrm{V}_{2}\mathrm{O}_{8}$  in the presence of the weak transverse magnetic field 
 can be understood  in the frame of  properly modified analytic perturbation-theory-technique developed  in  \cite{Rut18,Rut22}. Here we apply these techniques to the XXZ spin chain Hamiltonian in the gapped antiferromagnetic phase 
 perturbed not only by the staggered longitudinal, but also by the transverse uniform and transverse staggered magnetic fields. 
The necessity to account for the effective transverse staggered field stems from the fact, that the latter
 arises  \cite{Kimura13,Faur17} in the magnetic ion chains  in the crystal  $\mathrm{BaCo}_{2}\mathrm{V}_{2}\mathrm{O}_{8}$
 upon application of an external uniform transverse field due to the 
non-diagonal $g_{xy}$-component of the Land\'e tensor.

As in the previous papers,  \cite{Rut18,Rut22}, we perform perturbative calculations of the meson energy spectra in two different 
asymptotic regimes: (i) in the extreme anisotropic (Ising) limit, and (ii)  at weak magnetic fields  
at generic values of the anisotropy parameter. In the second regime, the dynamical structure factors (DSF) of the local spin operators
are calculated as well. The obtained meson spectra display in both regimes qualitatively similar non-linear dependences on the transverse magnetic field with avoided crossings of the neighboring in energy dispersion curves.

The rest of the paper is organized as follows. In the next section, we describe the Hamiltonian of the XXZ spin chain
perturbed by  magnetic fields of different nature: staggered and uniform, transverse and longitudinal. We recall there also the well-known 
 discrete-symmetry properties of the XXZ spin chain at zero magnetic field. 
Section  \ref{sec:Is} contains the perturbative calculation of the meson energy spectra in the XXZ spin chain in the limit of strong anisotropy
$\Delta\to-\infty$ for arbitrary  fixed values of the staggered longitudinal and mutually orthogonal staggered and uniform transverse magnetic fields.

In Secs. \ref{Sect:Gr}-\ref{DStF}, the anisotropy parameter is taken at a generic value in the domain $\Delta<-1$, corresponding to the 
gapped antiferromagnetic phase. In these sections, exploiting integrability of the XXZ spin chain at zero magnetic field, 
we use the components of the applied magnetic fields as small parameters in perturbative calculations, which are  
performed in two steps. First, we keep the staggered longitudinal field at zero value, and study in
Secs. \ref{Sect:Gr}, \ref{Sec:kink}, and \ref{two-k}, the effect of the weak transverse uniform and staggered magnetic fields
on the ground states, one-, and two-kink excitations, respectively. 
Next, we switch on the weak staggered 
longitudinal magnetic field inducing the confinement of kinks, which become coupled into the meson bound states.
In Section \ref{sec_Mes}, we describe, how the classification and symmetry properties of the resulting meson states are effected 
by the presence of the mutually orthogonal uniform and staggered transverse magnetic fields. The meson energy spectra
for this magnetic field configuration are studied in Secs.  \ref{HA},  \ref{Heur} by means of the semiclassical perturbative 
technique \cite{Rut18}. The effect of the weak transverse magnetic fields on the DSF of  local spin operators 
in the  confinement regime is studied in Section \ref{DStF}. The obtained analytical results  are compared  in 
Section~\ref{CompExp} with the 
results of the inelastic neutron scattering experiments 
 on the antiferromagnetic crystals $\mathrm{BaCo}_{2}\mathrm{V}_{2}\mathrm{O}_{8}$ reported by Faure {\it et el.} \cite{Faur17}.
Concluding remarks are presented in Section \ref{Conc}. Finally, there  are two Appendixes. In Appendix \ref{Ap1}, 
we collect the well-known explicit formulas for the two-kink scattering amplitudes and for the two-kink
form factors of the spin operators. Appendix \ref{AppB} contains the details
of some technical calculations relegated from Section \ref{Heur}.
\section{Model}
In this section, we introduce  several Hamiltonians of the XXZ spin-1/2 chain model, which is deformed
in different ways by  external magnetic fields. The models defined by these Hamiltonians will be studied in the subsequent sections
by means of different perturbative techniques.   

The  most general  Hamiltonian of the infinite XXZ spin-1/2 chain in the presence of both uniform 
and staggered magnetic fields can be written in the form:
\begin{equation}\label{Ham}
\mathcal{H}_{XXZ}= \mathcal{H}_0+V,
\end{equation}
where
\begin{align}\label{H0}
& \mathcal{H}_0=-\frac{1}{2}\sum_{j=-\infty}^\infty\!\!\left(\sigma_j^x\sigma_{j+1}^x+\sigma_j^y\sigma_{j+1}^y+
\Delta\,
\sigma_j^z\sigma_{j+1}^z
\right), \\\label{V}
&V=-\sum_{j=-\infty}^\infty\sum_{\mathfrak{a}=x,y,z}\!\!\
\left[
 (-1)^j h_{1\mathfrak{a}}+h_{2\mathfrak{a}}
\right]\sigma_j^\mathfrak{a}.
\end{align}
Here the index $j$ enumerates the spin-chain sites, $\sigma_j^{\mathfrak{a}}$ are
the Pauli matrices, $\mathfrak{a} = x, y, z$,  and $\Delta$ is the anisotropy parameter. The $``-"$sign in front of the 
right-hand side in \eqref{H0} is the
subject of convention, since it can be changed to $``+"$ by a certain unitary transformation of the Hamiltonian, see 
equations \eqref{Uodd}, \eqref{Hama2} below.
In our choice  of this sign  we  follow the convention
 widely accepted in the literature devoted to the algebraic approach to the XXZ spin-chain model, see e.g. \cite{Jimbo94,LukTer03}. 
 The anisotropy constant will be taken throughout this paper  in the interval $\Delta<-1$, and     
 parametrized in the usual way: 
\begin{align}
{\Delta}=({q}+{q}^{-1})/2=-\cosh \eta, \\
q=-\exp(- \eta)\in(-1,0),\quad \eta>0.
\end{align}

We will use also the notation  $S^\mathfrak{a}$  for the projection of the total spin operator on the $\mathfrak{a}$-axis:
\begin{equation}
S^\mathfrak{a}=\frac{1}{2}\sum_{j=-\infty}^\infty \sigma_j^\mathfrak{a}.
\end{equation}

The XXZ spin chain at zero magnetic field  defined by the Hamiltonian \eqref{H0} is integrable. 
In the gapped antiferromagnetic phase at $\Delta<-1$, it has two ground states $|vac\rangle^{(1)}$ and 
$|vac\rangle^{(0)}$, which display  
Ne\'el-type order: 
\begin{align}\label{vac1}
&\phantom{.}^{(1)} \langle vac| \sigma_j^z|vac\rangle^{(1)} =(-1)^j\bar{\sigma},\\
&\phantom{.}^{(0)} \langle vac| \sigma_j^z|vac\rangle^{(0)}=-(-1)^j\bar{\sigma},
\end{align}
with  the staggered spontaneous magnetization  \cite{Baxter1973,Baxter1976,Iz99}
\begin{equation}\label{sig}
\bar{\sigma}(\eta)=\prod_{n=1}^\infty \left(
\frac{1-e^{-2 n \eta}}{1+e^{-2 n\eta}}
\right)^2.
\end{equation}
The elementary excitations in this regime are the kinks interpolating 
between these two vacuums. 

The XXZ spin chain 
\eqref{Ham} remains 
also integrable, if  solely  the longitudinal uniform magnetic field $h_{2z}$ is applied. In the presence of any other
magnetic field $h_{i\mathfrak{a}}$, with $i\ne2$, and $\mathfrak{a}\ne z$,  model \eqref{Ham} becomes non-integrable.
In the latter case, it can be studied either by direct numerical methods, or by different analytic perturbative techniques.

The confinement regime in model \eqref{Ham} takes place upon application of the staggered longitudinal 
magnetic field $h_{1z}>0$. In the case, when all other components $h_{i\mathfrak{a}}$ in \eqref{V} are zero, the meson energy spectra  in model  \eqref{Ham} were studied in \cite{Shiba80,Rut18,Rut22}. 
It was shown  in \cite{Rut22}, that the meson states $|\pi_{s,\iota, n}(P)\rangle$ in this case can be classified by the quasimomentum
$P\in[0,\pi)$, the spin $s=0,\pm1$, the parity $\iota=0,\pm$, and the natural number $n=1,2,\ldots$. 
The quantum numbers  $\iota$  and $s$ are not independent: 
$\iota=0$ for $s=\pm1$, and $\iota=\pm $, for  $s=0$.
The meson states 
$|\pi_{s,\iota, n}(P)\rangle$  satisfy  the 
following equations (see equations (125) in \cite{Rut22}):
\begin{subequations}\label{mesA}
\begin{align}\label{mes}
&\mathcal{H}_1(h_{1z})|\pi_{s,\iota,n}(P)\rangle=E_{\iota,n}(P)|\pi_{s,\iota,n}(P)\rangle,\\\label{pT1}
&T_1^2|\pi_{s,\iota,n}(P)\rangle=e^{2 i P}|\pi_{s,\iota,n}(P)\rangle,\\
&S^z|\pi_{s,\iota,n}(P)\rangle=s|\pi_{s,\iota,n}(P)\rangle.
\end{align}
\end{subequations}
Here $T_1$ is the one-site translation operator defined below by equation \eqref{Tra1}, and the Hamiltonian $\mathcal{H}_1(h_{1z})$ 
is defined as follows:
\begin{equation}\label{Hame1}
\mathcal{H}_1(h_{1z})=\mathcal{H}_0- \sum_{j=-\infty}^\infty[ (-1)^j h_{1z} \sigma_j^z+c(h_{1z})].
\end{equation}
The constant $c(h_{1z})$ in the right-hand side is chosen in such a way, that the ground-state energy of the Hamiltonian \eqref{Hame1}
vanishes. 

It is easy to understand, how  the   application of the uniform longitudinal magnetic field $h_{2z}$ effects  the meson energy spectra. 
Really, since 
\[
[S^z,\mathcal{H}_1(h_{1z})]=0,
\]
the Hamiltonian 
\begin{equation}\label{Hame2}
\mathcal{H}_2(h_{1z},h_{2z})=\mathcal{H}_1(h_{1z})-2 h_{2z} \, S^z
\end{equation}
has the same set of eigenstates, as $\mathcal{H}_1(h_{1z})$, and 
\begin{equation}\label{Hame3}
\mathcal{H}_2(h_{1z},h_{2z})|\pi_{s,\iota,n}(P)\rangle=[E_{\iota,n}(P)-2h_{2z} s]|\pi_{s,\iota,n}(P)\rangle.
\end{equation}
Therefore, the uniform longitudinal magnetic field $h_{2z}$ has no effect on the energies of the meson modes  with $s=0$, 
while the two modes with $s=\pm1$, which were degenerate in energy at $h_{2z}=0$, get the linear Zeeman splitting at $h_{2z}>0$.  
Such longitudinal field dependences of the meson modes energies were  indeed observed in the neutron scattering \cite{Faur17} and 
high-resolution terahertz spectroscopic \cite{Wang19} experiments  on the compound $\mathrm{BaCo}_{2}\mathrm{V}_{2}\mathrm{O}_{8}$.

In the rest of this paper we will concentrate on the case of the zero uniform longitudinal magnetic field in the Hamiltonian 
\eqref{Ham}-\eqref{V}, $h_{2z}=0$.
The subsequent analysis of the tuning effect of the transverse magnetic field 
on the spinon confinement will be limited to the case, in which the  uniform and  staggered transverse magnetic fields are mutually orthogonal. 
The reason is twofold. First, the classification of the meson states and perturbative calculations of their energy spectra become  easer in this case.
Second, according to \cite{Kimura13,Faur17}, it is relevant to the experimental situation in the compound 
$\mathrm{BaCo}_{2}\mathrm{V}_{2}\mathrm{O}_{8}$. 

So, in  the study of the spinon confinement, we will restrict our attention 
 in this paper to the Hamiltonian \eqref{Ham}-\eqref{V} with $h_{2z}=h_{1x}=h_{2y}=0$. In order to simplify notations,
we rewrite this Hamiltonian in the equivalent form:
\begin{equation}\label{Hama}
\mathcal{H}(\Delta,\mathbf{h}_t,h_z)= \mathcal{H}_0(\Delta)+V_t(\mathbf{h}_t)+V_l(h_z),
\end{equation}
\begin{align}\label{Vt}
&{V}_t(\mathbf{h}_t)=-\sum_{j=-\infty}^\infty\!\!\left[
h_{2}\sigma_j^x
+(-1)^j h_{1}\sigma_j^y
\right],\\\label{Vl}
&V_l(h_z)=-h_z\sum_{j=-\infty}^\infty (-1)^j\sigma_j^z.
\end{align}
where $\mathbf{h}_t=h_2 \mathbf{e}_x+h_1 \mathbf{e}_y$.

To conclude this section, we  describe, following  essentially 
Lukyanov and Terras   \cite{LukTer03}, the set of the  discrete-symmetry operators, which will be important for the subsequent analysis.

 The Hamiltonians \eqref{Ham} and \eqref{Hama} act in the vector space $\mathcal{L}=\otimes_{j=-\infty}^\infty\mathbb{C}_j^{2}$
 spanned by the basis states
\begin{equation}\label{bas}
|\mathfrak{E}\rangle=\otimes_{j=-\infty}^\infty e_{j,s_j}=\ldots\otimes e_{-1,s_{-1}}
\otimes e_{0,s_0}\otimes e_{1,s_1}\otimes \ldots,
\end{equation}
with $s_j=\pm1$, such that
\[
\sigma_j^z |\mathfrak{E}\rangle =s_j |\mathfrak{E}\rangle.
\]

The discrete-symmetry  operators are defined by their action on the basis vectors $|\mathfrak{E}\rangle$ as follows.
\begin{enumerate}
\item The  translation (shift) operator by one chain site $T_1$: 
\begin{equation}\label{Tra1}
T_1|\mathfrak{E}\rangle=\otimes_{j=-\infty}^\infty e_{j,s_{j+1}}.
\end{equation}
\item
`Charge conjugation operators' $\mathbb{C}_\mathfrak{a}=\otimes_{j=-\infty}^\infty \sigma_j^{\mathfrak{a}}$,
with $\mathfrak{a}=x,y,z$. In particular, the operator $\mathbb{C}_x$ acts on the basis state \eqref{bas} as:
\[
\mathbb{C}_x|\mathfrak{E}\rangle=\otimes_{j=-\infty}^\infty e_{j,-s_{j}}.
\]
\item We shall use two  modified  translation operators by one chain site: 
\begin{align}\label{mTr}
&\widetilde{T}_1=T_1 \mathbb{C}_x,\\
\label{brTr}
&\breve{T}_1=T_1 \mathbb{C}_y.
\end{align}
\item The time inversion is the anti-unitary operators $\mathbb{T}$, that  acts trivially on the basis states:
$
\mathbb{T} |\mathfrak{E}\rangle=|\mathfrak{E}\rangle.
$
The following equality 
\begin{equation}
\mathbb{T} (c | \psi\rangle) =c^*  \mathbb{T}| \psi\rangle,
\end{equation}
holds for any  $| \psi\rangle \in \mathcal{L}$, and a complex number $c$.
\item Two spatial reflection operators $\mathbb{P}_{ev}$,  and $\mathbb{P}_{odd}=T_1 \mathbb{P}_{ev}$:
\begin{align}
\mathbb{P}_{odd}|\mathfrak{E}\rangle=\otimes_{j=-\infty}^\infty e_{j,s_{-j}},\\
\mathbb{P}_{ev}|\mathfrak{E}\rangle=\otimes_{j=-\infty}^\infty e_{j,s_{1-j}}.
\end{align}
\end{enumerate}
These operators act on the Pauli matrices as follows:
\begin{subequations}
\begin{align}
&T_1^{-1}\sigma_j^\mathfrak{a} \,T_1=\sigma_{j+1}^\mathfrak{a},\\
&\mathbb{C}_x\sigma_j^\mathfrak{a}\, \mathbb{C}_x=e^{i \pi d_{\mathfrak{a}}}\sigma_j^\mathfrak{a},\\
&\mathbb{C}_y\sigma_j^\mathfrak{a} \,\mathbb{C}_y=e^{i \pi \breve{d}_{\mathfrak{a}}}\sigma_j^\mathfrak{a},\\\label{wTs}
&\widetilde{T}_1^{-1}\sigma_j^\mathfrak{a}\, \widetilde{T}_1=e^{i \pi d_{\mathfrak{a}}}\sigma_{j+1}^\mathfrak{a},\\
&\breve{T}_1^{-1}\sigma_j^\mathfrak{a}\, \breve{T}_1=e^{i \pi \breve{d}_{\mathfrak{a}}}\sigma_{j+1}^\mathfrak{a},\\
&\mathbb{T} \sigma_j^x \,\mathbb{T} =\sigma_{j}^x, \quad
\mathbb{T} \sigma_j^y \mathbb{T} =-\sigma_{j}^y, \quad
\mathbb{T} \sigma_j^z \,\mathbb{T} =\sigma_{j}^z,\\
&\mathbb{P}_{odd}\, \sigma_j^\mathfrak{a}\,\mathbb{P}_{odd}=\sigma_{-j}^\mathfrak{a}, 
\quad \mathbb{P}_{ev} \,\sigma_j^\mathfrak{a}\,\mathbb{P}_{ev}=\sigma_{1-j}^\mathfrak{a},
\end{align}
\end{subequations}
where $d_x=0$,  $d_y=d_z=1$, $\breve{d}_y=0$, and $\breve{d}_x=\breve{d}_z=1$.

The Hamiltonian $\mathcal{H}_0$ of the infinite XXZ spin chain at zero magnetic field commutes 
with the operator $S^z$, and with all discrete symmetry operators listed above. In the antiferromagnetic phase $\Delta<-1$,  some of these symmetries are spontaneously broken, 
and the two ground states $|vac\rangle^{(1)}$, $|vac\rangle^{(0)}$ of the Hamiltonian $\mathcal{H}_0$ have the following properties:
\begin{subequations}\label{Simm}
\begin{align} \label{vac}
&\mathcal{H}_0|vac\rangle^{(\mu)}=E_{vac}^{(0)}|vac\rangle^{(\mu)},
\\\label{svac}
&\phantom{.}^{{(1)}}\!\langle vac| \sigma_0^z|vac\rangle^{(1)}=\bar{\sigma}=-\phantom{.}^{{(0)}}\!\langle vac| \sigma_0^z|vac\rangle^{(0)},\\
&T_1|vac\rangle^{(\mu)}=|vac\rangle^{(1-\mu)}, \\
&\mathbb{C}_x|vac\rangle^{(\mu)}=|vac\rangle^{(1-\mu)}, \\
&\mathbb{P}_{ev} |vac\rangle^{(\mu)}=|vac\rangle^{(1-\mu)},\\\label{CT}
&\mathbb{C}_x\mathbb{T}|vac\rangle^{(\mu)}=|vac\rangle^{(1-\mu)}, \\
&\widetilde{T}_1|vac\rangle^{(\mu)}=|vac\rangle^{(\mu)},\\
&\mathbb{T}|vac\rangle^{(\mu)}=|vac\rangle^{(\mu)}, \\
&\mathbb{P}_{odd} |vac\rangle^{(\mu)}=|vac\rangle^{(\mu)}, 
\end{align} 
\end{subequations}
where $\mu=0,1$.
The ground state energy $E_{vac}^{(0)}$ of the Hamiltonian $\mathcal{H}_0$ is proportional to the number 
of sites in the spin chain, which becomes infinite in the thermodynamic limit.

\section{Ising limit $\Delta\to-\infty$ \label{sec:Is}}
In this section we describe the perturbative calculation of the meson energy spectra in the confinement regime
 for the model defined by the 
Hamiltonian \eqref{Hama} in the strong-anisotropy (Ising) limit 
$-\Delta\gg1$  to  linear order in $|\Delta|^{-1}$. 
To this end, we use the strong-coupling expansion method developed by Ishimura and Shiba \cite{Shiba_80}.
Though the strong-anisotropy condition $|\Delta|^{-1}\ll1$ is not satisfied in real quasi-1D antiferromagnetic crystals, the results obtained
by means of the strong-coupling expansion provide a useful insight into the qualitative picture of the spinon confinement 
tuned by  transverse magnetic field.

In the Ising limit $\Delta\to-\infty$, it is convenient to rescale the Hamiltonian \eqref{Hama} and to add to it a suitable 
(infinite in the thermodynamic limit)  constant:
\begin{align}\label{HI}
\mathcal{H}_I(\varepsilon, \mathbf{h}_t,h_z)=|\Delta|^{-1}\,\mathcal{H}(\Delta, \mathbf{h}_t,h_z)+Const\\
=\mathcal{H}_I^{(0)}+\varepsilon V_I,\nonumber
\end{align}
where 
\begin{align}\label{HI0}
&\mathcal{H}_I^{(0)}=\frac{1}{2}\sum_{j=-\infty}^\infty
(\sigma_j^z\sigma_{j+1}^z+1),\\
&V_I=- \sum_{j=-\infty}^\infty(\sigma_j^+\sigma_{j+1}^-+\sigma_j^-\sigma_{j+1}^+)\\
&-\sum_{j=-\infty}^\infty\left[ h_2\, \sigma_j^x+h_1(-1)^j \,\sigma_j^y
\right]\nonumber\\\nonumber
&-h_z\sum_{j=-\infty}^\infty\left[
(-1)^j \,\sigma_j^z-1
\right].
\end{align}
Here $\varepsilon=|\Delta|^{-1}$ is the small parameter, and 
$\sigma_j^\pm= \frac{1}{2}(\sigma_j^x\pm i\sigma_j^y)$.

At $\varepsilon=0$, the Hamiltonian \eqref{HI} has two Ne\'el vacua $|0\rangle^{(\mu)}$, $\mu=0,1$, with zero energy: 
\begin{equation}
\mathcal{H}_I^{(0)}|0\rangle^{(\mu)}=0, \quad \mu=0,1,
\end{equation}
where
\begin{subequations}
\begin{align}
|0\rangle^{(1)}:  \quad \ldots\downarrow{\color{blue}\
\underline{\stackrel{0}{\uparrow}}}\stackrel{1}{\downarrow}\stackrel{2}{\uparrow}\downarrow\ldots,\\
|0\rangle^{(0)}: \quad \ldots\uparrow{\color{blue}\underline{\stackrel{0}{\downarrow}}}\stackrel{1}{\uparrow}\stackrel{2}{\downarrow}\uparrow\ldots.
\end{align}
\end{subequations}

\subsection{One-kink sector}
Let us consider the localized kink states $|\mathbf{K}_{\mu \nu}(j)\rangle$, which interpolate
between  vacua $|0\rangle^{(\mu)}$ to the left, and $|0\rangle^{(\nu)}$ to the right of the bond $( j,j+1)$. 
For example, the state $|\mathbf{K}_{10}(2)\rangle$ looks like as follows:
    \begin{equation}
|\mathbf{K}_{10}(2)\rangle:\quad \ldots\downarrow{\color{blue}\
\underline{\stackrel{0}{\uparrow}}}\stackrel{1}{\downarrow}
{\color{dgreen} {\stackrel{2}{\uparrow}}}\,\,{\color{red}\mid}\,{\color{dgreen}{\stackrel{3}{\uparrow}}
}\downarrow\,\uparrow\ldots.
\end{equation}
The  states $|\mathbf{K}_{\mu \nu}(j)\rangle$ are the eigenvectors of the zero-order Hamiltonian \eqref{HI0}, which are characterized by the same (unit) eigenvalue:
\begin{equation}
\mathcal{H}_I^{(0)}|\mathbf{K}_{\mu\nu}(j)\rangle=|\mathbf{K}_{\mu\nu}(j)\rangle,
\end{equation}
and  are normalized by the 
condition
\begin{equation}\label{nc}
\langle \mathbf{K}_{\nu\mu}(j) |\mathbf{K}_{\mu'\nu'}(j')\rangle=\delta_{\mu\mu'}\delta_{\nu\nu'}\delta_{jj'}.
\end{equation}
They transform  under the action of the operators $T_1$, $\mathbb{C}_x$, and $\widetilde{T}_1$ in the  following way:
\begin{eqnarray}
&&{T}_1 |\mathbf{K}_{\mu\nu}(j)\rangle=|\mathbf{K}_{\nu\mu}(j-1)\rangle,\\
&&\mathbb{C}_x  |\mathbf{K}_{\mu\nu}(j)\rangle=|\mathbf{K}_{\nu\mu}(j)\rangle,\\
&&\widetilde{T}_1 |\mathbf{K}_{\mu\nu}(j)\rangle=|\mathbf{K}_{\mu\nu}(j-1)\rangle.\label{T1K1}
\end{eqnarray}

Denote by $\mathcal{P}^{(1)}$ the projection operator onto the subspace $\mathcal{L}^{(1)}$  of the  one-kink states, and by 
$\mathcal{H}_1(\varepsilon, \mathbf{h}_t,h_z)$ the restriction of the Hamiltonian $ \mathcal{H}_I(\varepsilon, \mathbf{h}_t,h_z)$ to 
$\mathcal{L}^{(1)}$:
\begin{equation}\label{H1I}
\mathcal{H}_1(\varepsilon, \mathbf{h}_t,h_z)=\mathcal{P}^{(1)} \mathcal{H}_I(\varepsilon, \mathbf{h}_t,h_z) \mathcal{P}^{(1)}.
\end{equation}
The Hamiltonian  \eqref{H1I} acts at $h_z=0$ on the basis  localized kink states $|\mathbf{K}_{\mu\nu}(j)\rangle$ as follows:
\begin{align}
&\mathcal{H}_1(\varepsilon, \mathbf{h}_t,0)|\mathbf{K}_{\mu\nu}(j)\rangle=|\mathbf{K}_{\mu\nu}(j)\rangle\\\nonumber
&-
\varepsilon[|\mathbf{K}_{\mu\nu}(j+2)\rangle+|\mathbf{K}_{\mu\nu}(j-2)\rangle]\\
&-
\varepsilon h_2[|\mathbf{K}_{\mu\nu}(j+1)\rangle+|\mathbf{K}_{\mu\nu}(j-1)\rangle]\nonumber\\\nonumber
&-i(-1)^{\mu}\varepsilon h_1[|\mathbf{K}_{\mu\nu}(j+1)\rangle-|\mathbf{K}_{\mu\nu}(j-1)\rangle].
\end{align}
The one-kink Bloch states
\begin{equation}
|K_{\mu\nu}^{I}(p)\rangle=e^{ip}\sum_{j=-\infty}^\infty e^{i j p}|\mathbf{K}_{\mu\nu}(j)\rangle
\end{equation}
with $p\in (-\pi,\pi)$ diagonalize the Hamiltonian $\mathcal{H}_1(\varepsilon, \mathbf{h}_t,0)$ and the modified translation operator 
$\widetilde{T}_1$:
\begin{align}\label{eigH}
\mathcal{H}_1\Big|_{h_z=0} |K_{\mu\nu}^{I}(p)\rangle=\omega_{\mu\nu}(\varepsilon,p,\mathbf{h}_t)|K_{\mu\nu}^{I}(p)\rangle,\\
\widetilde{T}_1|K_{\mu\nu}^{I}(p)\rangle=e^{ip}|K_{\mu\nu}^{I}(p)\rangle.
\end{align}
where
\begin{align}\label{omA}
\omega_{\mu\nu}(\varepsilon,p,\mathbf{h}_t)=1-2 \varepsilon \cos (2p)\\\nonumber
 -2 \varepsilon [h_2 \cos p+
(-1)^\mu h_1 \sin p] 
\end{align}
is the kink dispersion law.
\subsection{Two-kink sector}
The two-kink subspace $\mathcal{L}^{(2)}$ is spanned by the basis of 
localized states $|\mathbf{K}_{\mu\nu}(j_1)\mathbf{K}_{\nu\mu}(j_2)\rangle$, 
with $j_1<j_2$.
Denote by $\mathcal{P}^{(2)}$ the projection operator onto this subspace, and by $\mathcal{H}_2(\varepsilon,  \mathbf{h}_t,h_z)$ 
the restriction of the Hamiltonian to $\mathcal{L}^{(2)}$:
\begin{equation}\label{H2I}
\mathcal{H}_2(\varepsilon,  \mathbf{h}_t,h_z)=\mathcal{P}^{(2)}\mathcal{H}_I(\varepsilon,  \mathbf{h}_t,h_z) \mathcal{P}^{(2)}.
\end{equation}

Let us define the Bloch state $|\Psi(P)\rangle$ in the subspace $\mathcal{L}^{(2)}$ as
follows:
\begin{equation}\label{psP}
|\Psi(P)\rangle=\sum_{j_1=-\infty}^\infty e^{i P j_1}\sum_{j=1}^\infty e^{i P j/2}\psi(j )|\mathbf{K}_{10}(j_1)\mathbf{K}_{01}(j_1+j)\rangle.
\end{equation}
Due to \eqref{T1K1}, it satisfies equation 
\begin{equation}
\widetilde{T}_1|\Psi(P)\rangle=e^{iP} |\Psi(P)\rangle.
\end{equation}
We require also, that the Bloch state $|\Psi(P)\rangle$ is the eigenstate of the reduced Hamiltonian \eqref{H2I}:
\begin{equation}\label{H2P}
\mathcal{H}_2(\varepsilon, \mathbf{h}_1,h_z)|\Psi(P)\rangle=E(P) |\Psi(P)\rangle,
\end{equation}
with the eigenvalue 
\begin{equation}\label{La}
E(P)=2+\varepsilon\,\Lambda(P).
\end{equation}
Due to \eqref{psP}-\eqref{La} the wave function $\psi(j )$ must satisfy the  fourth order linear
difference equation at $j\in \mathbb{N}$:
\begin{align}\label{SL}
(2 h_z j -\Lambda) \psi(j )- 2 \cos P \,\,[\psi(j+2 )+\psi(j-2 )]\\\nonumber
-2 h_2 \cos \frac{P}{2}\,[\psi(j+1 )+\psi(j-1 )]\\
+2 i h_1 \cos \frac{P}{2}\,\,[\psi(j+1 )-\psi(j-1 )]=0. \nonumber
\end{align}
Its solution must satisfy the Dirichlet boundary condition at the 
left boundary:
\begin{equation}\label{DBC}
\psi(-1 )=\psi(0 )=0, 
\end{equation}
and vanish at $j\to+ \infty$. 

For the  theory of  linear difference equations see the monograph by G. Teschl \cite{teschl2000jacobi}.
\subsubsection{Exact solution of the discrete Sturm-Liouville problem}
The discrete Sturm-Liouville problem \eqref{SL}, \eqref{DBC} can be solved exactly. 
 Indeed, let us define the generating function $\phi(z)$ of the complex variable $z$
\begin{equation}\label{GF}
\phi(z)=\sum_{j=1}^\infty \psi(j) z^j. 
\end{equation}
The Taylor series \eqref{GF} must converge at $|z|\le 1$.

\begin{figure}
\centering
\subfloat{
\includegraphics[width=.75\linewidth]{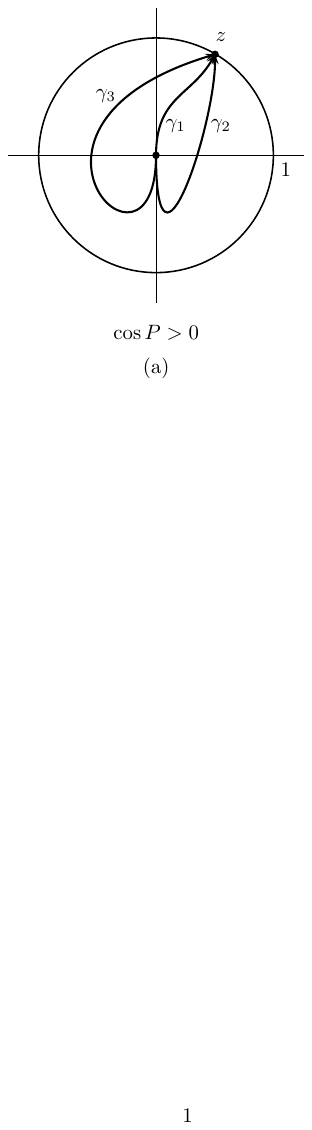}}

\subfloat{
\includegraphics[width=.75\linewidth]{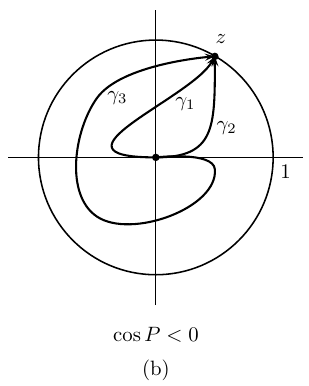}}

\caption{Integration paths $\gamma_1$,  $\gamma_2$, and $\gamma_3$ in  \eqref{phz} 
connecting the points $w=0$ and $w=z$ in
the $w$-complex plane: (a) for $\cos P>0$, (b) for  $\cos P<0$. \label{fig1}}
\end{figure}

Equations \eqref{SL}, \eqref{DBC} lead to the following first-order linear ordinary differential equation for the 
generating function $\phi(z)$:
\begin{align}\label{OD}
(-\Lambda+2 h_z z\partial_z) \phi(z)+ \epsilon_I(z|P)\phi(z)\\\nonumber
=-2 \cos P\,\, [\psi(1) z^{-1}+\psi(2)]\\\nonumber
- 2 (h_2-i h_1) \, \psi(1)\cos \frac{P}{2},
\end{align}
where
\begin{align}\label{ep}
 \epsilon_I(z|P)=-2(z^2+z^{-2}) \cos P\\\nonumber
 -2 [h_2 (z+z^{-1})+ i h_1 (z-z^{-1}) ] \cos \frac{P}{2} ,
\end{align}
and 
\begin{equation}
\psi(1)=\phi'(0), \quad  \psi(2)=\frac{\phi''(0)}{2}.
\end{equation}
Besides, the generating function $\phi(z)$ must vanish at the origin  due to \eqref{DBC}: $\phi(0)=0$. 
The appropriate partial solution of the differential equation \eqref{OD} 
reads:
\begin{align}\label{phz}
&\phi(z)=- \frac{1}{h_z}\int_0^z\frac{dw}{w}
\exp\left\{
\frac{i}{2 h_z}[\mathcal{F}_I(w|\Lambda)-\mathcal{F}_I(z|\Lambda)]\right\}\\\nonumber
&\times
\left[
\psi(1) \left( (h_2-i h_1)\, \cos \frac{P}{2}+\frac{\cos P}{w}  \right)+\psi(2) \cos P
\right],
\end{align}
where
\begin{align}\label{FF}
\mathcal{F}_I(z|\Lambda)=i \bigg[\Lambda \ln z+
(z^2-z^{-2}) \cos P\\\nonumber
+2 h_2(z-z^{-1}) \cos \frac{P}{2}+2 i h_1(z+z^{-1}) \cos \frac{P}{2}
\bigg].
\end{align}
Note, that 
\begin{equation}\label{eps}
i z\mathcal{F}_I'(z|\Lambda)=\epsilon_I(z|P)-\Lambda.
\end{equation}

\begin{figure}
\centering
\subfloat{
\includegraphics[width=.75\linewidth]{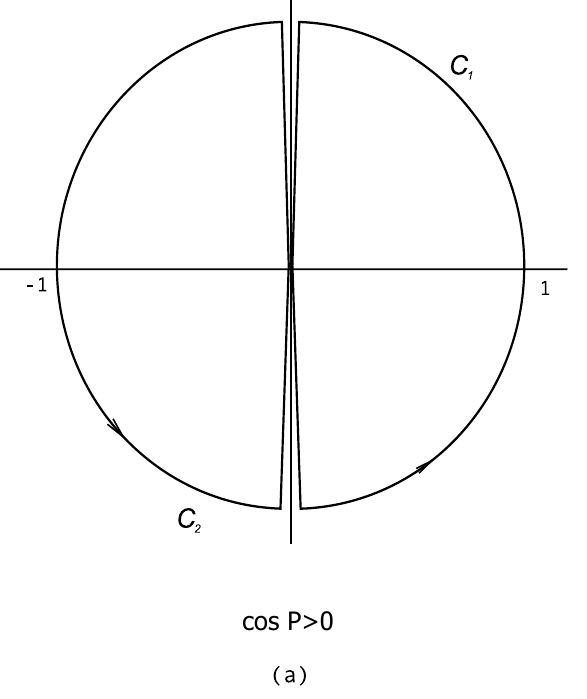}}

\subfloat{
\includegraphics[width=.75\linewidth]{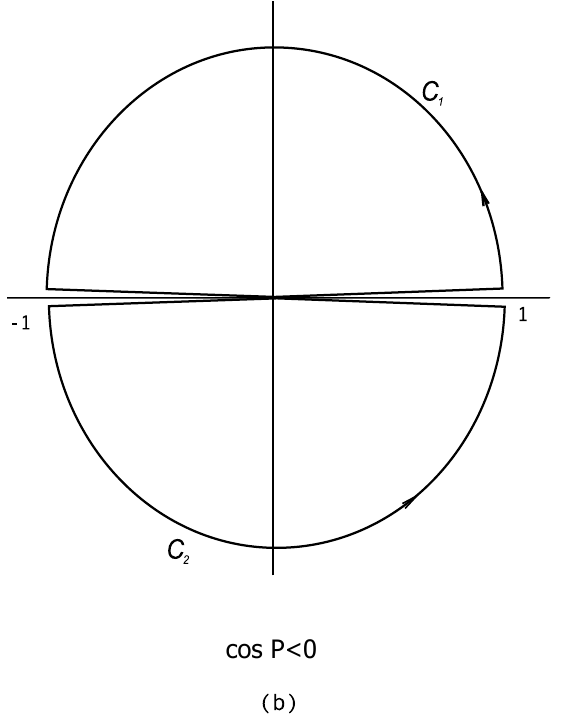}}

\caption{Integration contours $C_1$ and $C_2$ in \eqref{InPs}, \eqref{aver}: (a) for $\cos P>0$, (b) for  $\cos P<0$. \label{InP}}
\end{figure}

The integrand in the integral in the right-hand side of \eqref{phz} has the essential singularity at $w=0$, that arises from the 
second order pole of the function $\mathcal{F}_I(w|\Lambda)$ determined by \eqref{FF}:
\begin{equation}
\mathcal{F}_I(w|\Lambda)=-i \frac{\cos P}{w^2}+O(w^{-1}).
\end{equation}
The integral in \eqref{phz} converges, if  the integration path  approaches the origin $w=0$ along the 
line $\mathrm{Re}\,\frac{\cos P}{w^2}<0$. Depending on the sign of $\cos P$, the appropriate allowed integration path in
 \eqref{phz} approaches the origin either along the imaginary axis for $\cos P>0$, or along the real axis for $\cos P<0$, 
see Figure \ref{fig1}. In both cases, the integrals in the right-hand side of  \eqref{phz} performed along the 
topologically non-equivalent paths 
$\gamma_1$, $\gamma_2$ and $\gamma_3$ shown in Figure~\ref{fig1} must give the same result. This leads to two constraints
\begin{align}\label{InPs}
&\int_{C_j}\frac{dw}{w}\exp\left\{
\frac{i}{2 h_z}[\mathcal{F}_I(w|\Lambda)\right\}\\\nonumber
&\times\left[
\psi(1) \left( (h_2-i h_1)\, \cos \frac{P}{2}+\frac{\cos P}{w}  \right)+\psi(2) \cos P
\right]=0,
\end{align}
where $j=1,2,$ and the integration contours $C_1=\gamma_2-\gamma_1$, $C_2=\gamma_1-\gamma_3$ 
are shown in Figure '\ref{InP}.

Let us introduce notations:  $f_1(w)=1$, $f_2(w)=w^{-1}$,
\begin{equation}\label{aver}
\langle\ldots \rangle_j=\int_{C_j}\frac{dw}{w} \ldots \,\exp\left\{
\frac{i}{2 h_z}[\mathcal{F}_I(w|\Lambda)]
\right\},
\end{equation}
with $j=1,2$, and  
\begin{equation}\label{Wij}
W_{ji}=\langle f_i\rangle_j .
\end{equation}
Then, constraints \eqref{InPs} can be written as  a system of two linear uniform equations on the parameters
$Y_1$ and $Y_2$:
\begin{align}
W_{11} Y_1+W_{12} Y_2=0,\\\nonumber
W_{21} Y_1+W_{22} Y_2=0,
\end{align}
where 
\begin{align}
&Y_1=\psi(1)  (h_2-i h_1)\, \cos \frac{P}{2}+\psi(2) \cos P,\\\nonumber
&Y_2=\psi(1) \cos P.
\end{align}
This system has non-trivial solutions provided the following equality holds: 
 \begin{equation}\label{Kr}
W_{11}W_{22}-W_{12}W_{21}=0. 
\end{equation}
For given values of  $h_1,h_2,h_z$, and $P$,  the solutions of the 
transcendent equation  \eqref{Kr} on the parameter $\Lambda$ determine the discrete spectrum 
$\{\Lambda_n\}_{n=1}^\infty$, of the  Sturm-Liouville problem \eqref{SL}, \eqref{DBC}.
This completes calculation of the meson energy spectra for the model \eqref{Hama} in the Ising limit $\Delta\to-\infty$
to the linear order in the small parameter $\varepsilon =1/|\Delta|$:
\begin{equation}\label{EnergyE}
E_n(P,\Delta, \mathbf{h}_t,h_z)=2+\varepsilon \Lambda_n(P, \mathbf{h}_t,h_z)+O(\varepsilon^2).
\end{equation}

\subsubsection{Limit $\mathbf{h}_t=0$}
In the limit $\mathbf{h}_t=0$,  the reduction of  the function \eqref{FF}  to the form
\begin{equation}\label{FF0}
\mathcal{F}_I(z|\Lambda)\Big|_{\mathbf{h}_t=0}=i \left[
\Lambda \ln z+(z^2-z^{-2}) \cos P \right]
\end{equation}
leads to the following equalities:
\[
W_{21}=W_{11} \exp\left(-
\frac{i \Lambda \pi}{2 h_z}
\right), \quad 
W_{12}=-W_{22} \exp\left(
\frac{i \Lambda \pi}{2 h_z}
\right),
\]
and equation \eqref{Kr} simplifies to:
 \begin{equation}\label{Kr1}
2 \,W_{11}W_{22}=0. 
\end{equation}

Furthermore, the integrals $W_{11}$, $W_{22}$ admit at $\mathbf{h}_t=0$ the explicit representations in terms of the
Bessel function:
\begin{align}
\frac{W_{11}}{\pi i}=\begin{cases} \,J_\nu \left(\frac{\cos P}{h_z}\right), & \text{for }\cos P>0,\\
 \exp\left(
-\frac{i \pi \Lambda}{4 h_z}
\right) J_\nu \left(\frac{-\cos P}{h_z}\right),& \text{for } \cos P<0,
\end{cases}\\
\frac{W_{22}}{\pi i}=\begin{cases} - \exp\left(
-\frac{i \pi \Lambda}{2 h_z}
\right) 
J_{\nu-1/2} \left(\frac{\cos P}{h_z}\right), & \text{for }\cos P>0,\\
i \exp\left(
\frac{i \pi \Lambda}{4 h_z}
\right) J_{\nu-1/2} \left(\frac{-\cos P}{h_z}\right),& \text{for } \cos P<0,
\end{cases}
\end{align}
where $\nu=-\frac{\Lambda}{4 h_z}$. Accordingly, the dispersion laws of the meson states at $\mathbf{h}_t=0$ in the 
Ising limit $\Delta\to-\infty$ are determined by the solutions of the equation:
\begin{align}
J_{-\frac{\Lambda}{4 h_z}} \left(\frac{|\cos P|}{h_z}\right)\,\,J_{-\frac{1}{2}-\frac{\Lambda}{4 h_z}} \left(\frac{|\cos P|}{h_z}\right)=0
\end{align}
in agreement with \cite{Rut18}. It was shown in \cite{Rut18}, that the dispersion laws of the mesons with zero $z$-projection 
$s=0$ of the spin are
determined by solutions of equation 
\begin{equation}
J_{-\frac{\Lambda}{4 h_z}} \left(\frac{|\cos P|}{h_z}\right)=0,
\end{equation}
while the energies of the mesons with $s=\pm 1$   is determined by solutions of equation 
\begin{equation}
J_{-\frac{1}{2}-\frac{\Lambda}{4 h_z}} \left(\frac{|\cos P|}{h_z}\right)=0.
\end{equation}
\subsubsection{Meson energy spectra in the Ising limit at $\mathbf{h}_t\ne0$}
Application of the transverse magnetic field $\mathbf{h}_t$ breaks conservation of the
$z$-projection of the total spin and leads to the hybridization of the meson modes with $s=0$, and $s=\pm 1$.
\begin{figure}
\includegraphics[width=1\linewidth]{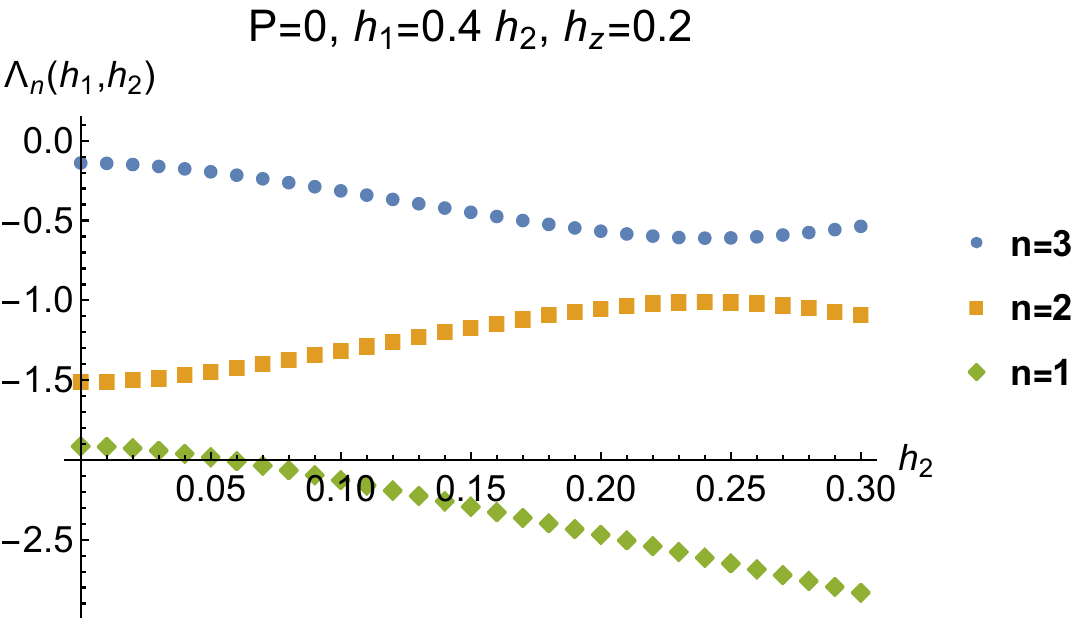}
\caption{The  energies  of three lightest mesons versus $h_2$ due to \eqref{Kr} 
and \eqref{EnergyE} at $h_1=0.4\,h_2$ 
and  fixed $P=0$, and $h_z=0.2$.  
 \label{3modes}}
\end{figure}

Obtained results are illustrated in Figure \ref{3modes}, which displays the evolution of 
 the energies  of three lightest meson modes with increasing transverse magnetic field  $h_2$ at the fixed ratio $h_1/h_2=0.4$, and fixed $P=0$ and $h_z=0.2$. 
Hybridizaton of the first longitudinal and the second transverse modes leads to the avoided crossing 
of their dispersion curves, that takes place at $h_2\approx 0.25$ for the chosen values of other parameters.

The described above qualitative  evolution of the meson energies with increasing 
 transverse magnetic field was indeed observed by Faure {\it et al.} \cite{Faur17} in the inelastic neutron scattering experiments in 
 the crystal $\mathrm{BaCo}_{2}\mathrm{V}_{2}\mathrm{O}_{8}$, see Figure  3 in \cite{Faur17}. 
However, Faure  {\it et al.} give the value $\varepsilon=0.53$ for the inverse anisotropy parameter $\varepsilon=|\Delta|^{-1}$ in this crystal, which is far from the strong anisotropic regime $\varepsilon\ll1$. Therefore,  the results obtained above in the limit $\varepsilon\to0$ cannot describe  quantitatively  the experimentally relevant regime.

In subsequent Secs. \ref{Sect:Gr}-\ref{Heur}, we will present the alternative perturbative scheme, which is free from the above 
shortcoming. It applies to the whole interval  of the anisotropy constant $\Delta<-1$, and exploits the staggered longitudinal 
magnetic field $h_z$ as a small parameter. In the more simple case of zero transverse magnetic field, this perturbative technique  was
already used for calculation of the meson dispersion laws in the XXZ spin chain in papers \cite{Rut18,Rut22}. However, 
at $\mathbf{h}_t=0$, one could benefit from the fact, that the XXZ model at $h_z=0$ is integrable. This is not the case anymore
in  presence  of the transverse  magnetic field $\mathbf{h}_t\ne0$. To overcome this difficulty, we perform the perturbative
calculations in two steps, as it was noticed in the Introduction. First, in Secs. \ref{Sect:Gr}-\ref{two-k} we concentrate on the deconfinement regime at $h_z=0$,
 and determine  the  deformations of the 
antiferromagnetic vacua, one-, and two-kink excitations by the weak transverse magnetic field having both uniform and staggered 
components. Then  the weak staggered longitudinal field $h_z$ is switched on inducing  the kink confinement. The energies 
of their bound states in this regime are calculated in Secs. \ref{sec_Mes},\ref{Heur} following the strategy developed in \cite{Rut18,Rut22}.
\section{Ground-state energy at $h_z=0$ \label{Sect:Gr}}
Let us return to the Hamiltonian \eqref{Ham}, and put in it $h_{1z}=h_{2z}=0$.
At non-zero $h_{ix}, h_{iy}$, $i=1,2$, the interaction term $V$ given by \eqref{V} does not commute with the 
total spin operator $S^z$, and with all listed in \eqref{Simm} discrete symmetry operators, except of $\mathbb{P}_{odd}$, and  $\mathbb{C}_x\mathbb{T}$:
\begin{align}\label{Podd}
\mathbb{P}_{odd}\, V = V\, \mathbb{P}_{odd},\\\label{CTsym}
     \mathbb{C}_x\mathbb{T}\, V= V\,\mathbb{T}\mathbb{C}_x.
\end{align}
It follows from equations \eqref{CT}  and  \eqref{CTsym}, that the application of the transverse magnetic fields 
$h_{ix}, h_{iy}$, $i=1,2$ does not lift the degeneracy between the two deformed antiferromagnetic vacua $|Vac(\mathfrak{h}_t)\rangle^{(1)}$ 
and $|Vac(\mathfrak{h}_t)\rangle^{(0)}$,
\begin{align*} 
&(\mathcal{H}_0 +V)|Vac(\mathfrak{h}_t)\rangle^{(1)}=E_{vac}(\mathfrak{h}_t)|Vac(\mathfrak{h}_t)\rangle^{(1)}, \\
 &(\mathcal{H}_0 +V)|Vac(\mathfrak{h}_t)\rangle^{(0)}=E_{vac}(\mathfrak{h}_t)|Vac(\mathfrak{h}_t)\rangle^{(0)},\\
 & \mathbb{C}_x\mathbb{T} |Vac(\mathfrak{h}_t)\rangle^{(\mu)}=|Vac(\mathfrak{h}_t)\rangle^{(1-\mu)}, \;\mu=0,1,
\end{align*} 
where $\mathfrak{h}_t=\langle h_{1x},h_{2x},h_{1y},h_{2y}\rangle$.

The ground-state energy $E_{vac}(\mathfrak{h}_t)$ admits the Rayleigh-Schr\"odinger expansion in the components
of the transverse magnetic field. The leading correction to the ground state energy $E_{vac}^{(0)}$ is of the second 
 order. In the two-kink approximation, it can be written as \cite{LL3}:
\begin{align} \nonumber
&{E_{vac}^{(2)}(\mathfrak{h}_t)}=-\sum_{s=\pm1/2 }\int_{-\pi/2}^{\pi/2} \frac{dp_1}{\pi}\int_{-\pi/2}^{p_1} \frac{dp_2}{\pi} \frac{1}{\omega(p_1)+\omega(p_2)} \\\label{E2}
&\phantom{.}^{(1)} \langle  vac|V| K_{10}(p_1)K_{01}(p_2)\rangle_{ss}\\\nonumber
&\times \phantom{.}_{ss} \langle K_{10}(p_2)K_{01}(p_1)|V|vac\rangle^{(1)}.
\end{align} 
Here $| K_{10}(p_1)K_{01}(p_2)\rangle_{s_1s_2}$ denotes the two-kink Bloch state characterized by the quasimomenta $p_1,p_2$ and 
spins $s_1,s_2$ of two kinks. Some properties of these states are collected in Appendix \ref{Ap1}. The kink dispersion law
$\omega(p)$  is explicitly known due to Johnson, Krinsky, and McCoy \cite{McCoy73}:
\begin{equation}\label{dl}
\omega({p},\eta)=I \,\sqrt{1-k^2 \cos^2 {p}},
\end{equation}
where
\begin{equation}\label{Iet}
I=\frac{2  K}{\pi}\sinh \eta,
\end{equation}
and $K$ [$K'$] is the complete elliptic integral of modulus $k$ [$k'=\sqrt{1-k^2}$] such that 
\begin{equation}\label{KpK}
\frac{K'}{K }= \frac{\eta}{\pi}.
\end{equation}

The perturbing operator $V$ can be represented as
\begin{equation}\label{Vm}
V=\sum_{m=-\infty}^\infty \widetilde{T}_1^{-2m} v_0\,\widetilde{T}_1^{2m}, 
\end{equation}
where
\begin{align*} 
v_0=-[(h_{1x}+h_{2x}
)\sigma_0^x+(h_{1y}+h_{2y})\sigma_0^y]\\
-[(-h_{1x}+h_{2x}
)\sigma_1^x+(-h_{1y}+h_{2y})\sigma_1^y].
\end{align*} 
After substitution of \eqref{Vm} into \eqref{E2} and summation over $m$ using  equality \eqref{tilT},
one obtains in the thermodynamic limit 
\begin{align} \label{E2a}
&\lim_{N\to\infty}\frac{E_{vac}^{(2)}(\mathfrak{h}_t)}{N}=-\frac{1}{2}\sum_{s=\pm1/2 }\int_{-\pi/2}^{\pi/2} \frac{dp}{\pi}\frac{1}{2\omega(p)}\times\\\nonumber
& \phantom{.}^{(1)} \langle vac| v_0| K_{10}(p)K_{01}(-p)\rangle_{ss}\\\nonumber
&\times
 \phantom{.}_{ss} \langle K_{10}(-p)K_{01}(p)|v_0|vac\rangle^{(1)},
\end{align} 
where $N$ is the number of  sites in the spin chain.
The matrix elements of the operator $v_0$ in the right-hand side can be expressed in terms of the two-kink 
form factors $X^{0}(\xi_1,\xi_2), X^{1}(\xi_1,\xi_2)$ given in equations \eqref{XXF} in Appendix \ref{Ap1}:
\begin{align*}
&\phantom{.}^{(1)} \langle vac| v_0| K_{10}(p)K_{01}(-p)\rangle_{-1/2,-1/2}=\\
&-\frac{\sinh \eta}{\omega(p)}\{(h_{1x}-i h_{1y})[X^{1}(\xi,\xi^{-1})-X^{0}(\xi,\xi^{-1})]\\
&+(h_{2x}-i h_{2y})[X^{1}(\xi,\xi^{-1})+X^{0}(\xi,\xi^{-1})]\},\\
&\phantom{.}^{(1)} \langle vac| v_0| K_{10}(p)K_{01}(-p)\rangle_{1/2,1/2}=\\
&\frac{\sinh \eta}{\omega(p)}\{(h_{1x}+i h_{1y})[X^{1}(\xi,\xi^{-1})-X^{0}(\xi,\xi^{-1})]\\
&-(h_{2x}+i h_{2y})[X^{1}(\xi,\xi^{-1})+X^{0}(\xi,\xi^{-1})]\},
\end{align*}
where 
\begin{equation}\label{xip}
\xi:=\xi(p)=- i e^{i\alpha(p)}, 
\end{equation}
and  $\alpha$ is the kink rapidity corresponding to the momentum $p$. Note, that the kink momentum $p$ and energy
$\omega$
can be parametrized in terms the Jacobi elliptic functions of the rapidity variable $\alpha$:
\begin{align}\label{pe}
&{p}(\alpha)=-\frac{\pi}{2}+\mathrm{am}\,\left(\frac{2 K \alpha}{\pi},k\right),\\
&\omega(\alpha)=I \, \mathrm{dn}\,\left(\frac{2 K \alpha}{\pi},k\right)= \sinh\eta\,\, \frac{d p(\alpha)}{d\alpha}.\label{le}
\end{align}

Let us proceed to the polar coordinates in the magnetic field components:
\begin{align}
h_{1x}=h_1 \cos \varphi_1, \quad h_{1y}=h_1 \sin \varphi_1,\\
h_{2x}=h_2 \cos \varphi_2, \quad h_{2y}=h_2 \sin \varphi_2.
\end{align}
The second order correction \eqref{E2a} then can be represented in the form
\begin{equation}
\lim_{N\to \infty}\frac{E_{vac}^{(2)}(\mathfrak{h}_t)}{N}=-\frac{\chi_1}{2} h_1^2-\frac{\chi_2}{2} h_2^2
\end{equation}
where $\chi_1$ and $\chi_2$ are the magnetic  susceptibilities corresponding to the 
staggered and uniform transverse magnetic fields, respectively:
\begin{align}
&\chi_1= \frac{A_+^2}{A_-^2}\, \chi_2,\\
&\chi_2=\frac{\sinh \eta}{\pi}
 \int_{0}^{\pi}\frac{dp}{\omega^2(p)}
|X^1(\xi,\xi^{-1})+X^0(\xi,\xi^{-1})|^2,
\end{align}
with $\xi$ given by \eqref{xip}, and
\begin{align}\label{Apm}
A_+(\eta)=2 \vartheta_4(0|e^{-\eta})\,\vartheta_2(i\eta/\pi|e^{-4\eta}),\\\nonumber
A_-(\eta)=2 \vartheta_3(0|e^{-\eta})\,\vartheta_2(i\eta/\pi|e^{-4\eta}).
\end{align}
Here  $\vartheta_i(u|p)$, with $i=1,2,3,4$,  denotes  the elliptic theta-functions defined by equations \eqref{theta}.

Note, that the ratio ${A_+^2}/{A_-^2}$ is equal to the  complementary elliptic modulus $k'$:
\begin{align*}
\frac{A_+^2}{A_-^2}=\left[\frac{\vartheta_4(0|e^{-\eta})}{\vartheta_3(0|e^{-\eta})}\right]^2=\left[\frac{\vartheta_2(0|e^{-\pi^2/\eta})}{\vartheta_3(0|e^{-\pi^2/\eta})}\right]^2\\\nonumber
=k'(\eta)=\sqrt{1-k(\eta)^2}.
\end{align*}
\section{First order correction to the kink energy at $h_z=0$\label{Sec:kink}}
At zero magnetic field, the  XXZ model is determined by the Hamiltonian 
\eqref{H0}. At $\Delta<-1$, the infinite chain \eqref{H0} has two antiferromagnetic ground states 
$|vac\rangle^{(1)}$, and $|vac\rangle^{(0)}$.
The one-kink subspace $\mathcal{L}^{(1)}$ has two topological sectors $\mathcal{L}^{(1)}_{10}$
and $\mathcal{L}^{(1)}_{01}$, which are spanned by the basis vectors $|K_{10}(p)\rangle_{s}$, and $|K_{01}(p)\rangle_{s}$, 
respectively, with $p\in (0,\pi)$, and $s=\pm1/2$.
The defining equations for these one-kink Bloch states read:
\begin{subequations}\label{kinkBloch}
\begin{align}\label{T1a}
&\widetilde{T}_1 |{K_{\mu\nu}}(p)\rangle_s=e^{ i p}\, |{K_{\mu\nu}}(p)\rangle_{-s},\\
&S^z |{K_{\mu\nu}}(p)\rangle_s= s  |{K_{\mu\nu}}(p)\rangle_s,\\
&(\mathcal{H}_0-E_{vac}^{(0)}) |{K_{\mu\nu}}(p)\rangle_s= \omega(p)
|{K_{\mu\nu}}(p)\rangle_s,\\
& \phantom{.}_{s} \langle K_{\nu\mu}(p)|K_{\mu'\nu'}(p')\rangle_{s'}=\pi \delta_{\mu\mu'} \delta_{\nu\nu'}\delta_{s,s'}\delta(p-p'),
\end{align}
\end{subequations}
where $\omega(p)$ is given by \eqref{dl}.
Note also the relation \cite{Rut22}
\begin{equation}\label{Kpi1}
|{K_{\mu\nu}}(p+\pi)\rangle_s=\varkappa(\mu,s)|{K_{\mu\nu}}(p)\rangle_s,
\end{equation}
where $\varkappa(0,1/2)=\varkappa(1,-1/2)=1$, and $\varkappa(1,1/2)=\varkappa(0,-1/2)=-1$. 
Formula \eqref{Kpi1} allows one to extend the above definition of the one-kink states $|{K_{\mu\nu}}(p)\rangle_s$ 
from the interval $p\in(0,\pi)$ to
the whole real axis of the momentum $p\in \mathbb{R}$.
All one-kink Bloch states $|{K_{\mu\nu}}(p)\rangle_s$ have the same dispersion law \eqref{dl}.

If $h_{1z}=h_{2z}=0$, application of the weak transverse uniform and staggered magnetic fields 
$\mathfrak{h}_t=\langle h_{1x},h_{2x},h_{1y},h_{2y}\rangle$ deform the two antiferromagnetic ground states $|vac\rangle^{(\mu)}$, 
with $\mu=0,1$ into the vacua $|Vac(\mathfrak{h}_t)\rangle^{(\mu)}$, which remain degenerate in energy. 
The lowest in energy excitations form the subspace $\mathcal{L}^{(1)}$, which splits into two sectors   $\mathcal{L}^{(1)}_{10}$
and $\mathcal{L}^{(1)}_{01}$ formed by kink Bloch states  $|\mathbb{K_{\mu\nu}}(p|\mathfrak{h}_t)\rangle_a$, 
interpolating between the two antiferromagnetic vacua. These kink Bloch states are defined as  solutions of the 
eigenvalue problem:
\begin{align}\label{T2}
&T_1^2 |\mathbb{K_{\mu\nu}}(p|\mathfrak{h}_t)\rangle_a=e^{2 i p}\, |\mathbb{K_{\mu\nu}}(p|\mathfrak{h}_t)\rangle_a,\\
&[\mathcal{H}_0+V-E_{vac}(\mathfrak{h}_t)] |\mathbb{K_{\mu\nu}}(p|\mathfrak{h}_t)\rangle_a\\\nonumber
&= \Omega_{\mu\nu}^{(a)}(p|\mathfrak{h}_t)
|\mathbb{K_{\mu\nu}}(p|\mathfrak{h}_t)\rangle_a,
\end{align}
 with $a=1,2$ and $p\in (0,\pi)$. 
 
 Note, that 
 we have used the two-site translation operator $T_1^2$ in \eqref{T2} [instead  of the
 one-site modified translation $\widetilde{T}_1$ in 
 equation \eqref{T1a}] in order 
 to define the kink quasi-momentum $p$. The reason is that the operator $\widetilde{T}_1$ does not commute
 with $V$ for generic values of the transverse magnetic field components at $h_{1z}=h_{2z}=0$.
 
The one-kink  Bloch states $|\mathbb{K_{\mu\nu}}(p|\mathfrak{h}_t)\rangle_a$ admit 
a Taylor expansion in powers of the components 
of the
applied transverse magnetic field. The standard Rayleigh-Schr\"odinger perturbation theory arguments 
\cite{LL3} yield:
\begin{align}
&|\mathbb{K_{\mu\nu}}(p|\mathfrak{h}_t)\rangle_a\\\nonumber
&=U_{a1}|K_{\mu\nu}(p)\rangle_{1/2}
+U_{a2}|K_{\mu\nu}(p)\rangle_{-1/2}+
O(|\mathfrak{h}_t|),
\\
& \Omega_{\mu\nu}^{(a)}(p|\mathfrak{h}_t)=\omega(p)+\delta\Omega_{\mu\nu}^{(a)}(p|\mathfrak{h}_t) +O(|\mathfrak{h}_t|^2),
\end{align}
where $\delta\Omega_{\mu\nu}^{(a)}(p|\mathfrak{h}_t) \sim \mathfrak{h}_t$.
The matrix elements of the $2\times2$ unitary matrix $U_{ab}$ are, generally speaking, 
different for the topological sectors $\mathcal{L}^{(1)}_{10}$
and $\mathcal{L}^{(1)}_{01}$,  depend on the kink momentum $p\in(0,\pi)$,
and on the orientation in the 
$xy$-plane of the applied transverse magnetic fields, but do not depend on $|\mathfrak{h}_t|$.
In order to determine  $\delta\Omega_{\mu\nu}^{(a)}(p)$ and 
 $U_{ab}$, one has to diagonalize the matrix 
 $ \phantom{.}_{s} \langle K_{\nu\mu}(p)|V|K_{\mu\nu}(p')\rangle_{s'}$, which can be written as:
 \begin{align}
 \phantom{.}_{s} \langle K_{\nu\mu}(p)|V|K_{\mu\nu}(p')\rangle_{s'}=\pi \delta(p-p')\mathcal{V}_{ss'}(p),
\end{align}
 where the  Hermitian  $2\times2$ matrix $\mathcal{V}_{ss'}(p)$ is defined by the relation
 \begin{align}\label{Vss}
\mathcal{V}_{ss'}(p)= \phantom{.}_{s} \langle K_{\nu\mu}(p)|v_0|K_{\mu\nu}(p)\rangle_{s'}\\\nonumber
=\frac{\sinh \eta}{\omega(p)} \delta_{-s,s'}\,\phantom{.}_{s} \langle \mathcal{K}_{\nu\mu}(\xi)|v_0|\mathcal{K}_{\mu\nu}(\xi)\rangle_{s'}.
\end{align}
In the second line $ |\mathcal{K}_{\mu\nu}(\xi)\rangle_{s}$ denotes 
 the kink state parametrised by the multiplicative spectral parameter 
$\xi(\alpha)=-i e^{i\alpha}$. This state differs from $|K_{\mu\nu}(p)\rangle_{s}$ by the numerical factor
$\sqrt{p'(\alpha})$:
\begin{equation}
 |\mathcal{K}_{\mu\nu}(\xi)\rangle_{s}=\sqrt{\frac{\omega(p)}{\sinh \eta}}\,\,|K_{\mu\nu}(p)\rangle_{s}.
 \end{equation}
Using \eqref{wTs} and the crossing relation \cite{Jimbo94}
 \begin{align}
 \phantom{.}_{s'} \langle \mathcal{K}_{\nu\mu}(\xi')|\sigma_0^{\mathfrak{a}}|\mathcal{K}_{\mu\nu}(\xi)\rangle_{s}\\\nonumber
 =
 \phantom{.}^{(\nu)} \langle vac |\sigma_0^{\mathfrak{a}}| \mathcal{K}_{\nu\mu}(-q \xi') \mathcal{K}_{\mu\nu}(\xi)\rangle_{-s',s},
\end{align}
with $\mathfrak{a}=x,y,z$, the non-zero matrix elements in the second line of \eqref{Vss} can be expressed in terms of the  two-kink form factors
 $X^0(\xi_1,\xi_2), X^1(\xi_1,\xi_2)$ of the spin operators $\sigma_0^\pm$ described in Appendix \ref{Ap1}:
\begin{align}
\phantom{.}_{-1/2} \langle \mathcal{K}_{01}(\xi)|v_0|\mathcal{K}_{10}(\xi)
\rangle_{1/2}\\\nonumber
=-(h_1 e^{i \varphi_1}+h_2 e^{i \varphi_2})X^1(e^{-\eta}\xi,\xi)\\\nonumber
-(-h_1 e^{i \varphi_1}+h_2 e^{i \varphi_2})X^0(e^{-\eta}\xi,\xi),\\
\phantom{.}_{1/2}\langle \mathcal{K}_{01}(\xi)|v_0|\mathcal{K}_{10}(\xi)
\rangle_{-1/2}\\\nonumber
=-(h_1 e^{-i \varphi_1}+h_2 e^{-i \varphi_2})X^0(e^{-\eta}\xi,\xi)\\\nonumber
-(-h_1 e^{-i \varphi_1}+h_2 e^{-i \varphi_2})X^1(e^{-\eta}\xi,\xi),\\
\phantom{.}_{-1/2} \langle \mathcal{K}_{10}(\xi)|v_0|\mathcal{K}_{01}(\xi)
\rangle_{1/2}\\\nonumber
=-(h_1 e^{i \varphi_1}+h_2 e^{i \varphi_2})X^0(e^{-\eta}\xi,\xi)\\\nonumber
-(-h_1 e^{i \varphi_1}+h_2 e^{i \varphi_2})X^1(e^{-\eta}\xi,\xi),\\
\phantom{.}_{1/2}\langle \mathcal{K}_{10}(\xi)|v_0|\mathcal{K}_{01}(\xi)
\rangle_{-1/2}\\\nonumber
=-(h_1 e^{-i \varphi_1}+h_2 e^{-i \varphi_2})X^1(e^{-\eta}\xi,\xi)\\\nonumber
-(-h_1 e^{-i \varphi_1}+h_2 e^{-i \varphi_2})X^0(e^{-\eta}\xi,\xi).
\end{align}

The following explicit formulas hold for the form factors $X^{0,1}(e^{-\eta}\xi,\xi)$ that stand in the above expressions:
\begin{align}
X^{1}(e^{-\eta}\xi,\xi)+X^{0}(e^{-\eta}\xi,\xi)= A_+ \cos[p(\alpha)],\\
X^{1}(e^{-\eta}\xi,\xi)-X^{0}(e^{-\eta}\xi,\xi)=i  A_- \sin[p(\alpha)],
\end{align}
where $p(\alpha)$ is the kink momentum parametrized by the rapidity variable according to equation \eqref{pe}, 
and  the amplitudes $A_\pm(\eta)$ are given by \eqref{Apm}.

As the result, the matrix elements $\mathcal{V}_{s,s'}(p)$ in the case $\mu=1,\nu=0$ reduce to  the form:
\begin{align*} 
&\mathcal{V}_{1/2,-1/2}(p)= \frac{\sinh \eta}{\omega(p)} [i e^{-i \varphi_1} h_1  A_-\sin p
-e^{-i \varphi_2} h_2 A_+ \cos p],\\
&\mathcal{V}_{-1/2,1/2}(p)= \frac{\sinh \eta}{\omega(p)} [-i e^{i \varphi_1} h_1 A_-\sin p
-e^{i \varphi_2}h_2  A_+ \cos p],\\
&\mathcal{V}_{1/2,1/2}(p)=\mathcal{V}_{-1/2,-1/2}(p)=0.
\end{align*}
In the case $\mu=0,\nu=1$, the matrix elements $\mathcal{V}_{s,s'}(p)$  read instead:
\begin{align*} 
&\mathcal{V}_{1/2,-1/2}(p)= \frac{\sinh \eta}{\omega(p)} [-i e^{-i \varphi_1} h_1  A_-\sin p
-e^{-i \varphi_2} h_2 A_+ \cos p],\\
&\mathcal{V}_{-1/2,1/2}(p)= \frac{\sinh \eta}{\omega(p)} [i e^{i \varphi_1} h_1 A_-\sin p
-e^{i \varphi_2}h_2  A_+ \cos p],\\
&\mathcal{V}_{1/2,1/2}(p)=\mathcal{V}_{-1/2,-1/2}(p)=0.
\end{align*}
 \begin{figure}[htb]
\centering
\includegraphics[width=\linewidth, angle=00]{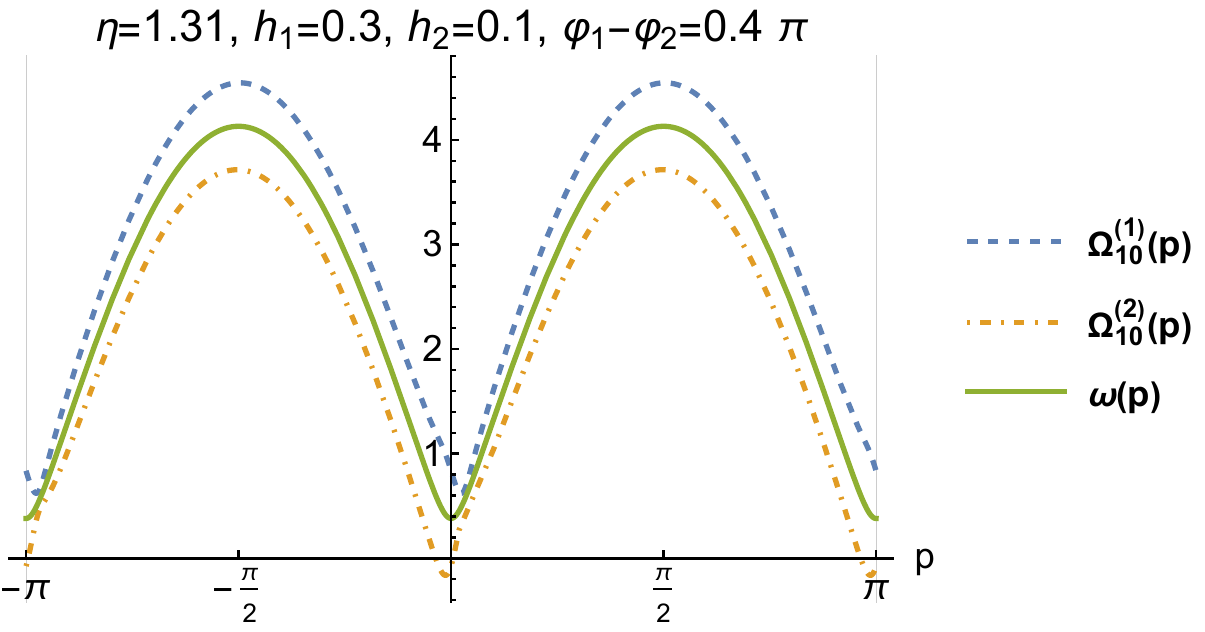}
\caption{\label{fig:dlaw}  Dashed blue and dot-dashed orange lines: dispersion laws \eqref{Om} of kinks perturbed by transverse magnetic fields. Solid green line: kink dispersion law \eqref{dl} at zero magnetic field.} 
\end{figure}

Diagonalization of the matrix $\mathcal{V}_{ss'}(p)$ allows one to determine  the kink 
dispersion law to the first order in $|\mathfrak{h}_t|$:
\begin{align}\label{Om}
&\Omega_{10}^{(a)}(p|\mathfrak{h}_t)=\omega(p)\pm \frac{\sinh \eta}{\omega(p)}\\\nonumber
&\times \big[(h_1 A_- \sin p)^2+ (h_2 A_+ \cos p)^2 \\\nonumber
&-h_1 h_2 A_+A_- \sin(2 p)\sin(\varphi_1-\varphi_2)\big]^{1/2}+O(|\mathfrak{h}_t|^2),\\
&\Omega_{01}^{(a)}(p|\mathfrak{h}_t)=\Omega_{10}^{(a)}(-p|\mathfrak{h}_t),\label{Oma}
\end{align}
with $a=1,2$, and $p\in(0,\pi)$.

 \begin{figure}[htb]
\centering
\subfloat[]
{\label{fig:2a}\includegraphics[width=\linewidth, angle=00]{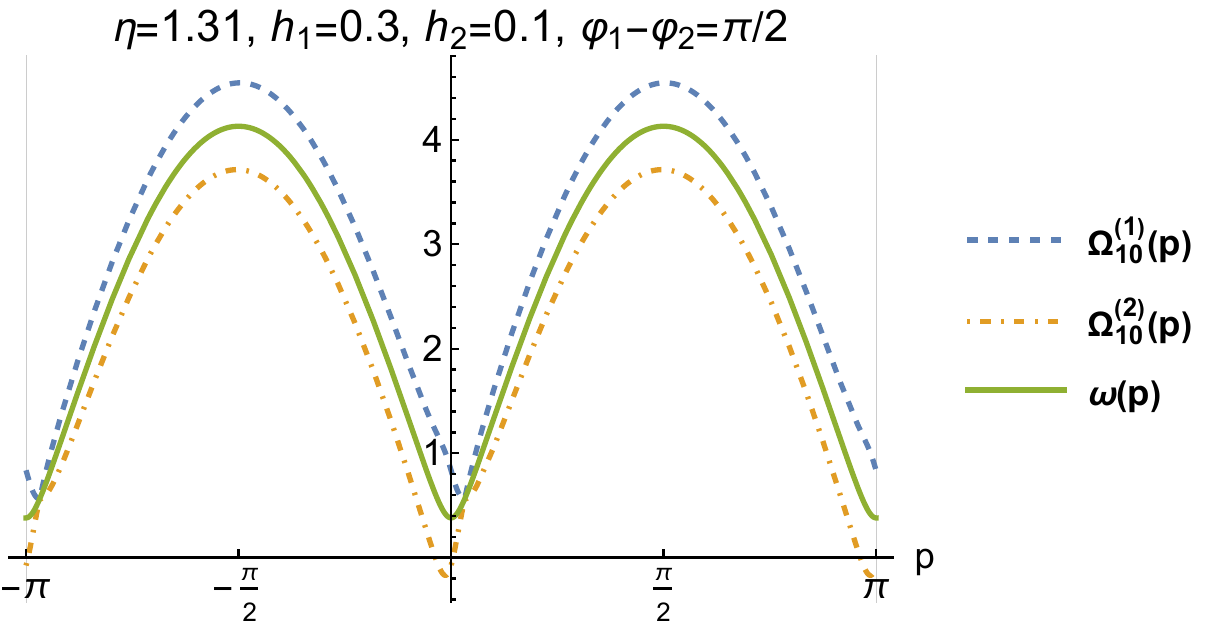}}

\subfloat[]
{\label{fig:2b}\includegraphics[width=\linewidth, angle=00]{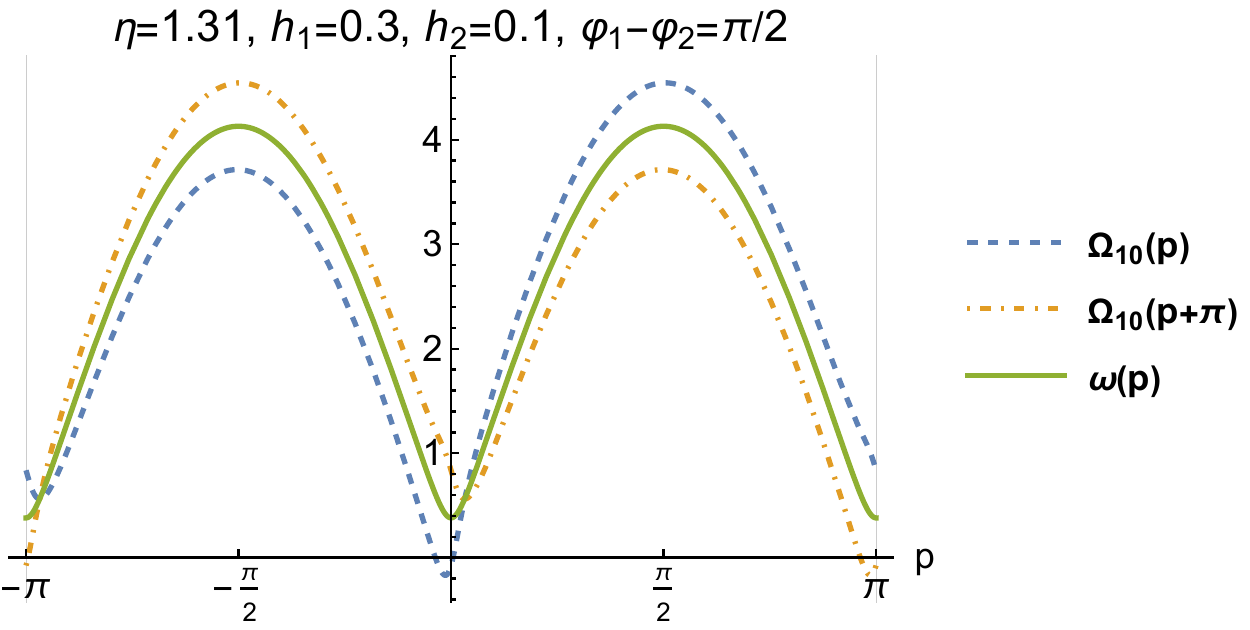}}
\caption{\label{fig:dlaw2}  Dashed blue and dot-dashed orange lines: dispersion laws [\eqref{OM} in (a),  and \eqref{Om2} in (b)] of kinks perturbed by the mutually orthogonal 
transverse staggered and uniform magnetic fields. Solid green line: the kink dispersion law \eqref{dl} at zero magnetic field.} 
\end{figure}

Figure \ref{fig:dlaw} illustrates  the dispersion laws \eqref{Om} of two kink modes 
$\Omega_{10}^{(a)}(p|\mathfrak{h}_t)$, $a=1,2$, in the topological sector $\mathcal{L}_{10}^{(1)}$
at $\eta=1.31$, $h_1=0.3$, $h_2=0.1$, and $\varphi_1-\varphi_2=0.4 \pi$.  These large enough values of the magnetic fields
$h_1$ and $h_2$ have been chosen in order to provide a sufficient separation of three dispersion curves in Figure \ref{fig:dlaw}.
The drawback of this choice is that the kink frequency $\Omega_{10}^{(2)}(p)$ vanishes and becomes negative at certain values
of $p$. This does not happen at small  values of the  transverse magnetic fields required by the applicability of the Rayleigh-Schr\"odinger perturbation  theory.

At generic mutual orientations and  strengths  of the transverse uniform and staggered magnetic fields, the degeneracy between the two kink modes $\Omega_{10}^{(1)}(p)$ and $\Omega_{10}^{(2)}(p)$ is lifted at all $p$.
The situation is different, however, in two cases: (i) if the staggered and uniform transverse magnetic fields are mutually 
orthogonal; (ii)  if either the staggered, or the uniform transverse magnetic field vanishes. 

Let us address the first case, and  put $\varphi_1-\varphi_2=\pi/2$ in equations \eqref{Om}, \eqref{Oma}, which reduce 
then  to the form:
\begin{eqnarray}\label{OM}
&&\Omega_{10}^{(a)}(p|\mathfrak{h}_t)=\omega(p)\pm \frac{\sinh \eta}{\omega(p)} \\ \nonumber
&&\times \big|h_1 A_- \sin p-h_2 A_+ \cos p \big|+O(|\mathfrak{h}_t|^2),\\
&&\Omega_{01}^{(a)}(p|\mathfrak{h}_t)=\Omega_{10}^{(a)}(-p|\mathfrak{h}_t),\label{Om1a}
\end{eqnarray}
with $p\in(0,\pi)$.

As one can see from Figure \ref{fig:2a}, the gap between the two modes $\Omega_{10}^{(1)}(p|\mathfrak{h}_t)$, and 
$\Omega_{10}^{(2)}(p|\mathfrak{h}_t)$ vanishes at a certain value of the kink momentum $p$.
The physical reason of this partly restored degeneracy is the additional symmetry of the Hamiltonian. Indeed, at
$h_{1z}=h_{2z}=0$,  
$\varphi_1=\pi/2$
and $\varphi_2=0$, the interaction operator \eqref{V} reduces to the operator $V_t(\mathbf{h}_t)$ given by 
\eqref{Vt}:
\[
V\Big|_{h_{1z}=h_{2z}=0,\,\varphi_1=\pi/2,\varphi_2=0}=V_t(\mathbf{h}_t),
\]
where $\mathbf{h}_t=h_2 \mathbf{e}_x+h_1 \mathbf{e}_y$.
 The latter operator commutes with
 the modified translation 
$\widetilde{T}_1$:
\begin{equation}\label{Vta}
[V_t(\mathbf{h}_t),\widetilde{T}_1]=0.
\end{equation}
This allows us to redefine the kink quasi-momentum by means of  the relation
\begin{equation} \label{T1w}
\widetilde{T}_1 |\mathbb{K_{\mu\nu}}(p|\mathbf{h}_t)\rangle=e^{ i p}\, |\mathbb{K_{\mu\nu}}(p|\mathbf{h}_t)\rangle,
\end{equation}
instead of \eqref{T2},  and to let the kink quasi-momentum   run through the whole interval $p\in(0,2\pi)$ in equation \eqref{T1w}. 
As a result, one can 
 consider only the single one-kink mode in the extended Brillouin zone $p\in(-\pi,\pi)$ in each topological sector $\mathcal{L}_{10}^{(1)}$ and $\mathcal{L}_{01}^{(1)}$.  The one-kink states $|\mathbb{K_{\mu\nu}}(p|\mathbf{h}_t)\rangle$ with $p\in (0,2\pi)$ 
 form the basis in the one-kink subspace $\mathcal{L}_{\mu\nu}^{(1)}$. These basis states will be  normalized by the 
 condition
 \[
 \langle \mathbb{K_{\nu\mu}}(p|\mathbf{h}_t)|\mathbb{K_{\mu'\nu'}}(p'|\mathbf{h}_t)\rangle = 2 \pi \delta_{\mu\mu'} \delta_{\nu\nu'}\delta(p-p'),
 \]
with  $p,p'\in (0,2\pi)$, and $\mu\ne \nu$.

In the limit $\mathbf{h}_t\to 0$, the one-kink states  $|\mathbb{K_{\mu\nu}}(p|\mathbf{h}_t)\rangle$ can be  related with the previously introduced states $|K_{\mu\nu}(p)\rangle_s$. Really, it follows from \eqref{T1a} and \eqref{T1w}, that 
we can put without loss of generality
\begin{equation}\label{limK}
\lim_{\mathbf{h}_t\to 0}|\mathbb{K_{\mu\nu}}(p|\mathbf{h}_t)\rangle=
|K_{\mu\nu}(p)\rangle_{1/2}+|K_{\mu\nu}(p)\rangle_{-1/2},
\end{equation}
for $0<p<\pi$.  For these one-kink state we shall use the notation $|K_{\mu\nu}(p)\rangle$:
\begin{equation}\label{KK}
|K_{\mu\nu}(p)\rangle:=\lim_{\mathbf{h}_t\to 0}|\mathbb{K_{\mu\nu}}(p|\mathbf{h}_t)\rangle.
\end{equation}

Exploiting equation \eqref{Kpi1}, we can extend  formula \eqref{limK} from the interval $p\in(0,\pi)$ to the whole real axis
$p\in \mathbb{R}$. In particular, we get this way:
\begin{align}
&|{K}_{10}(p+\pi)\rangle=-|K_{10}(p)\rangle_{1/2}+|K_{10}(p)\rangle_{-1/2},\\
&|{K}_{01}(p+\pi)\rangle=|K_{01}(p)\rangle_{1/2}-|K_{01}(p)\rangle_{-1/2}.
\end{align}  

The Bloch states $|\mathbb{K_{\mu\nu}}(p|\mathbf{h}_t)\rangle$  are the eigenstates of the Hamiltonian $(\mathcal{H}_0+V_t)$:
\begin{equation}\label{dlK}
[\mathcal{H}_0+V_t-E_{vac}(\mathbf{h}_t)] |\mathbb{K_{\mu\nu}}(p|\mathbf{h}_t)\rangle= \Omega_{\mu\nu}(p|\mathbf{h}_t)
|\mathbb{K_{\mu\nu}}(p|\mathbf{h}_t)\rangle.
\end{equation}
Their  dispersion laws $\Omega_{\mu\nu}(p|\mathbf{h}_t)$ are the $2\pi$-periodical functions of $p$.
To the first order in $\mathbf{h}_t$ these functions  read as:
\begin{align}\label{Om2}
&\Omega_{10}(p|\mathbf{h}_t)=\omega(p)+\delta\Omega_{10}(p|\mathbf{h}_t)+O(|\mathbf{h}_t|^2),\\\nonumber
&\Omega_{01}(p|\mathbf{h}_t)=\omega(p)+\delta\Omega_{01}(p|\mathbf{h}_t)+O(|\mathbf{h}_t|^2),
\end{align}
where 
\begin{align}\label{delOm}
&\delta\Omega_{10}(p|\mathbf{h}_t)=\frac{\sinh \eta}{\omega(p)}\left(h_1 A_- \sin p-h_2 A_+ \cos p \right),\\
&\delta\Omega_{01}(p|\mathbf{h}_t)=\delta\Omega_{10}(-p|\mathbf{h}_t).\nonumber
\end{align}
The plot of the function $\Omega_{10}(p|\mathbf{h}_t)$ at $\eta=1.31$, $h_1=0.3$, $h_2=0.1$ is shown in Figure \ref{fig:2b}.

The situation is very similar in the second case, when either the staggered, or the uniform transverse magnetic field vanishes.
At $h_{2x}>0$, $h_{1x}=h_{1y}=h_{2y}=0$, and, as well, at $h_{1x}=h_{2y}=h_{2x}=0$, $h_{1y}>0$,  the interaction term 
$V_t$ commutes with $\widetilde{T}_1$, and one can use \eqref{T1w} in order to define the kink quasi-momentum $p\in(-\pi,\pi)$.
Accordingly, one can still use equations \eqref{Om2}, \eqref{delOm} to determine the kink dispersion 
laws in the extended Brillouin zone $p\in(-\pi,\pi)$.

\section{Two-kink states at  $h_z=0$, $\mathbf{h}_t\ne{0}$ \label{two-k}}
The two-kink subspace $\mathcal{L}^{(2)}$ is the direct sum of two subspaces 
\[
\mathcal{L}^{(2)}=\mathcal{L}_{11}^{(2)}\oplus \mathcal{L}_{00}^{(2)}.
\]
At a generic $\mathfrak{h}_t\ne{0}$, the basis in the first subspace $\mathcal{L}_{11}^{(2)}$ is formed by the vectors
$ |\mathbb{K}_{10}(p_1|\mathfrak{h}_t)\mathbb{K}_{01}(p_2|\mathfrak{h}_t)\rangle_{a_1,a_2}$, and the basis in the second space
$\mathcal{L}_{00}^{(2)}$  is formed by the vectors
$ |\mathbb{K}_{01}(p_1|\mathfrak{h}_t)\mathbb{K}_{10}(p_2|\mathfrak{h}_t)\rangle_{a_1,a_2}$. In both bases $a_{i}=1,2$, and
$0<p_2<p_1<\pi$.

These basis states must satisfy the following relations:
\begin{align}\label{T2A}
&T_1^2 |\mathbb{K_{\mu\nu}}(p_1|\mathfrak{h}_t)\mathbb{K_{\nu\mu}}(p_2|\mathfrak{h}_t)
\rangle_{a_1,a_2}=\\\nonumber
&e^{2 i (p_1+p_2)}\,  |\mathbb{K_{\mu\nu}}(p_1|\mathfrak{h}_t)\mathbb{K_{\nu\mu}}(p_2|\mathfrak{h}_t)
\rangle_{a_1,a_2},\\
&[\mathcal{H}_0+V_t-E_{vac}(\mathfrak{h}_t)]|\mathbb{K_{\mu\nu}}(p_1|\mathfrak{h}_t)\mathbb{K_{\nu\mu}}(p_2|\mathfrak{h}_t)
\rangle_{a_1,a_2}=\\\nonumber
&[ \Omega_{\mu\nu}^{(a_1)}(p_1|\mathfrak{h}_t)+ \Omega_{\nu\mu}^{(a_2)}(p_1|\mathfrak{h}_t)]
|\mathbb{K_{\mu\nu}}(p_1|\mathfrak{h}_t)\mathbb{K_{\nu\mu}}(p_2|\mathfrak{h}_t)
\rangle_{a_1,a_2},
\end{align}
where $\Omega_{\mu\nu}^{(a)}(p|\mathfrak{h}_t)$ are the kink dispersion laws determined by equations \eqref{Om}, \eqref{Oma}.

If the  transverse staggered and uniform magnetic fields are mutually orthogonal, we can parametrize them by
the two-component vector $\mathbf{h}_t=\mathbf{e}_x h_2+\mathbf{e}_y h_1$. In this case a  different, more convenient 
classification of two-kink 
basis Bloch states is possible. Since equation \eqref{Vta} holds at $\varphi_1=\pi/2$
and $\varphi_2=0$, we can define in this case the alternative basis of the two-kink  states 
by the relations:
\begin{subequations}\label{2keq}
\begin{align}\label{T1K}
&\widetilde{T}_1 |\mathbb{K_{\mu\nu}}(p_1|\mathbf{h}_t)
\mathbb{K}_{\nu\mu}(p_2|\mathbf{h}_t)\rangle\\\nonumber
&=e^{ i (p_1+p_2)}|\mathbb{K}_{\mu\nu}(p_1|\mathbf{h}_t)
\mathbb{K}_{\nu\mu}(p_2|\mathbf{h}_t)\rangle,\\\label{EnOm}
&[\mathcal{H}_0+V_t-E_{vac}(\mathbf{h}_t)]|\mathbb{K}_{\mu\nu}(p_1|\mathbf{h}_t)\mathbb{K}_{\nu\mu}(p_2|\mathbf{h}_t)
\rangle\\\nonumber
&=[ \Omega_{\mu\nu}(p_1|\mathbf{h}_t)+ \Omega_{\nu\mu}(p_2|\mathbf{h}_t)]
|\mathbb{K}_{\mu\nu}(p_1|\mathbf{h}_t)\mathbb{K}_{\nu\mu}(p_2|\mathbf{h}_t)
\rangle,\\
&\langle \mathbb{K}_{\mu\nu}(p_2|\mathbf{h}_t)\mathbb{K}_{\nu\mu}(p_1|\mathbf{h}_t)
|\mathbb{K}_{\mu\nu}(p_1'|\mathbf{h}_t)\mathbb{K}_{\nu\mu}(p_2'|\mathbf{h}_t)\rangle\\\nonumber
&=4\pi^2 \delta(p_1-p_1')\delta(p_2-p_2'),
\end{align}
\end{subequations}
where $0<p_2<p_1<2\pi$, and $0<p_2'<p_1'<2\pi$.

We shall use the following notation for the two-kink states at $\mathbf{h}_t=0$:
\begin{equation}\label{K2}
|K_{\mu\nu}(p_1)K_{\nu\mu}(p_2)\rangle:=\lim_{\mathbf{h}_t\to 0}
|\mathbb{K_{\mu\nu}}(p_1|\mathbf{h}_t)\mathbb{K_{\nu\mu}}(p_2|\mathbf{h}_t)
\rangle
\end{equation} 

\begin{figure}[htb]
\centering
\includegraphics[width=\linewidth, angle=00]{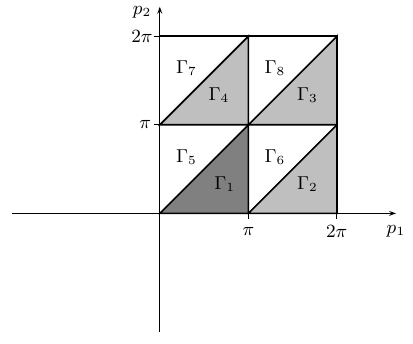}
\caption{\label{pR} Triangular regions in the plane of momenta $p_1,p_2$,  which parametrize
the two-kink states.} 
\end{figure}

These two-kink basis states can be related with the two-kink
basis states 
$|{K_{\mu\nu}}(p_1){K_{\nu\mu}}(p_2)\rangle_{s_1,s_2}$ characterized by the $z$-projections 
of the kink spins $s_1,s_2$.
Namely, 
\begin{align}\label{Kbb}
|K_{\mu\nu}(p_1)K_{\nu\mu}(p_2)\rangle\\\nonumber
=\sum_{s_1=\pm 1/2}\sum_{s_2=\pm 1/2}|{K_{\mu\nu}}(p_1){K_{\nu\mu}}(p_2)\rangle_{s_1,s_2},
\end{align}
for $\langle p_1,p_2\rangle \in \Gamma_1$, where the triangular region $\Gamma_1$ is shown in Figure \ref{pR}.
By means of equalities \eqref{Kpi},
one can extend formula \eqref{Kbb} to the three other triangular regions $\Gamma_2$,
$\Gamma_3$, $\Gamma_4$  in Figure  \ref{pR}.
Extension of equation \eqref{Kbb} from the region $ \Gamma_1$ into the region $ \Gamma_5$ can be performed,
in turn, by means of the Faddeev-Zamolodchikov commutation relation
\begin{align}\label{Sma}
 &  |{K}_{\mu\nu}(p_1){K}_{\nu\mu}(p_2)\rangle_{s_1 s_2}=\sum_{s_1',s_2'=\pm1/2 }{S}_{s_1s_2}^{s_1's_2'}(p_1,p_2) \\\nonumber
&  \times    |{K}_{\mu\nu}(p_2){K}_{\nu\mu}(p_1)\rangle_{s_2' s_1'},
  \end{align}
applied to the right-hand side of  \eqref{Kbb}. Finally, exploiting again equality \eqref{Kpi}, one can extend formula \eqref{Kbb}  from the 
triangular region $ \Gamma_5$ into the regions $\Gamma_6$, $\Gamma_7$, and $\Gamma_8$ shown in Figure  \ref{pR}.
As the result, equation \eqref{Kbb} allows one to determine the two-kink basis state 
$|{K_{\mu\nu}}(p_1){K_{\nu\mu}}(p_2)\rangle$ for any real $\langle p_1,p_2\rangle \in \mathbb{R}^2$. 

In the basis \eqref{Kbb}, the Faddeev-Zamolodchikov commutations  relations \eqref{Sma} reduce to the form:
\begin{align}\label{scat}
&|K_{\mu\nu}(p_1)K_{\nu\mu}(p_2)\rangle \\\nonumber
&=\frac{w_0(p_1,p_2)+w_+(p_1,p_2)}{2}
|K_{\mu\nu}(p_2)K_{\nu\mu}(p_1)\rangle\\\nonumber
&+\frac{-w_0(p_1,p_2)+w_+(p_1,p_2)}{2}
|K_{\mu\nu}(p_2+\pi)K_{\nu\mu}(p_1+\pi)\rangle,
 \end{align} 
 where  the scattering amplitudes $w_0(p_1,p_2)$ and $w_+(p_1,p_2)$ 
 are given by equations \eqref{wio}
 in Appendix \ref{Ap1}.
The commutation relation  \eqref{scat} holds for any $p_1,p_2\in \mathbb{R}$.
The projector operator 
$\mathcal{P}_{\mu\mu}^{(2)}$  on the subspace $\mathcal{L}_{\mu\mu}^{(2)}$ can be written at $\mathbf{h}_t=0$ as:
\begin{align}\label{Pr2a}
\mathcal{P}_{\mu\mu}^{(2)}=\sum_{s_1=\pm1/2\atop{s_2=\pm1/2}}\iint_{\Gamma_1} \frac{dp_1dp_2}{\pi^2}
|K_{\mu\nu}(p_1)K_{\nu\mu}(p_2)\rangle_{s_1s_2}\\\nonumber
\times\phantom{.}_{s_2 s_1} \langle K_{\mu\nu}(p_2) K_{\nu\mu}(p_1) |\\\label{Pr2b}
=\frac{1}{2}\int_{0}^{2\pi} \frac{dp_1}{2\pi}\int_{0}^{2\pi} \frac{dp_2}{2\pi}
|K_{\mu\nu}(p_1)K_{\nu\mu}(p_2)\rangle\\\nonumber
\times \langle K_{\mu\nu}(p_2) K_{\nu\mu}(p_1) |.
 \end{align} 
 
 Application of a small enough, but finite transverse magnetic field $\mathbf{h}_t=h_2 \mathbf{e}_x+h_1 \mathbf{e}_y$ not only modifies 
 the dispersion laws of kinks due to \eqref{dlK}-\eqref{delOm}, but also effects their 
mutual scattering.  

Let us consider for given $P$ and $E$  all solutions of the equations 
 \begin{align}\label{enmom}
&\exp[i(p_1+p_2)]=\exp(i P),\\\nonumber
&\Omega_{\mu\nu}(p_1)+\Omega_{\nu\mu}(p_2)=E,
 \end{align}
 lying in the interval $\langle p_1,p_2\rangle\in (0,2\pi)$.
 We denote such a solution  $\langle p_{1,in},p_{2,in}\rangle$, if 
 \begin{equation}
 \partial_{p_1}\Omega_{\mu\nu}(p_1)- \partial_{p_2}\Omega_{\nu\mu}(p_2)>0
 \end{equation}
 at this solution, 
 and use the notation $\langle p_{1,out},p_{2,out}\rangle$ for the solutions, in which the opposite inequality 
  \begin{equation}
 \partial_{p_1}\Omega_{\mu\nu}(p_1)- \partial_{p_2}\Omega_{\nu\mu}(p_2)<0
 \end{equation}
 holds.  If there are $m$ in- and out- solutions 
 of equations \eqref{enmom}, we distinguish them by the index $i=1,\ldots,m$.
 Using these notations, the elastic two-kink scattering can be characterized by the scattering matrix
 $W_{i'}^i(E,P|\mathbf{h}_t)$ by means of the following ``deformed Faddeev-Zamolodchikov commutation 
 relation":
  \begin{align}\label{defFZ}
 &|\mathbb{K_{\mu\nu}}(p_{1,in}^{i}|\mathbf{h}_t)\mathbb{K_{\nu\mu}}(p_{2,in}^{i}|\mathbf{h}_t)\rangle\\
 &=\sum_{i'=1}^m W_{i'}^i(E,P|\mathbf{h}_t) |\mathbb{K_{\mu\nu}}(p_{1,out}^{i'}|\mathbf{h}_t)\mathbb{K_{\nu\mu}}(p_{2,out}^{i'}|\mathbf{h}_t)\rangle.\nonumber
  \end{align}
Of course, at $\mathbf{h}_t\ne 0$, the $n$-particle scattering matrices with $n>2$ do not factorize in the two-particle ones, since the transverse magnetic field breaks integrability of the XXZ spin-chain model.

At a small $|\mathbf{h}_t|$, the scattering matrix  $W_{i'}^i(E,P|\mathbf{h}_t)$  should
 analytically depend  on $h_1,h_2$, and 
the  initial terms of its Taylor expansion in these parameters
could be, in principal, determined by means of the standard perturbation theory. At $\mathbf{h}_t=0$, 
 equation \eqref{defFZ} must reduce    to \eqref{scat}. 

Using equation \eqref{defFZ}, one can extend definition of the two-kink
basis states $|\mathbb{K_{\mu\nu}}(p_1|\mathbf{h}_t)\mathbb{K_{\nu\mu}}(p_2|\mathbf{h}_t)
\rangle$ from the triangular region $0<p_2<p_1<2\pi$ to the square 
$ p_1,p_2 \in (0,2\pi)$. Subsequent application of the periodicity relation
\begin{align*}
&|\mathbb{K_{\mu\nu}}(p_1|\mathbf{h}_t)\mathbb{K_{\nu\mu}}(p_2|\mathbf{h}_t)
\rangle=|\mathbb{K_{\mu\nu}}(p_1+2\pi|\mathbf{h}_t)\mathbb{K_{\nu\mu}}(p_2|\mathbf{h}_t)
\rangle\\
&=|\mathbb{K_{\mu\nu}}(p_1|\mathbf{h}_t)\mathbb{K_{\nu\mu}}(p_2+2\pi|\mathbf{h}_t)
\rangle
\end{align*}
 allows one to define such two-kink Bloch states for any real $\langle p_1,p_2\rangle\in 
 \mathbb{R}^2$.

It is natural to expect, that the  projector operator $\mathcal{P}_{\mu\mu}^{(2)}(\mathbf{h}_t)$
 onto the two-kink subspace 
 $\mathcal{L}_{\mu\mu}^{(2)}(\mathbf{h}_t)$
 admits  at a small enough, but finite $\mathbf{h}_t=h_2 \mathbf{e}_x+h_1 \mathbf{e}_y$ the representations
 \begin{align}\nonumber
&\mathcal{P}_{\mu\mu}^{(2)}(\mathbf{h}_t)=\int_{0}^{2\pi} \frac{dp_1}{2\pi}\int_{0}^{p_1} \frac{dp_2}{2\pi}
|\mathbb{K_{\mu\nu}}(p_1|\mathbf{h}_t)\mathbb{K_{\nu\mu}}(p_2|\mathbf{h}_t)\rangle\\\nonumber
&\times \langle \mathbb{K_{\mu\nu}}(p_2|\mathbf{h}_t) \mathbb{K_{\nu\mu}}(p_1|\mathbf{h}_t) |\\\label{P112}
&=\frac{1}{2}\int_{0}^{2\pi} \frac{dp_1}{2\pi}\int_{0}^{2\pi} \frac{dp_2}{2\pi}
|\mathbb{K_{\mu\nu}}(p_1|\mathbf{h}_t)\mathbb{K_{\nu\mu}}(p_2|\mathbf{h}_t)\rangle\\
&\times \langle \mathbb{K_{\mu\nu}}(p_2|\mathbf{h}_t) \mathbb{K_{\nu\mu}}(p_1|\mathbf{h}_t) |,\nonumber
 \end{align} 
which are analogous to \eqref{Pr2a}, \eqref{Pr2b}.
\section{Meson states at $h_z>0$ 
\label{sec_Mes}}
Proceeding to the XXZ spin-chain model \eqref{Ham} with  non-zero staggered longitudinal magnetic field, we 
will restrict our attention to the case of the Hamiltonian \eqref{Hama}, in which the staggered and uniform transverse fields  are mutually orthogonal.  

The first reason in favor of  this choice is that  the Hamiltonian \eqref{Hama} commutes with the modified translation 
operator \eqref{mTr}:
\begin{equation}\label{com}
[\mathcal{H}(\Delta,\mathbf{h}_t,h_z),\widetilde{T}_1]=0.
\end{equation}
This makes 
the perturbative analysis of the kink-confinement in model \eqref{Hama} 
more simple, than in the original one \eqref{Ham}. The second reason is that  the XXZ model with mutually orthogonal staggered 
and uniform transverse magnetic fields has been used by Faure {\it et al.} \cite{Faur17} for the interpretation of their 
results of the neutron scattering study of the meson energy spectra in the quasi-one-dimensional magnetic crystal 
$\mathrm{BaCo}_{2}\mathrm{V}_{2}\mathrm{O}_{8}$ in the kink confinement regime in the presence of the external
uniform transverse  magnetic field. The effective staggered transverse magnetic field orthogonal to the applied 
external 
uniform transverse field is induced in this compound due to the off-diagonal components of the anisotropic $g$-tensor 
\cite{Kimura13}.
 
 The application of the  longitudinal staggered magnetic field $h_z$ explicitly breaks  the $\mathbb{Z}_2$ symmetry of the model
 decreasing the energy of the vacuum  $|Vac(\mathbf{h}_t,h_z)\rangle^{(1)}$, which becomes the true ground state, 
 and increasing the energy of the vacuum $|Vac(\mathbf{h}_t,h_z)\rangle^{(0)}$, which becomes  metastable.
 The true (and also, the false) vacuum $|Vac(\mathbf{h}_t,h_z)\rangle^{(1)}$ remains invariant with respect to the action of 
 modified translation operator:
 \begin{equation}\label{vacT}
\widetilde{T}_1|Vac(\mathbf{h}_t,h_z)\rangle^{(1)}=|Vac(\mathbf{h}_t,h_z)\rangle^{(1)}.
 \end{equation}
The margin between the energies of the true and false vacuums opened by the staggered 
longitudinal field $h_z$ leads to the confinement of kinks $|\mathbb{K_{\mu\nu}}(p|\mathbf{h}_t)\rangle$ into the meson bound states 
$|\pi_n(P|\mathbf{h}_t,h_z)\rangle$. The defining relations for these meson bound states read:
\begin{subequations}\label{mes0}
\begin{align} \label{mes}
&\widetilde{T}_1|\pi_n(P|\mathbf{h}_t,h_z)\rangle=e^{i P}|\pi_n(P|\mathbf{h}_t,h_z)\rangle,\\\label{mesP}
&(\mathcal{H}+C)|\pi_n(P|\mathbf{h}_t,h_z)\rangle=E_n(P|\mathbf{h}_t,h_z)|\pi_n(P|\mathbf{h}_t,h_z)\rangle,
\end{align}
\end{subequations}
where the Hamiltonian $\mathcal{H}$ is given by \eqref{Hama}, $E_n(P|\mathbf{h}_t,h_z)$ is dispersion law of the $n$-th meson mode,  $0<P<2\pi$, $n=1,2,\ldots$,   
and the numerical constant $C$ is chosen in such a way, that 
\[
(\mathcal{H}+C)|Vac(\mathbf{h}_t,h_z)\rangle^{(1)}=0.
\]

At $h_z=0$, the meson states $|\pi_n(P|\mathbf{h}_t,h_z)\rangle$ decouple into some
linear combinations of two-kink states:
\[
|\pi_n(P|\mathbf{h}_t,0)\rangle\to |\mathbb{K}_{10}(p_1|\mathbf{h}_t)\mathbb{K}_{01}(p_2|\mathbf{h}_t)\rangle,
\]
with $\exp[i(p_1+p_2)]=\exp(i P)$.

On the other hand, it will be shown later, that at $\mathbf{h}_t=0$, the meson states 
 $|\pi_n(P|\mathbf{h}_t,h_z)\rangle$ transform into the
 meson states $|\pi_{s,\iota,m}(P|h_z)\rangle$ with $s=\pm1$, $\iota =0,\pm$, introduced in \cite{Rut22}. Namely, 
 for $0<P<\pi$  and odd $n=2m-1$:
\begin{align}
&|\pi_{2m-1}(P|\mathbf{0},h_z)\rangle\cong|\pi_{1,0,m}(P|h_z)\rangle+|\pi_{-1,0,m}(P|h_z)\rangle,\\
&|\pi_{2m-1}(P+\pi|\mathbf{0},h_z)\rangle\cong|\pi_{1,0,m}(P|h_z)\rangle-|\pi_{-1,0,m}(P|h_z)\rangle,\nonumber
\end{align}
while for $0<P<\pi$  and even $n=2m$:
\begin{align} 
&|\pi_{2m}(P|\mathbf{0},h_z)\rangle \cong
|\pi_{0,+,m}(P|h_z)\rangle,\\\nonumber
&|\pi_{2m}(P+\pi|\mathbf{0},h_z)\rangle \cong|\pi_{0,-,m}(P|h_z)\rangle,
\end{align}
where $\cong$ denotes the equality up to some numerical  factor, and $m=1,2,\ldots$.

It follows from \eqref{mes} and \eqref{mesP}, that
the meson dispersion law $E_n(P|\mathbf{h}_t,h_z)$ is the $2\pi$-periodical function of the meson momentum $P$,
and we can set without loss of generality:
\begin{equation}
|\pi_n(P+2 \pi l|\mathbf{h}_t,h_z)\rangle=|\pi_n(P|\mathbf{h}_t,h_z)\rangle,
\end{equation} 
for $l\in \mathbb{Z}$. The function $E_n(P|\mathbf{h}_t,h_z)$ is also even with respect to the  $P$-inversion:
\[
E_n(P|\mathbf{h}_t,h_z)= E_n(-P|\mathbf{h}_t,h_z).
\]

As in the previously studied case  \cite{Rut22} $\mathbf{h}_t=0$, the eigenvalue problem \eqref{mes0} is very difficult, 
since the interaction term $V_l(h_z)$ given by \eqref{Vl} does not conserve the number of kinks. The key simplification 
is achieved by means of the two-kink approximation \cite{Rut22}, which implies replacement of the 
exact eigenvalue problem \eqref{mes0} by its projection onto the 
two-kink subspace $\mathcal{L}_{11}^{(2)}(\mathbf{h}_t)$:
\begin{subequations}\label{mes2}
\begin{align} \label{mes2a}
&\widetilde{T}_1|\tilde{\pi}_n(P|\mathbf{h}_t,h_z)\rangle=e^{i P}|\tilde{\pi}_n(P|\mathbf{h}_t,h_z)\rangle,\\\label{mesP2}
&\mathcal{H}^{(2)}|\tilde{\pi}_n(P|\mathbf{h}_t,h_z)\rangle=\widetilde{E}_n(P|\mathbf{h}_t,h_z)|\tilde{\pi}_n(P|\mathbf{h}_t,h_z)\rangle,
\end{align}
\end{subequations}
where $|\tilde{\pi}_n(P|\mathbf{h}_t,h_z)\rangle\in \mathcal{L}_{11}^{(2)}(\mathbf{h}_t)$, and 
\begin{equation}\label{H2}
\mathcal{H}^{(2)}=\mathcal{P}_{11}^{(2)}(\mathbf{h}_t)\mathcal{H}\mathcal{P}_{11}^{(2)}(\mathbf{h}_t).
\end{equation}
Tildes in $\tilde{\pi}_n(P|\mathbf{h}_t,h_z), \widetilde{E}_n(P|\mathbf{h}_t,h_z)$ distinguish solutions of equations \eqref{mes2} from 
those of the exact eigenvalue problem \eqref{mes0}.
The meson states $|\tilde{\pi}_n(P|\mathbf{h}_t,h_z)\rangle$ will be normalized by the condition
\begin{equation}\label{norm}
\langle\tilde{\pi}_{n}(P|\mathbf{h}_t,h_z) |\tilde{\pi}_{n'}(P'|\mathbf{h}_t,h_z)\rangle=2 \pi \delta_{nn'}
\delta(P-P'),
\end{equation}
for $0<P,P'<2\pi$. These states form a basis in the two-kink subspace $\mathcal{L}_{11}^{(2)}(\mathbf{h}_t)$,
and the projector \eqref{P112} on this subspace admits the expansion: 
\begin{equation}\label{P112a}
\mathcal{P}_{11}^{(2)}  (\mathbf{h}_t )=\sum_{n=1}^\infty \int_0^{2 \pi}\frac{dP}{2\pi} |\tilde{\pi}_n(P|\mathbf{h}_t,h_z)\rangle
\langle\tilde{\pi}_n(P|\mathbf{h}_t,h_z)|.
\end{equation}

The  two-kink meson state $|\tilde{\pi}_n(P|\mathbf{h}_t,h_z)\rangle$ can be characterized by the wave-function:
\begin{align}\label{Phiphi2}
&\Phi_n(p_1,p_2|P,\mathbf{h}_t,h_z)=\\
&\langle \mathbb{K}_{10}(p_2|\mathbf{h}_t) \mathbb{K}_{01}(p_1|\mathbf{h}_t) |\tilde{\pi}_n(P|\mathbf{h}_t,h_z)\rangle,\nonumber
\end{align} 
It follows also from \eqref{T1K}, \eqref{mes2a}, that
\begin{equation}
\Phi_n(p_1,p_2|P,\mathbf{h}_t,h_z)=e^{i(P-p_1-p_2)}\Phi_n(p_1,p_2|P,\mathbf{h}_t,h_z).
\end{equation}
For $(p_1+p_2),P\in (0,2\pi)$, this allows one to represent the meson wave function 
$\Phi_n(p_1,p_2|P,\mathbf{h}_t,h_z)$ in the form
\begin{align}\label{Phph}
\Phi_n(p_1,p_2|P,\mathbf{h}_t,h_z)=2\pi\,\delta(p_1+p_2-P)\\\nonumber
\times\phi_n\!\left(\frac{p_1-p_2}{2}|P,\mathbf{h}_t,h_z\right).
\end{align}

The reduced wave function $\phi_n\!\left(p|P,\mathbf{h}_t,h_z\right)$ in the right-hand side
is the $2\pi$-periodical function of $p$. The normalization conditions following from \eqref{norm} reads:
\begin{equation}\label{normphi}
\int_{-\pi}^{\pi}\frac{dp}{2\pi} \,\phi_n^*\!\left(p|P,\mathbf{h}_t,h_z\right)\phi_{n'}\!\left(p|P,\mathbf{h}_t,h_z\right)=
2 \delta_{nn'}.
\end{equation}
By means of  the procedure described in \cite{Rut22}, one can show, that this 
function must solve  the Bethe-Salpeter  integral equation:
\begin{align}\label{BS}
[\mathcal{E}(p|P)-\widetilde{E}_n(P)]\phi_n(p|P)=\frac{h_z}{2}\int_{-\pi}^\pi \frac{dp'}{2\pi}\phi_n(p'|P)\\\nonumber
\times\langle \mathbb{K}_{10}(p_2) \mathbb{K}_{01}(p_1) |
(\sigma_0^z -\langle \sigma_0^z \rangle)|\mathbb{K}_{10}(p_1') \mathbb{K}_{01}(p_2')\rangle,
\end{align}
where
\begin{align}\label{EpP}
&\mathcal{E}(p|P,\mathbf{h}_t)=\Omega_{10}(p_1|\mathbf{h}_t)+\Omega_{01}(p_2|\mathbf{h}_t)\\\nonumber
&=\epsilon(p|P)+\delta \mathcal{E}(p|P,\mathbf{h}_t)+ O(|\mathbf{h}_t|^2),\\\label{eeps}
&\epsilon(p|P)=\omega(P/2+p)+\omega(P/2-p),\\\label{vare}
&\delta \mathcal{E}(p|P,\mathbf{h}_t)=
\delta \Omega_{10}(p+P/2|\mathbf{h}_t)+\delta \Omega_{01}(-p+P/2|\mathbf{h}_t),\\\label{lsig}
&\langle \sigma_0^z \rangle= \phantom{.}^{(1)} \!\langle Vac(\mathbf{h}_t,0)|\sigma_0^z |Vac(\mathbf{h}_t,0)\rangle^{(1)},
\end{align}
$\delta \Omega_{\mu\nu}(p|\mathbf{h}_t)$ are given 
by \eqref{delOm}, and $p_{1,2}=P/2\pm p$, $p_{1,2}'=P/2\pm p'$.
We have omitted the explicit dependancies of  functions  in equation \eqref{BS} 
on $\mathbf{h}_t$ and $h_z$.

The function $\epsilon(p|P)$ defined by \eqref{eeps} determines the effective kinetic energy 
of the relative motion of two kinks at zero magnetic field. It satisfies a number of symmetry relations
\begin{align}\label{symE1}
&\epsilon({p}|P)=\epsilon({p}|P+2\pi)=\epsilon(-{p}|P)=\epsilon({p}|-P)\\\nonumber
&=\epsilon({p}+\pi|P)=\epsilon({p}-\pi/2|P-\pi),
\end{align}  
following directly from \eqref{eeps}, \eqref{dl}. Application of the transverse magnetic field 
$\mathbf{h}_t$ deforms the effective kinetic energy to the function
$\mathcal{E}(p|P,\mathbf{h}_t)$ defined by \eqref{EpP}, which is much less 
symmetric, than $\epsilon({p}|P)$. The linear part $\delta \mathcal{E}(p|P,\mathbf{h}_t)$ of 
this deformation satisfies relations
\begin{align}
& \delta \mathcal{E}(p|P,\mathbf{h}_t)= \delta \mathcal{E}(p|P+2\pi,\mathbf{h}_t),\\
& \delta \mathcal{E}(p+\pi|P,\mathbf{h}_t)=-\delta \mathcal{E}(p|P,\mathbf{h}_t),\\
&\delta \mathcal{E}(-p|P,\mathbf{h}_t)\ne\delta \mathcal{E}(p|P,\mathbf{h}_t), \\\label{piE}
&\delta \mathcal{E}(p|\pi,\mathbf{h}_t)=0,
\end{align}
that follow from \eqref{vare}, \eqref{delOm}.
From \eqref{piE} and $2\pi$-periodicity of the effective kinetic energy in $P$, we get:
\begin{equation}\label{Eeps}
\mathcal{E}(p|P,\mathbf{h}_t)=\epsilon(p|P)+O(|\mathbf{h}_t|^2)
\end{equation}
at $P=  \pi+2\pi n$, with $n\in \mathbb{Z}$.

In the limit $\mathbf{h}_t=0$, equation \eqref{BS} reduces to the form
\begin{align}\label{BS0}
[\epsilon(p|P)-\widetilde{E}_n(P)]\phi_n(p|P)=\frac{h_z}{2}\int_{-\pi}^\pi \frac{dp'}{2\pi}\phi_n(p'|P)\\\nonumber
\times\langle {K}_{10}(p_2) {K}_{01}(p_1) |
(\sigma_0^z -\bar{\sigma} )|{K}_{10}(p_1') {K}_{01}(p_2')\rangle,
\end{align}
where 
 $\bar{\sigma}$ is the staggered spontaneous magnetization at zero magnetic field given by \eqref{sig}, 
and the two-kink basis states $|{K}_{10}(p_1) {K}_{01}(p_2)\rangle$ 
are defined by \eqref{K2}.

It follows from \eqref{scat}, and \eqref{Phiphi2},   that the reduced wave function 
$\phi_{n}\!\left(p|P,\mathbf{h}_t,h_z\right)$ satisfies  at $\mathbf{h}_t=0$ the reflection property:
\begin{align}\label{refl}
\phi_{n}(p|P)=\frac{w_0^*+w_+^*}{2}\phi_{n}(-p|P)\\\nonumber
+\frac{-w_0^*+w_+^*}{2}\phi_{n}(-p+\pi|P),
\end{align}
where
\begin{align}
w_0=w_0(p+P/2,-p+P/2), \\\nonumber
w_+=w_+(p+P/2,-p+P/2)
\end{align}
 are the two-kink scattering amplitudes \eqref{wio}.
 
Let us show, that  the integral equation \eqref{BS0} represents in the compact form all three Bethe-Salpeter equations (230) of paper \cite{Rut22}, which were derived there for the wave-functions of the mesons with different spins $s=0,\pm1$, and parities $\iota=0,\pm$. Really, using equation \eqref{Kbb}, the integral kernel in the second line of  \eqref{BS0}
can be rewritten as
\begin{align}
\langle {K}_{10}(p_2) {K}_{01}(p_1) |
(\sigma_0^z -\bar{\sigma} )|{K}_{10}(p_1') {K}_{01}(p_2')\rangle=\\\nonumber
4\bar{\sigma}\left[
e^{i(p'-p)}G_0(p,p'|P)+G_+(p,p'|P)\right],
\end{align}
where $G_0(p,p'|P)$ and $G_+(p,p'|P)$ were defined by equations (112) and (228) in \cite{Rut22}, respectively.
Since due to equation (233)  in \cite{Rut22}, the kernels $G_0(p,p'|P)$ and $G_+(p,p'|P)$ are  $\pi$-periodic 
functions of the variables $p$ and $p'$, 
the integral equation \eqref{BS0} decouples in two separate equations for the meson wave functions 
$\phi_n(p|P)$ with different transformation properties upon 
the translation $p\to p+\pi$. Namely, the $\pi$-antiperiodic  solution of \eqref{BS0}
satisfies equations
\begin{align}
&\phi_n(p+\pi|P)=-\phi_n(p|P),\\\label{BS1}
&[\epsilon(p|P)-\widetilde{E}_n(P)]e^{ip}\phi_n(p|P) \\ \nonumber
&=2 h_z \bar{\sigma}\int_{-\pi/2}^{\pi/2} \frac{dp'}{\pi} e^{ip'}\phi_n(p'|P)\, G_0(p,p'|P),
\end{align}
whereas for the  $\pi$-periodic  solution of \eqref{BS0} we get instead:
\begin{align}
&\phi_n(p+\pi|P)=\phi_n(p|P),\\\label{BS2}
&[\epsilon(p|P)-\widetilde{E}_n(P)]\phi_n(p|P)\\
&=2 h_z \bar{\sigma}\int_{-\pi/2}^{\pi/2} \frac{dp'}{\pi}\phi_n(p'|P)\,
 G_+(p,p'|P).\nonumber
\end{align}
Finally, 
it follows from the equality
\[
G_+(p,p'|P+\pi)=G_-(p+\pi/2,p'+\pi/2|P),
\]
that the $\pi$-periodic  solution $\phi_n(p|P+\pi)$ of the integral equation \eqref{BS0} must satisfy equalities
\begin{align}
&\phi_n(p+\pi|P+\pi)=\phi_n(p|P+\pi),\\\label{BS3}
&[\epsilon(p|P)-\widetilde{E}_n(P+\pi)]\phi_n(p|P+\pi)\\
&=2 h_z \bar{\sigma}\int_{-\pi/2}^{\pi/2} \frac{dp'}{\pi}\phi_n(p'|P+\pi)\,
 G_-(p,p'|P).\nonumber
\end{align}
Equations \eqref{BS1}, \eqref{BS2}, and \eqref{BS3} coincide with 
the Bethe-Salpeter equation (230) derived in \cite{Rut22} for the cases 
$\iota=0,+,-$, respectively.

The most consistent, but technically demanding  approach for the  calculation of the meson energy spectrum in  model \eqref{Hama}  in the presence
of the transverse magnetic field  $\mathbf{h}_t\ne 0$ requires  the perturbative solution of the Bethe-Salpeter equation \eqref{BS}. In the 
 case $\mathbf{h}_t=0$, this strategy was realized in \cite{Rut22}. 
 In what follows, we shall use the less rigorous, but more simple  heuristic semiclassical approach. 
 
 To conclude this section, let us comment on the two-kink approximation \eqref{mes2}, \eqref{H2}, which is essential in derivation of the 
 Bethe-Salpeter equation \eqref{BS} and will be implicitly exploited in the subsequent semiclassical calculations. The same approximation
 was used by Fonseca and Zamolodchikov \cite{FonZam2003,FZ06}
in derivation of the Bethe-Salpeter equation for the Ising field theory.  The kink confinement in this model is induced by the uniform magnetic field
 $h>0$. It was shown in \cite{FZ06}, that in the case of the IFT,   the two-kink approximation is asymptotically 
 exact in the weak confinement regime at $h\to+0$, and the leading `multi-kink' correction 
 (the correction beyond the two-kink approximation) to the meson masses is quadratic in $h$. In analogy with IFT, we expect, that exact meson 
 energies $E_n(P|\mathbf{h}_t,h_z)$ coincide with their two-kink approximations $\widetilde{E}_n(P|\mathbf{h}_t,h_z)$ to the linear order in $h_z\to0$, 
 and
 \begin{equation}\label{difE}
E_n(P|\mathbf{h}_t,h_z)-\widetilde{E}_n(P|\mathbf{h}_t,h_z)=O(h_z^2).
 \end{equation}
 Since the subsequent perturbative analysis of the meson energy spectra will be restricted to the linear order in $h_z$, we will neglect the difference
 \eqref{difE} and drop tildes in $\widetilde{E}_n(P|\mathbf{h}_t,h_z)$.
 \section{Heuristic approach to the kink confinement problem \label{HA}}
 It turns out, that even in the frame of the heuristic approach, the perturbative calculations of the meson energy spectra for the model \eqref{Hama}-\eqref{Vl}
remain   rather involved.  By this reason, we present in this section an informal non-technical introduction to our heuristic 
semiclassical approach, which will be applied in  Section \ref{Heur} for the calculation of the meson energy
 spectra.
 
 We start from the more simple case  studied in \cite{Rut22}   of the XXZ model \eqref{Hama}-\eqref{Vl} 
 in a weak staggered longitudinal field $h_z>0$, and zero transverse field
$\mathbf{h}_t=0$. 

At the first step, one 
 treats the two kinks in the spin chain as classical spinless particles  moving along the line and attracting one another with a linear potential. Their Hamiltonian is  taken in the form
 \begin{equation}\label{Hk}
 H(x_1,x_2,p_1,p_2)=\omega(p_1)+\omega(p_2)+f\cdot(x_2-x_1),
 \end{equation}
 where $\omega(p)$ is the kink dispersion law \eqref{dl}, and $f=2 h_z \bar{\sigma}$ is the string tension.
 The kink spatial coordinates $x_1,x_2\in \mathbb{R}$ are subjected to the constraint 
 \begin{equation}\label{x1x2}
 -\infty<x_1<x_2<\infty,
  \end{equation}
that
results from the  local "hard-sphere interaction"  of two particles.

\begin{figure}[ht]
\centering
\includegraphics[width=.95\linewidth, angle=00]{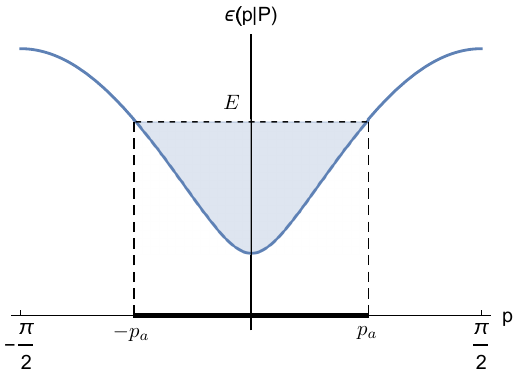}
\caption{\label{fig:eps} The $p$-dependence of the effective kinetic energy $\epsilon(p|P)$ determined by 
\eqref{eeps} at a fixed $P$ satisfying \eqref{PPc}. The kinematically allowed region $(-p_a,p_a)$  of the momentum $p$  for 
the given energy $E$ of two particles is shown as well.} 
\end{figure}

After the canonical transformation
\begin{subequations}\label{Canon}
\begin{align}
X=\frac{x_1+x_2}{2}, \quad x=x_1-x_2,\\
P=p_1+p_2, \quad p=\frac{p_1-p_2}{2},
\end{align}
\end{subequations}
the Hamiltonian \eqref{Hk} takes the form
\begin{equation}\label{Hk1}
H(p,x|P)=\epsilon(p|P)-f\, x,
\end{equation}
where $\epsilon(p|P)$  is given by \eqref{eeps}, and  $x<0$. The  effective kinetic energy of two kinks 
$\epsilon(p|P)$ is a $\pi$-periodical even function of $p$. 
At small enough absolute values of the total momentum 
\begin{equation}\label{PPc}
|P|<P_c(\eta), 
\end{equation}
with 
\begin{equation}\label{Pc1}
P_c(\eta)=\mathrm{arccos}\,\frac{1-k'(\eta)}{1+k'(\eta)}, 
\end{equation}
it monotonically increases
with $p$ in the interval $(0,\pi/2)$, as it is shown in Figure~\ref{fig:eps}.

The total energy-momentum conservation laws read:
\begin{align}
&\epsilon(p(t)|P)-f\, x(t)=E=\mathrm{const},\\
&P(t)=\mathrm{const}.
\end{align}
The classical evolution in the ``center of mass frame"  in the time interval between  
two successive  particle collisions is determined by the canonical equations of motion:
\begin{subequations}\label{can}
\begin{align}
&\dot{X}(t)=\frac{\partial \epsilon(p|P)}{\partial P},\\
&\dot{P}(t)=0,\\\label{xt}
&\dot{x}(t)=\frac{\partial \epsilon(p|P)}{\partial p},\\
&\dot{p}(t)=f.\label{pt}
\end{align}
\end{subequations}

\begin{figure}[ht]
\centering
\includegraphics[width=.95\linewidth, angle=00]{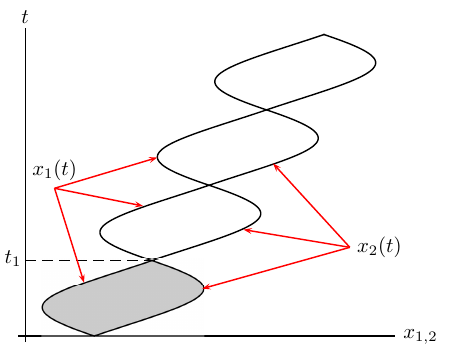}
\caption{\label{fig:dn1} World paths $x_1(t)$ and $x_2(t)$ of two particles 
for their classical evolution determined by equation \eqref{Canon}, \eqref{can}.} 
\end{figure}

If the condition \eqref{PPc} is satisfied, and the energy $E$ of two particles lies in the interval 
$\epsilon(0|P)<E<\epsilon(\pi/2|P)$, 
the momentum $p$ characterising their relative motion varies in the lacuna $-p_a<p<p_a$ shown in 
Figure~\ref{fig:eps}. The functions $x(t)$ and $p(t)$ are periodical in $t$ with the period $t_1=2 p_a/f$, and 
\begin{equation}\label{p1t}
p(t)=-p_a+ \{{t}/{t_1}\}\,t_1\,f, \quad \text{at   }t>0,
\end{equation}
where
$\{ z\}$ denotes the fractional part of $z$.
Figure \ref{fig:dn1} illustrates the world paths $x_1(t)<x_2(t)$ of two particles for the simple Hamiltonian dynamics
determined by equations \eqref{Canon}, and \eqref{can}.

The second step of the  calculation of the meson energy spectra $E_n(P)$ in our heuristic approach requires 
quantization of the simple two-particle  Hamiltonian dynamics  described above. Two approximate
quantization schemes have been uses  in \cite{Rut22}. The canonical quantization is  appropriate for calculations 
of energies of `lightest' mesons with small $n=1,2,\ldots$, while the semiclassical
quantization procedure is best suited for calculation of energies of highly excited mesons with $n\gg1$. 
It turns out, however,  the obtained in \cite{Rut22}  predictions for the masses of the lightest mesons in the XXZ 
spin-chain model perturbed by the staggered longitudinal field, 
which are  given by the semiclassical and low-energy expansions, are numerically very close to each other.
The high efficiency of the semiclassical expansions for prediction the energies of lightest mesons in several 
other models exhibiting confinement were also noticed in papers  \cite{Tak14,Kor16,Lagnese_2020,Mus22}, 
in which the meson energies were studied by direct numerical methods. By this reason, only 
the semiclassical quantization procedure will be considered in the present work.

It is easy to show \cite{Rut08a}, that if  two particles  behave  as free fermions at 
$f=0$,  application of the Bohr-Sommerfeld  rule 
\cite{LL3} to their classical dynamics determined by equations \eqref{can} with $f>0$  leads to the
quantization of the area $S_n$  of the dashed region shown
in Figure \ref{fig:dn1}:
\begin{equation}\label{semicS}
f\, S_n=2\pi \left( n-\frac{1}{4} \right),
\end{equation}
where $n\gg1$ is the number of the semiclassical energy level.

If the two Fermi-particles are not free, but strongly interact at short distances, formula \eqref{semicS} requires 
only a minor modification. Really, the short-range interaction between  two particle induces their  
 elastic scattering. Suppose, that the latter can be characterized at $f=0$ by the commutation relation
\begin{equation}\label{scEl}
|p_1p_2\rangle=-\exp[i \,\theta(p_1,p_2)]\,|p_2p_1\rangle,
\end{equation}
where $p_1> p_2$ are the momenta of two colliding particles,  and $\theta(p_1,p_2)$ denotes their scattering phase.
At $f>0$, the semiclassical evolution of two particles can be viewed as a combination of the classical movement along the 
world paths $x_1(t),x_2(t)$ shown in Figure \ref{fig:dn1} with their quantum scattering at their meeting points,
to which they arrive having the momenta $p_1=P/2+p_a,$, and  $p_2=P/2-p_a$. The elastic quantum scattering
of two particles must be taken into account in the semiclassical quantization condition \eqref{semicS} by adding 
 the scattering phase $\theta(p_1,p_2)$ to its left-hand side: 
\begin{equation}\label{semicS2}
f\, S_n+\theta(P/2+p_a,-P/2-p_a)=2\pi \left( n-\frac{1}{4} \right).
\end{equation}
This semiclassical quantization condition can be rewritten in the explicit  form:
\begin{align}\label{EnSem0}
&  2E_{n}(P)\,p_a  -\int_{-p_a}^{p_a}dp \, \epsilon(p|P)=\\\nonumber
&  f\left[2\pi  \,(n-1/4)-\theta\left(P/2+p_a,P/2-p_a\right)\right].
 \end{align}

In the above heuristic analysis based on the Hamiltonian \eqref{Hk}, 
the kinks were treated as the spinless particles. However, the one-kink
Bloch states $|K_{ab}(p)\rangle_{s}$ of the unperturbed XXZ model, which are 
defined by equations \eqref{kinkBloch},
carry the spin index $s=\pm1/2$. Accordingly, the two-particle scattering of such 
kinks is determined at zero magnetic field by  three scattering phases $\theta_\iota(p_1,p_2)$, 
which are distinguished by the parity index $\iota=0,\pm$, and explicitly given by
equations \eqref{scph} in Appendix \ref{Ap1}. Then, one gets instead of \eqref{EnSem0}
three quantization conditions:
\begin{align}\label{EnSem1a}
&  2E_{\iota,n}(P)\,p_a  -\int_{-p_a}^{p_a}dp \, \epsilon(p|P)=\\\nonumber
&  f\left[2\pi  \,(n-1/4)-\theta_\iota\left(P/2+p_a,P/2-p_a\right)\right], 
 \end{align}
 which determine the energies $E_{\iota,n}(P)$ of the meson modes with $\iota=0,\pm$.

The result \eqref{EnSem1a}   for the semiclassical meson energy spectra in the XXZ model 
\eqref{Hama}-\eqref{Vl} in the  presence 
of a weak staggered longitudinal magnetic field $h_z$ was obtained in \cite{Rut18,Rut22} and validated in 
\cite{Rut22} by the alternative derivation in the frame of the more rigorous 
and systematic approach exploiting the 
Bethe-Salpeter equation. 

Now let us switch on the transverse magnetic field  $\mathbf{h}_t=h_2 \mathbf{e}_x+h_1 \mathbf{e}_y$
in the XXZ model Hamiltonian \eqref{Hama}-\eqref{Vl}, and address the main question of our interest:
how tuning the   transverse magnetic field 
effects the meson energies in the weak confinement regime, which takes place in the presence of a weak
staggered longitudinal field  $h_z>0$? In order to clarify this issue, we intend to apply the 
heuristic semiclassical technique outlined above. To this end, three quantities characterising the model 
\eqref{Hama}-\eqref{Vl} at $\mathbf{h}_t\ne 0$ are required.
\begin{enumerate}
\item The dispersion laws $\Omega_{\mu\nu}(p|\mathbf{h}_t)$  of the kink topological excitations 
$|\mathbb{K_{\mu\nu}}(p|\mathbf{h}_t)\rangle$ in the deconfined phase at $h_z=0$. 
To the first order in $\mathbf{h}_t$, these dispersion laws
are given by equations \eqref{Om2}, \eqref{delOm} derived in section \ref{Sec:kink}.
\item
The string tension 
\begin{equation}\label{strt}
\mathfrak{f}=2 h_z \langle \sigma_0^z\rangle+O(h_z^3), 
\end{equation}
that determines the attractive force between two kinks at large distances
in the weak confinement regime at $h_z>0$. Here $\langle \sigma_0^z\rangle$ is the  given by \eqref{lsig}
 staggered spontaneous magnetization  in the presence of the 
transverse magnetic field $\mathbf{h}_t$ at $h_z=0$.
\item
The two-kink scattering matrix in the deconfined phase $h_z=0$, 
which at nonzero transverse magnetic field is determined by the commutation relation \eqref{defFZ}. At $\mathbf{h}_t=0$, 
this commutation relation  reduces to \eqref{scat}.
\end{enumerate}
Note, that the one-kink Bloch states $|\mathbb{K_{\mu\nu}}(p|\mathbf{h}_t)\rangle$ determined by equations
 \eqref{T1w}, \eqref{dlK}
describe spinless 
topological excitations, and their energies  $\Omega_{\mu\nu}(p|\mathbf{h}_t)$ are the  $2\pi$-periodical functions of the quasimomentum $p$.

Following our previous strategy, we treat the two kinks as  classical particles moving in the line and attracting one another
with a linear potential. Their effective Hamiltonian will be taken in the form, which is analogous to  \eqref{Hk}:
 \begin{equation}\label{Hktr}
 H_{tr}(x_1,x_2,p_1,p_2)=\Omega_{10}(p_1|\mathbf{h}_t)+\Omega_{01}(p_2|\mathbf{h}_t)+\mathfrak{f}\cdot(x_2-x_1),
 \end{equation}
where the particle spatial coordinates $x_1,x_2$ satisfy \eqref{x1x2}.

At $h_z=0$, the string tension $\mathfrak{f}$ vanishes, and the 
right-hand side of \eqref{Hktr} reduces to the energy 
 \begin{equation}
\Omega_{10}(p_1|\mathbf{h}_t)+\Omega_{01}(p_2|\mathbf{h}_t) 
 \end{equation}
 of the two-kink Bloch state $|\mathbb{K}_{10}(p_1|\mathbf{h}_t)
\mathbb{K}_{01}(p_2|\mathbf{h}_t)\rangle$, in agreement with \eqref{EnOm}.

\begin{figure}
\centering
\subfloat[]
{\label{eprof1}\includegraphics[width=.9\linewidth]{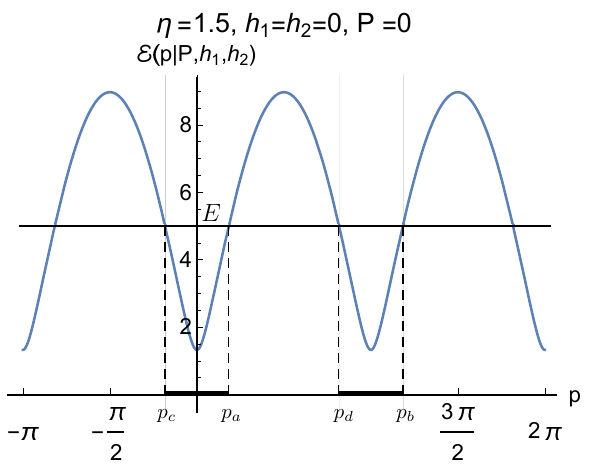}}

\subfloat[]
{\label{eprof2}\includegraphics[width=.9\linewidth]{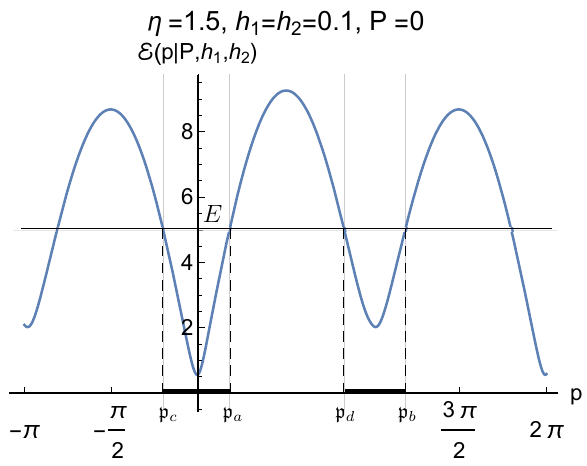}}
\caption{Effective kinetic energy profile $\mathcal{E}(p|P,\mathbf{h}_t)$ given by \eqref{EpP}
at $\eta=1.5$, $ P=0$  at  $h_1=h_2=0$ (a); and  at   $h_1=h_2=0.1$(b). \label{eprof}} 
\end{figure}
 Analysis of  classical dynamics of two particles determined by the Hamiltonian \eqref{Hktr}  is straightforward.
 After the canonical transformation \eqref{Canon}, 
the latter  takes  the form
 \begin{equation}\label{Hktr}
H_{tr}(x,p|P)=\mathcal{E}(p|P,\mathbf{h}_t)-\mathfrak{f}\, x,
\end{equation}
where $\mathcal{E}(p|P,\mathbf{h}_t)$  is given by \eqref{EpP}, and  $x<0$.   The canonical equations of motion 
\eqref{can} now modify to:
\begin{subequations}\label{can2}
\begin{align}
&\dot{X}(t)=\frac{\partial \mathcal{E}(p|P,\mathbf{h}_t)}{\partial P},\\
&\dot{P}(t)=0,\\\label{xt2}
&\dot{x}(t)=\frac{\partial \mathcal{E}(p|P,\mathbf{h}_t)}{\partial p},\\
&\dot{p}(t)=\mathfrak{f}.\label{pt2}
\end{align}
\end{subequations}

At $\mathbf{h}_t=0$ the effective kinetic energy  
$\mathcal{E}(p|P,\mathbf{h}_t)$ reduces to 
the function
$
\mathcal{E}(p|P,\mathbf{0})=\epsilon(p|P), 
$
which is even and $\pi$-periodical in the $p$-variable. These symmetries are broken at  $\mathbf{h}_t\ne 0$,
 though the function $\mathcal{E}(p|P,\mathbf{h}_t)$ remains $2\pi$-periodical in $p$ at any $\mathbf{h}_t$
and $P$.

Figures \ref{eprof1} and \ref{eprof2} illustrate  the  profiles of the effective kinetic energy \eqref{EpP} at $ P=0$ in the cases $\mathbf{h}_t=0$ and  $\mathbf{h}_t\ne 0$, respectively. 
Two  kinematically allowed lacunas  in the $p$-variable, which are determined by the inequality
\begin{equation}
\mathcal{E}(p|P,\mathbf{h}_t)<E,
\end{equation} 
are also shown in these figures. Accordingly, for a given energy $E$ of two particles, 
 the canonical equations of motion \eqref{can2} admit two periodical phase trajectories, in which
the momentum $p$ varies in the left, and in the right lacuna, respectively. In the case $\mathbf{h}_t=0$ illustrated in 
Figure \ref{eprof1}, these two phase trajectories being shifted one with respect to  another by $\pi$ in the 
$p$-variable, describe, in fact, the same physical dynamics. In contrast, at $\mathbf{h}_t\ne0$ the $\pi$-periodicity 
 in the $p$-variable of the function  $\mathcal{E}(p|P,\mathbf{h}_t)$ is broken, 
and the phase trajectories corresponding to the lacunas $(\mathfrak{p}_c,\mathfrak{p}_a)$ and 
$(\mathfrak{p}_d,\mathfrak{p}_b)$ shown in Figure \ref{eprof2}, describe slightly different dynamics.

Existence of two non-equivalent periodical phase trajectories, which contributions interfere in the
quantum state of the system, strongly effects the semiclassical quantization of the
two-kink dynamics in the presence of the transverse magnetic field $\mathbf{h}_t\ne0$. The energy spectrum of the 
two-kink bound states cannot be determined in this case by means of the standard Bohr-Sommerfeld quantization 
rule. Instead, we will adopt to this end the well-known procedure, which was developed for calculation
of the energy spectra of  conducting electrons in normal metals in the regime of the magnetic breakdown, see e.g. \cite{KadSl72,Abr17}. 

According to this procedure, the wave function  of two kinks should be written as a linear combination
of two semiclassical exponents, which correspond to the classical phase trajectories associated with the left 
and right lacunas in Figure~\ref{eprof2}. This wave function describes the quantum state of two kinks at 
large enough separations $|x|=x_2-x_1$ between them. When two kinks meet together 
in the real space, they undergo the elastic two-channel quantum scattering, 
which, roughly speaking, induce the quantum jumps of the phase point  from one lacuna to another.
Sewing  at small $x_2-x_1$   the semiclassical  wave function  by means of the scattering matrix yields the 
secular equation, which determines the semiclassical energy spectrum of the two-kink bound states. 

The outlined above procedure of calculation of the semiclassical meson energy spectrum in the presence of the 
transverse magnetic field will be described in much details in the next section.
\section{Quantitative analysis of the kink confinement at a weak $h_z>0$\label{Heur}}
Let us introduce the Fourier coefficients $\psi_n(j|P)$
of the $2\pi$-periodical reduced wave function $\phi_n(p|P)$ determined by equation \eqref{Phph}:
\begin{align}\label{phps}
&\phi_n(p|P)=\sum_{j=-\infty}^\infty  e^{-i p j}\psi_n(j|P),\\\label{redps}
&\psi_n(j|P)=\int_0^{2\pi}\frac{dp}{2\pi} e^{i p j} \phi_n(p|P).
\end{align}
For negative $j$,  $\psi_n(j|P)$ can be viewed as  the reduced wave function of two kinks forming a meson, in the 
spatial coordinate representation. 
The integer variable $-j=j_2-j_1$ at $j<0$ has the physical meaning of the distance between two kinks 
"located near the points" $j_1$ and $j_2$, respectively, with $j_1<j_2$. Of course, if $\eta$ is not too large, these kinks
are not well localized, but instead are spread along the spin chain over the  widths of order $2\xi_c(\eta)$,
where 
\begin{equation}\label{corr}
\xi_c(\eta)=\frac{m(\eta)^{-1}}{2 } 
\end{equation}
is the correlation length, 
and 
\begin{equation}\label{mass}
m(\eta)=\frac{k'(\eta)}{k(\eta)}
\end{equation}
is the kink mass. 

At large negative $j<0$, such that $|j|\gg 2 \xi_c(\eta)$, the kinks forming a meson  interact one with another 
only by the linear attractive potential $\mathfrak{f}\,(j_2-j_1)=-\mathfrak{f} j$, where $\mathfrak{f}$ is the string tension 
\eqref{strt} in the presence of the transverse magnetic field $\mathbf{h}_t$.
Accordingly, the wave function $\psi(j|P)$ should satisfy at large negative $j$ the following equation:
\begin{equation}\label{Shdis1}
\hat{\mathcal{E}}\,\psi(j|P) - \mathfrak{f} \,j \,\psi(j|P) = E(P)\,\psi(j|P),
\end{equation}
where $\hat{\mathcal{E}}$ is the integral convolution operator of the "kinetic energy of two kinks":
\begin{equation}\label{conv}
\hat{\mathcal{E}}\,\psi(j) =\sum_{j'=-\infty}^\infty \psi(j')\int_0^{2\pi} \frac{dp}{2\pi}\,e^{i p (j-j')}\mathcal{E}(p|P).
\end{equation}
The kernel of this integral operator exponentially decays $\sim \exp\{-|j-j'|/[2\xi_c(\eta)]\}$ at large distances 
$|j-j'|\gg 2\xi_c(\eta)$.

The function $\mathcal{E}(p|P)$ that stands  in the right-hand side of \eqref{conv} is the familiar 
effective kinetic energy 
of two kinks  \eqref{EpP}, which properties were discussed in the previous section.

At a small enough $P\in (0,P_c(\eta)$) with $P_c(\eta)$ given by \eqref{Pc1},
 and at a fixed $E\in (\epsilon({0}|P),\epsilon({\pi/2}|P))$, 
equation 
\begin{equation}\label{kinEn1}
\epsilon(p|P)=E, 
\end{equation}
has in the interval  $-\pi/2<p<3/2\pi$ four solutions 
$p_c(E),p_a(E), p_d(E),  p_b(E)$, which are shown
in Figure \ref{eprof1}.  Due to the symmetry relations \eqref{symE1}, we get:
\begin{equation}
p_c=-p_a, \quad p_b=p_a+\pi, \quad p_d=\pi-p_a.
\end{equation}
Accordingly, the  two classically allowed intervals $(p_c, p_a)$  and
$(p_d,p_b)$ for  the momentum $p$ have the same  widths in this case. 

Application of the weak transverse magnetic field $\mathbf{h}_t$ deforms the kinetic energy profile, as it is shown in 
Figure~\ref{eprof2}.
As the result, the four solutions  $\mathfrak{p}_{c},\mathfrak{p}_{a}, \mathfrak{p}_{d}, \mathfrak{p}_{b}$, of the equation 
\begin{equation}\label{EqE1}
\mathcal{E}(p|P,\mathbf{h}_t)=E
\end{equation}
become slightly shifted with respect to their positions at $\mathbf{h}_t=0$. 
To the first order in  $|\mathbf{h}_t|$, we get from \eqref{EqE1}, \eqref{EpP}:
 \begin{align}\label{pab}
 &\mathfrak{p}_{a}(\mathbf{h}_t)=p_a -  \frac{\delta \mathcal{E}(p_a )|P,\mathbf{h}_t)}{\epsilon'(p_a|P)}+O(|\mathbf{h}_t|^2),\\\nonumber
 &\mathfrak{p}_{c}(\mathbf{h}_t)=p_c + \frac{\delta \mathcal{E}(-p_a |P,\mathbf{h}_t)}{\epsilon'(p_a |P)}+O(|\mathbf{h}_t|^2),\\\nonumber
 &\mathfrak{p}_{b}(\mathbf{h}_t)=p_b+ \frac{\delta \mathcal{E}(p_a|P,\mathbf{h}_t)}{\epsilon'(p_a|P)}+O(|\mathbf{h}_t|^2),\\\nonumber
 &\mathfrak{p}_{d}(\mathbf{h}_t)=p_d- \frac{\delta \mathcal{E}(-p_a |P,\mathbf{h}_t)}{\epsilon'(p_a |P)}+O(|\mathbf{h}_t|^2).
\end{align}
The widths of two kinematically allowed intervals $(\mathfrak{p}_c,\mathfrak{p}_a) $, and $(\mathfrak{p}_d,\mathfrak{p}_b) $ become
different at $\mathbf{h}_t\ne \mathbf{0}$.

At a small $\mathfrak{f}>0$, the approximate semiclassical solution of equation \eqref{Shdis1} can be written as
\begin{equation}\label{ps2aa}
\psi(j|P)=C_1 \psi^{(1)}(j|P)+C_2 \psi^{(2)}(j|P), 
\end{equation}
where
\begin{subequations}\label{ps12a}
\begin{align}\label{ps1}
\psi^{(1)}(j|P)=\int_{-\pi/2}^{\pi/2} \frac{dp}{2\pi}\,e^{i p j-i F_1(p,E|\mathbf{h}_t)/\mathfrak{f}},\\\label{pS2}
\psi^{(2)}(j|P)=\int_{\pi/2}^{3\pi/2} \frac{dp}{2\pi}\,e^{i p j-i F_2(p,E|\mathbf{h}_t)/\mathfrak{f}},
\end{align}
\end{subequations}
and
\begin{subequations}\label{F12a}
\begin{align}\label{F_1}
F_1(p,E|\mathbf{h}_t)=\int_0^p dp' [\mathcal{E}(p'|P,\mathbf{h}_t)- E ],\\
F_2(p,E|\mathbf{h}_t)=\int_\pi^p dp' [\mathcal{E}(p'|P,\mathbf{h}_t)- E ].
\end{align}
\end{subequations}

The integrals in \eqref{ps12a} are determined at small $\mathfrak{f}$ by their saddle-points asymptotics. 
Close to the scattering region at $-j\sim 2 \xi_c(\eta)$, this yields to the leading order in $\mathfrak{f}>0$:
\begin{equation}\label{ps2ba}
\psi(j|P)=B_{in,1} e^{i \mathfrak{p}_a j}+B_{in,2} e^{i \mathfrak{p}_b j}+B_{out,1} e^{i \mathfrak{p}_c j}+B_{out,2} e^{i \mathfrak{p}_d j}, 
\end{equation}
and
\begin{align}\label{C12AB2a}
C_1=B_{in,1}\sqrt{\frac{2 \pi \mathcal{E}'(\mathfrak{p}_a|P,\mathbf{h}_t)}{\mathfrak{f}}}\, e^{i F_1(\mathfrak{p}_a,E|\mathbf{h}_t)/\mathfrak{f}+i \pi/4}\\\nonumber
=B_{out,1}\sqrt{-\frac{2 \pi \mathcal{E}'(\mathfrak{p}_c|P,\mathbf{h}_t)}{\mathfrak{f}}}\, e^{i F_1(\mathfrak{p}_c,E|\mathbf{h}_t)/\mathfrak{f}-i \pi/4},\\\nonumber
C_2=B_{in,2}\sqrt{\frac{2 \pi \mathcal{E}'(\mathfrak{p}_b|P,\mathbf{h}_t)}{\mathfrak{f}}}\, e^{i F_2(\mathfrak{p}_b,E|\mathbf{h}_t)/\mathfrak{f}+i \pi/4}\\\nonumber
=B_{out,2}\sqrt{-\frac{2 \pi \mathcal{E}'(\mathfrak{p}_d|P,\mathbf{h}_t)}{\mathfrak{f}}}\, e^{i F_2(\mathfrak{p}_d,E|\mathbf{h}_t)/\mathfrak{f}-i \pi/4}. 
\end{align}

On the other hand, the amplitudes of out- and in-  plane waves  at $-j\gtrsim 2 \xi_c(\eta)$  in equation 
\eqref{ps2ba} must be related by the scattering condition. Neglecting the effect of the 
weak longitudinal staggered magnetic field $h_z$ on the  two-kink scattering at small distances, we 
obtain from \eqref{defFZ}:
\begin{align}\label{Bab}
B_{out,1}=W_1^1(E,P|\mathbf{h}_t) B_{in,1}+W_1^2(E,P|\mathbf{h}_t) B_{in,2}+O(h_z),\\\nonumber
B_{out,2}=W_2^1 (E,P|\mathbf{h}_t)B_{in,1}+W_2^2 (E,P|\mathbf{h}_t)B_{in,2}+O(h_z).
\end{align}
Combining \eqref{Bab} with \eqref{C12AB2a}, we arrive to the system of two uniform linear equations
on the amplitudes $B_{in,1}$, $B_{in,2}$:
\begin{align}\label{B12}
&B_{in,1}\sqrt{\mathcal{E}'(\mathfrak{p}_a|P,\mathbf{h}_t)}e^{i F_1(\mathfrak{p}_a,E|\mathbf{h}_t)/\mathfrak{f}+i \pi/4}\\\nonumber
&=[W_1^1(E,P|\mathbf{h}_t)  B_{in,1}+W_1^2 (E,P|\mathbf{h}_t)B_{in,2}]\\\nonumber
&\times \sqrt{-\mathcal{E}'(\mathfrak{p}_c|P,\mathbf{h}_t)}
e^{i F_1(\mathfrak{p}_c,E|\mathbf{h}_t)/\mathfrak{f}-i \pi/4},\\\nonumber
&B_{in,2}\sqrt{\mathcal{E}'(\mathfrak{p}_b|P,\mathbf{h}_t)}e^{i F_2(\mathfrak{p}_b,E|\mathbf{h}_t)/\mathfrak{f}+i \pi/4}\\\nonumber
&=[W_2^1(E,P|\mathbf{h}_t)  B_{in,1}+W_2^2 (E,P|\mathbf{h}_t)B_{in,2}]\\\nonumber
&\times \sqrt{-\mathcal{E}'(\mathfrak{p}_d|P,\mathbf{h}_t)}
e^{i F_2(\mathfrak{p}_d,E|\mathbf{h}_t)/\mathfrak{f}-i \pi/4}.
\end{align}
Equating the determinant of these equations to zero leads to the secular equation, which determines the meson energy spectrum $E_n(P| \mathbf{h}_t,h_z)$.
 
In the analysis described above, we did not require, that the components $h_1,h_2$ of the transverse magnetic field 
are infinitesimally  small. Now let us treat $h_1,h_2$ as small 
parameters  and simplify the secular equation in two steps.

On the first step, we replace in \eqref{B12} 
 the  scattering matrix $W_{i'}^i(E,P|\mathbf{h}_t)$ by the zero-order term in its
expansions in $\mathbf{h}_t$:
 \begin{align}\label{Ww}
 &W_1^1(E,P|\mathbf{h}_t)=\frac{w_0+w_+}{2} +O(\mathbf{h}_t),\\\nonumber
&W_1^2(E,P|\mathbf{h}_t)= \frac{-w_0+w_+}{2} +O(\mathbf{h}_t),\\\nonumber
&W_2^1(E,P|\mathbf{h}_t)=\frac{-w_0+w_+}{2} +O(\mathbf{h}_t), \\\nonumber
&W_2^2(E,P|\mathbf{h}_t)=\frac{w_0+w_+}{2} +O(\mathbf{h}_t),
  \end{align}
where 
\begin{align}\label{w0+}
&w_0=w_0(P/2+p_a,P/2-p_a), \\\nonumber
&w_+=w_+(P/2+p_a,P/2-p_a).
  \end{align}
We  also replace the string tension
$\mathfrak{f}=2 h_z \langle{\sigma}_0^z\rangle$ by its value $f=2 h_z \bar{\sigma}$ at $\mathbf{h}_t=0$, taking
into account that the linear 
in $\mathbf{h}_t$ term in the expansion of  $\langle{\sigma}_0^z\rangle$  vanishes:
$\langle{\sigma}_0^z\rangle=\bar{\sigma}+O(|\mathbf{h}_t|^2)$.
As the result, the secular equation takes the form:
  \begin{equation}\label{DDet}
 \det \mathbf{D}=0,
  \end{equation}
   where $\mathbf{D}$ is the $2\times 2$-matrix with the following entries:
\begin{align}\label{DD}
&D_{11}=\frac{w_0+w_+}{2}\\\nonumber
&-i \sqrt{\frac{\mathcal{E}'(\mathfrak{p}_a|P,\mathbf{h}_t)}{|\mathcal{E}'(\mathfrak{p}_c|P,\mathbf{h}_t)|}}\,
e^{i[ F_1(\mathfrak{p}_a,E|\mathbf{h}_t)] -F_1(\mathfrak{p}_c,E|\mathbf{h}_t)]/{f}},\\\nonumber
&D_{12}=D_{21}= \frac{-w_0+w_+}{2},\\\nonumber
&D_{22}=\frac{w_0+w_+}{2}\\\nonumber
&-i \sqrt{\frac{\mathcal{E}'(\mathfrak{p}_b|P,\mathbf{h}_t)}{|\mathcal{E}'(\mathfrak{p}_d|P,\mathbf{h}_t)|}}\,
e^{i[ F_2(\mathfrak{p}_b,E|\mathbf{h}_t)] -F_2(\mathfrak{p}_d,E|\mathbf{h}_t)]/{f}}.
\end{align}

At the second step, we hold the linear 
terms in $\mathbf{h}_t$ in the functions $F_{1,2}(\mathfrak{p}_j,E|\mathbf{h}_t)$, $j=a,b,c,d$:
 \begin{align}\label{F12ht}
&F_1(\mathfrak{p}_a,E|\mathbf{h}_t)=F_1(p_a,E|0)\\\nonumber
&+ \int_0^{p_a}dp\, \delta \mathcal{E}(p|P,\mathbf{h}_t)+O(|\mathbf{h}_t|^2),\\\nonumber
&F_1(\mathfrak{p}_c,E|\mathbf{h}_t)=-F_1(p_a,E|0)\\\nonumber
&- \int_{-p_a}^0dp\,  \delta \mathcal{E}(p|P,\mathbf{h}_t)+O(|\mathbf{h}_t|^2),\\\nonumber
&F_2(\mathfrak{p}_b,E|\mathbf{h}_t)=F_1(p_a,E|0)\\\nonumber
&- \int_0^{p_a}dp\, \delta \mathcal{E}(p|P,\mathbf{h}_t)+O(|\mathbf{h}_t|^2),\\\nonumber
&F_2(\mathfrak{p}_d,E|\mathbf{h}_t)=-F_1(p_a,E|0)\\\nonumber
&+ \int_{-p_a}^0dp\,  \delta \mathcal{E}(p|P,\mathbf{h}_t)+O(|\mathbf{h}_t|^2),
 \end{align}
 since these functions
are divided by the small parameter ${f}\sim h_z$ in the exponent factors in \eqref{DD}. 
 In the functions under the square roots in the right-hand side of \eqref{DD}, we put $\mathbf{h}_t=0$:
 \begin{align}\label{ht0}
&\mathcal{E}'(p_a|P,\mathbf{h}_t)=\epsilon'(p_a^{(0)}|P)+O(\mathbf{h}_t),\\\nonumber
&\mathcal{E}'(p_c|P,\mathbf{h}_t)=-\epsilon'(p_a^{(0)}|P)+O(\mathbf{h}_t),\\\nonumber
&\mathcal{E}'(p_b|P,\mathbf{h}_t)=\epsilon'(p_a^{(0)}|P)+O(\mathbf{h}_t),\\\nonumber
&\mathcal{E}'(p_d|P,\mathbf{h}_t)=-\epsilon'(p_a^{(0)}|P)+O(\mathbf{h}_t).
 \end{align}

Upon  substitutions \eqref{Ww}, \eqref{F12ht}, and \eqref{ht0}, 
equations \eqref{B12} reduce to the form: 
\begin{align}\label{B12a}
B_{in,1}\frac{w_0+w_+}{2}+B_{in,2}\frac{-w_0+w_+}{2}
=B_{in,1} \,e^{i \Lambda_0}\,Z_0 ,\\\nonumber
B_{in,1}\frac{-w_0+w_+}{2}+B_{in,2}\frac{w_0+w_+}{2}
=B_{in,2} \,e^{-i \Lambda_0}\,Z_0,
\end{align}
where
\begin{align}\label{w0p1}
&Z_0(E,P)=\exp[{2 i F_1(p_a,E|0)/f+i \pi/2}]\\\nonumber
&=\exp\left\{\frac{2 i}{f} \int_{0}^{p_a} dp [\epsilon(p|P)-E] +\frac{i \pi}{2}
\right\},\\\label{Lam1}
&\Lambda_0(E,P)=\frac{1}{f}\int_{-p_a}^{p_a} dp\, \delta\mathcal{E}(p|P,\mathbf{h}_t).
\end{align}
Using equations \eqref{vare}, \eqref{delOm}, the integral in the right-hand side 
of  \eqref{Lam1} can be rewritten in the explicit form, yielding:
\begin{equation}\label{Lam1a}
\Lambda_0(E,P)=- \frac{2 \sinh \eta }{f}A_+\, h_2 \int_{p_2}^{p_1} dp \, \frac{\cos p}{\omega(p)},
\end{equation}
where $p_{1,2}$ are the solutions of the equations
\begin{align}\label{PE}
p_1+p_2=P,\quad \omega(p_1)+\omega(p_2)=E,
\end{align}
such that $-\pi/2<p_2<p_1<\pi/2$.
Note, that parameter $\Lambda_0$ does not depend on the $y$-component of the transverse magnetic field $h_1$,
as one can see from \eqref{Lam1a}.

Equating the determinant of the system \eqref{B12a}   to zero yields the secular equation
\begin{equation}\label{Seq}
Z_0^2-(w_{0}+w_{+}) Z_0 \cos \Lambda_0+w_{0}w_{+}=0,
\end{equation}
that determines the meson energies $E_n(P)$. This
equation  can be rewritten in the more symmetric form, which is convenient for 
numerical calculations:
 \begin{widetext}
\begin{align}\label{sec}
\sin\left[\frac{2 F_1(p_a,E|0)}{f}-
\frac{\theta_0(P/2+p_a,P/2-p_a)+\theta_+(P/2+p_a,P/2-p_a)}{2}
\right]\\\nonumber
-\cos \left[ \frac{\theta_+(P/2+p_a,P/2-p_a)-\theta_0(P/2+p_a,P/2-p_a)}{2}  \right]\,\cos \Lambda_0=0.
 \end{align}
  \end{widetext}
For the ratio $B_{in,2}/B_{in,1}$, we get from \eqref{B12a}:
 \begin{equation}\label{rB}
\frac{B_{in,2}}{B_{in,1}}=\frac{2 Z_0 \exp(i\Lambda_0)-w_+-w_0}{w_+-w_0}.
\end{equation}
The absolute value of the coefficient $B_{in,1}$ can be fixed from the normalization condition \eqref{normphi},
as it is described in Appendix \ref{AppB}. The final result reads: 
 \begin{equation}\label{absB}
|B_{in,1}|^2=\frac{f}{2 p_a \epsilon'(p_a)(1+|B_{in,2}/B_{in,1}|^2)}.
\end{equation}
At zero transverse magnetic field $\mathbf{h}_t=0$, parameter $\Lambda_0$ vanishes,
and equation \eqref {Seq} has two solutions:
\begin{equation}\label{Zw}
Z_0=w_{0}, \quad  \text{and  } Z_0=w_{+}.
\end{equation} 
The first one leads to the equation
\begin{equation}\label{Aw1a}
\exp[2 i F_1(p_a,E|0)/f+i \pi/2]=w_{0},
\end{equation}
which determines in the first semiclassical regime the energy spectrum of the so-called {\it transverse meson modes} (T)  with spin $s=\pm 1$. 
 In the explicit form, the above equation reduced to the semiclassical quantization condition \eqref{EnSem1a}
with $\iota=0$, in agreement with equation (172) in \cite{Rut22}.

The second solution  in \eqref{Zw} leads to the equation 
\begin{equation}\label{Aw2a}
\exp[2 i F_1(p_a,E|0)/f+i \pi/2]=w_{+}.
\end{equation}
Its explicit form is given by equation \eqref{EnSem1a}  for the {\it longitudinal meson modes}  (L) with $s=0$ and parity $\iota=+$, in agreement with  \cite{Rut22}.

  The meson states in  equation \eqref{EnSem1a} are enumerated by the natural number  $n=1,2,\ldots$.
Note, that two equations  \eqref{Aw1a}, and \eqref{Aw2a}   can be joined into the single one,
that can be obtained from 
equation \eqref{sec} by putting $\Lambda_0=0$ in the latter. Solutions $E_n(P)$ of equation \eqref{sec} at $\Lambda_0=0$ determine the spectra of the meson modes in the first semiclassical regime at $\mathbf{h}_t=0$, and $P\in (-P_c(\eta),P_c(\eta))$. The solutions $E_n(P)$ with odd $n$ correspond to the transverse meson modes with spin $s=\pm1$, 
while solutions with even $n$ represent the  longitudinal meson modes with spin $s=0$: 
\begin{equation}\label{EnP}
E_n(P)=
\begin{cases}
E_{0,(n+1)/2}(P), & n \text{ odd},\\
E_{+,n/2}(P), & n \text{ even},
\end{cases}
\end{equation}
where $n=1,2,\ldots$, and $E_{0,n}(P)$, $E_{+,n}(P)$ are given by \eqref{EnSem1a}. 
  At $\mathbf{h}_t\ne0$, the factor $\cos \Lambda_0$ in the right-hand side of \eqref{sec} leads to the mixing of the transverse and longitudinal meson modes.
\begin{figure}
\centering
\subfloat[]
{\label{fig:Sec1}\includegraphics[width=.9\linewidth]{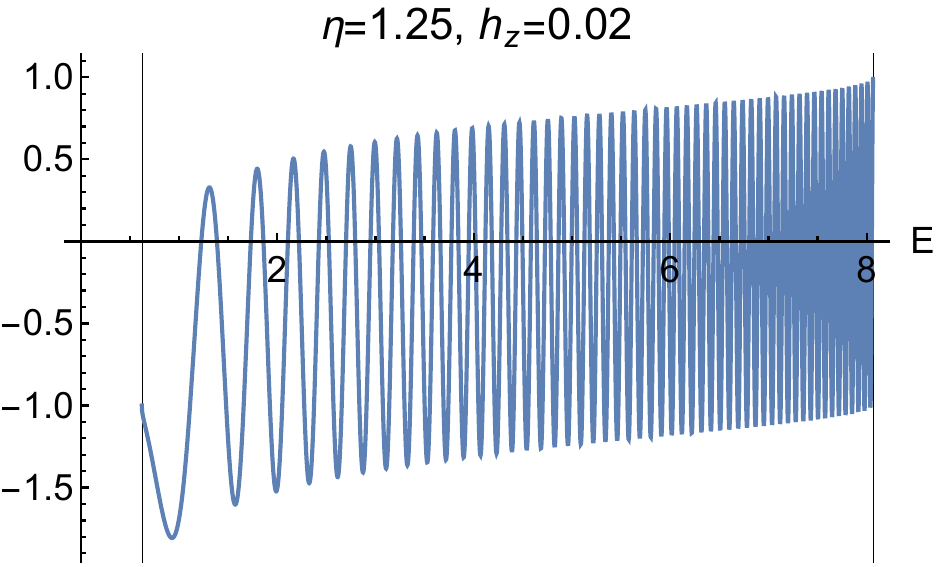}}

\subfloat[]
{\label{fig:Sec2}\includegraphics[width=.9\linewidth]{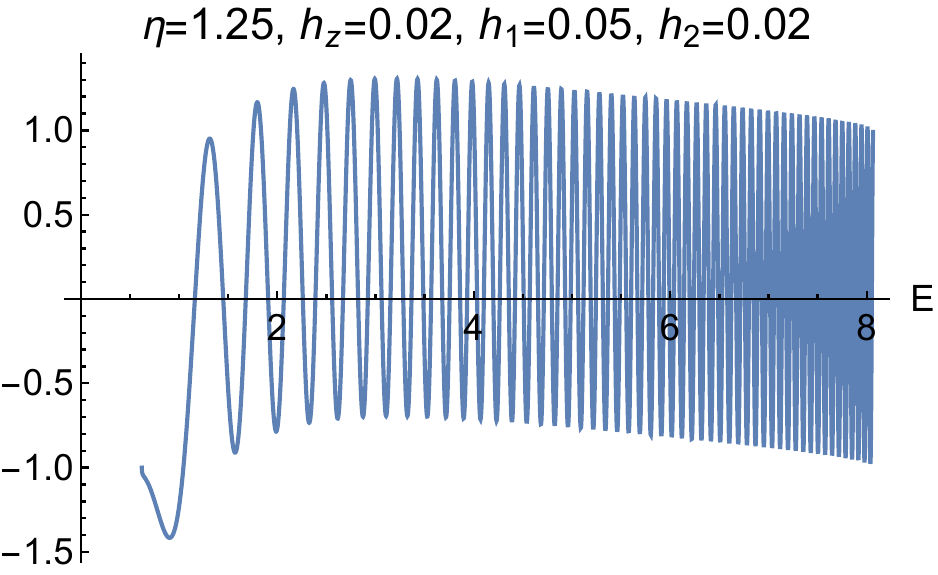}}

\subfloat[]
{\label{fig:SecA}\includegraphics[width=.9\linewidth]{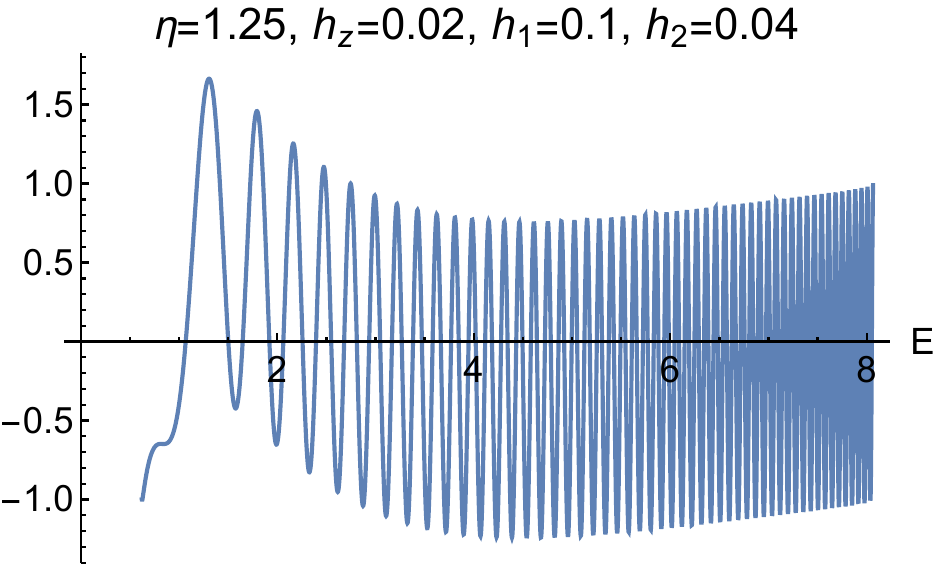}}

\caption{Plot of the left-hand side of \eqref{sec}  versus the energy $E$ of two kinks for $\eta=1.25$, $h_z=0.02$, and $P=0$ at three values of the  transverse magnetic field $y$-component $h_1$: (a) $h_1=0$ ; (b) $h_1=0.05$ ; (c) $h_1=0.1$. The $x$-component 
$h_2$ of the transverse magnetic fields is taken $h_2=0.4 h_1$ in all  cases.\label{fig:Sec}}
\end{figure}

Figures \ref{fig:Sec} display the variation of the left-hans side of the secular equation \eqref{sec} with the energy 
of two kinks $E$  for $P=0$ at three different values $h_1=0,\,0.05, \,0.01$ of the $y$-component of the transverse magnetic field.
The  the $x$-component of the transverse magnetic field is taken by $h_2=0.4 h_1$ in all three cases.
Two other parameters of the Hamiltonian \eqref{Hama} are taken at the fixed values $\eta=1.25$, and $h_z=0.02$. The meson energies $E_n(P)$, $n=1,2,\ldots$ in model   \eqref{Hama} at small enough $h_z, h_1,h_2$   are given in the semiclassical  approximation by zeroes of equation \eqref{sec}.

At $h_1=h_2=0$, the parameter $\Lambda_0$ in the secular equation \eqref{sec} 
vanishes, and $\cos\Lambda_0=1$. Figure \ref{fig:Sec1} displays the energy dependence of the left-hand side of \eqref{sec}  in the latter case.
This oscillating function is defined in the interval between the minimal
and maximal values of the two-kink effective kinetic energy  $\epsilon(p|P, \eta)$, which are indicated by vertical lines in Figure \ref{fig:Sec1}.  The zeroes of this function  are located at the energies  $E_n(P)$ of the transverse and 
longitudinal meson modes, which are given by equation \eqref{EnSem1a} with $\iota=0$, and $\iota=+$, respectively.
\begin{figure}
\centering
\includegraphics[width=.9\linewidth]{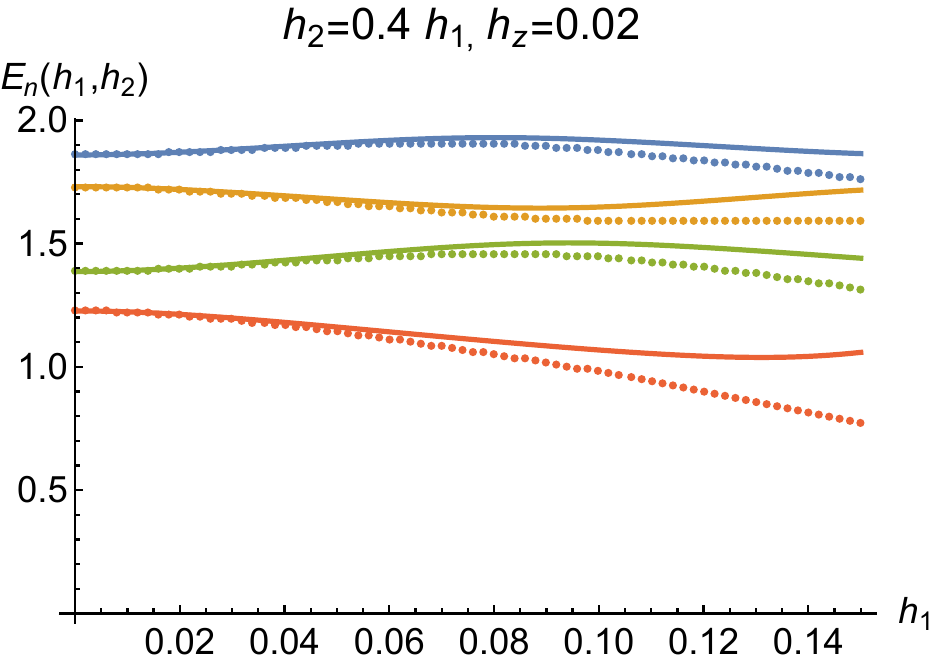}
\caption{Dependences of energies $E_n$ of four lowest meson modes on the $y$-component $h_1$  of the transverse magnetic field  at 
$\eta=1.25$, $P=0$, $h_z=0.02$, $h_2=0.4 h_1$ according to \eqref{sec} (solid curves), and \eqref{DDet} (dotted curves), respectively. \label{fig:Mag}}
\end{figure}

Application of the transverse magnetic field $\mathbf{h}_t$ breaks the symmetry 
between the left and right classically allowed lacunas $\mathfrak{P}^{(1)}=(\mathfrak{p}_c, \mathfrak{p}_a)$ and  $\mathfrak{P}^{(2)}=(\mathfrak{p}_d, \mathfrak{p}_b)$ in the $p$-variable, which are  shown in Figure \ref{eprof2}. As the result, the    wave
functions $\psi^{(1)}(j|P)$ and $\psi^{(2)}(j|P)$ defined by \eqref{ps12a}, which correspond to  the semiclassical
evolution $\langle x(t),p(t)\rangle $ of two kinks with   $p(t)$ in the lacunas $p(t)\in \mathfrak{P}^{(1)}$ and $p(t)\in \mathfrak{P}^{(2)}$, respectively, become different.  The quantum scattering of two kinks upon their collisions at
$|x|\lesssim \xi_c(\eta)$ provides  the effective ``hopping" of the two-kink classical phase point from one lacuna to another. This leads to 
the interference of two semiclassical wave functions $\psi^{(1)}(j|P)$ and $\psi^{(2)}(j|P)$, which, in turn, causes the $\mathbf{h}_t$-dependent modulation  of the function in the left-hand of the secular equation \eqref{sec} shown in Figures \ref{fig:Sec2}, \ref{fig:SecA}. 

Figure \ref{fig:Mag} shows the dependences of energies $E_n(P|\mathbf{h}_t,h_z)$  of four lightest meson modes at $P=0$
on the $y$-component $h_1$ of the transverse magnetic field $\mathbf{h}_t$.  Three other parameters of the Hamiltonian \eqref{Hama}
are taken the same as in Figure \ref{fig:Sec}: $\eta=1.25$, $h_2=0.4 h_1$, $h_z=0.02$. The solid lines in Figure \ref{fig:Mag} represent the meson energies obtained from the secular equation \eqref{sec}. The dotted curves display the meson 
energies, that were obtained from the  secular equation \eqref{DDet}, which we expect to be slightly more accurate.

Comparison of solid and dotted curves in Figure \ref{fig:Mag} allows one to estimate the accuracy of 
the described above calculations of the meson energy spectra. One can conclude from
Figure \ref{fig:Mag}, that the perturbative in $\mathbf{h}_t$ calculations of the meson energies are rather 
accurate at $0<h_1\lesssim 0.1$ for  chosen values of other parameters, but the accuracy rapidly decreases 
at larger $h_1\gtrsim 0.12$. 

As it was noticed above, at $\mathbf{h}_t=0$,  the states $|\tilde{\pi}_n(P)\rangle$ with odd $n=1,3,\ldots$ 
represent the transverse meson modes characterized by the spin polarization $s=\pm1$, while the states with $n=2,4,\ldots$ correspond to the longitudinal meson modes with $s=0$. Application of the 
transverse magnetic field $\mathbf{h}_t\ne0$ leads to the hybridization of the transverse and longitudinal meson modes. 
At very weak  transverse magnetic fields $\mathbf{h}_t$, the energies of the mesons with odd $n$ exhibit an downward variation with increasing $|\mathbf{h}_t|$,  while  the energies of the modes with even $n$ increase together with 
$|\mathbf{h}_t|$.  Upon further increase of the transverse magnetic field, the avoided crossing 
of the energies of the modes with $n=2m$ and $n=2m+1$ takes place. 
As one can see in Figure \ref{fig:Mag}, for the second and third meson modes, this avoided crossing takes
place at $h_1\approx 0.1$, at chosen values of other parameters.

\begin{figure}
\centering
\subfloat[]
{\label{eprofPi1}\includegraphics[width=.9\linewidth]{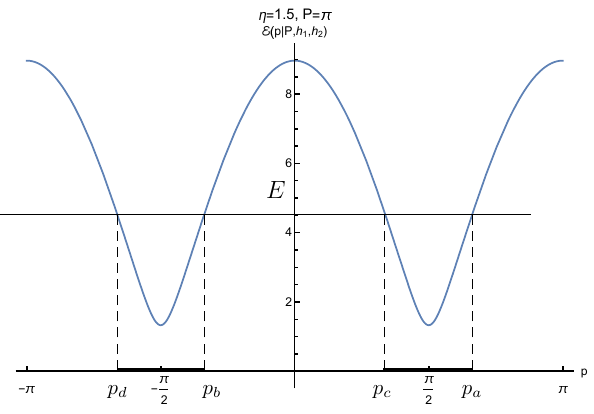}}

\subfloat[]
{\label{eprofPi2}\includegraphics[width=.9\linewidth]{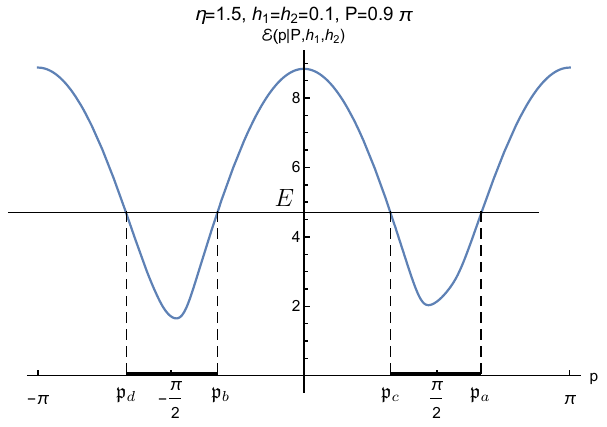}}
\caption{Effective kinetic energy profile $\mathcal{E}(p|P,\mathbf{h}_t)$ given by \eqref{EpP}
at $\eta=1.5$, and (a)  $P=\pi$  and arbitrary $h_1,h_2$;  (b)   $P=0.9\,\pi$ and  $h_1=h_2=0.1$. \label{eprofPi}} 
\end{figure}

Let us turn now to  the energy  spectra of mesons with momenta $P$ close to $\pi$. 
Perturbative calculations of these energy spectra can be performed be means 
 of the  described above heuristic semiclassical technique. The central role in this 
 calculation plays the effective kinetic energy of two kinks $\mathcal{E}(p|P,\mathbf{h}_t)$
 defined by \eqref{EpP}. Due to \eqref{piE},  the first variation with $\mathbf{h}_t$ of this function 
 vanishes at $P=\pi$:
\begin{equation}
\mathcal{E}(p|\pi,\mathbf{h}_t)=\epsilon(p|\pi)+O(|\mathbf{h}_t|^2).
\end{equation}

Figure \ref{eprofPi1} displays the plot of the function $\epsilon(p|\pi)$  at $\eta=1.5$. At a given 
$E\in \left(\epsilon(\pi/2|\pi),\epsilon(0|\pi)\right)$, equation $\epsilon(p|\pi)=E$ has  in the interval $(-\pi/2,\pi/2)$ four solutions
$p_i$, $i=a,b,c,d$, shown in this figure.
It follows from the symmetry properties
\eqref{symE1} of the function $\epsilon(p|P)$, that:
\begin{equation}
p_c=\pi-p_a, \quad p_b=p_a-\pi, \quad p_d=-p_a.
\end{equation}

Figure \ref{eprofPi2} illustrates the plot of the function $\mathcal{E}(p|P,\mathbf{h}_t)$  at $P=0.9\pi$, 
$h_1=h_2=0.1$, and $\eta=1.5$. The four solutions $\mathfrak{p}_i$, $i=a,b,c,d$ 
of the equation $\mathcal{E}(p|P,\mathbf{h}_t)=E$ shown in this figure are slightly shifted from their positions at $P=\pi$, and the widths of the lacunas $(\mathfrak{p}_d,\mathfrak{p}_b)$ and 
$(\mathfrak{p}_c,\mathfrak{p}_a)$ become different.

Subsequent calculations of the meson energy spectra at $P$ close to $\pi$ are very similar to the performed above 
calculations at small $P$. We define by  \eqref{redps} the reduced wave function  
$\psi_n(j|P)$  of two kinks in the spatial coordinate representation. At large negative $j$, this function must satisfy the integral equation \eqref{Shdis1} with the convolution operator \eqref{conv}. We again 
write the  approximate semiclassical solution of equation \eqref{Shdis1} in the form \eqref{ps2aa}, but now
\begin{subequations}\label{ps12aA}
\begin{align}\label{ps1}
\psi^{(1)}(j|P)=\int_{0}^{\pi} \frac{dp}{2\pi}\,e^{i p j-i F_1(p,E|\mathbf{h}_t)/\mathfrak{f}},\\\label{pS2A}
\psi^{(2)}(j|P)=\int_{-\pi}^{0} \frac{dp}{2\pi}\,e^{i p j-i F_2(p,E|\mathbf{h}_t)/\mathfrak{f}},
\end{align}
\end{subequations}
and
\begin{subequations}\label{F12aA}
\begin{align}\label{F_1A}
F_1(p,E|\mathbf{h}_t)=\int_{\pi/2}^p dp' [\mathcal{E}(p'|P,\mathbf{h}_t)- E ],\\
F_2(p,E|\mathbf{h}_t)=\int_{-\pi/2}^p dp' [\mathcal{E}(p'|P,\mathbf{h}_t)- E ].
\end{align}
\end{subequations}

Formula \eqref{ps2ba} still gives us the saddle-point asymptotics of the 
wave function $\psi(j|P)$ at $-j\sim  \xi_c(\eta)$, but now
the four solutions of equation \eqref{EqE1} are ordered as it is shown in Figure \ref{eprofPi2}:
\begin{equation}\label{pp}
-\pi<\mathfrak{p}_d<\mathfrak{p}_b<0<\mathfrak{p}_c<\mathfrak{p}_a<\pi.
\end{equation}
To the leading order in $|\mathbf{h}_t|$, the coefficients $B_{in,i}$, and $B_{out,i}$, $i=1,2$,  in equation \eqref{ps2ba}  must satisfy two linear equations, which are 
analogous to \eqref{B12a}:
\begin{align}\label{B12b}
B_{in,1}\frac{-w_0+w_+}{2}+B_{in,2}\frac{w_0+w_+}{2}
=B_{in,1} \,e^{i \Lambda_\pi}\,Z_\pi ,\\\nonumber
B_{in,1}\frac{w_0+w_+}{2}+B_{in,2}\frac{-w_0+w_+}{2}
=B_{in,2} \,e^{-i \Lambda_\pi}\,Z_\pi,
\end{align}
where
\begin{align}\label{w0p2A}
&Z_\pi(E,P)=\exp\left\{\frac{2 i}{f} \int_{\pi/2}^{p_a} dp [\epsilon(p|P)-E] +\frac{i \pi}{2}
\right\},\\\label{LamA}
&\Lambda_\pi(E,P)=\frac{1}{f}\int_{\pi-p_a}^{p_a} dp\, \delta\mathcal{E}(p|P,\mathbf{h}_t),
\end{align}
and $w_{0}$,   $w_{+}$ are the scattering amplitudes 
given by \eqref{w0+}.  Note, that 
\begin{equation}
Z_\pi(E,P)=Z_0(E,P-\pi).
\end{equation}
For the ratio $B_{in,2}/B_{in,1}$, we get from \eqref{B12b}:
 \begin{equation}\label{rBpi}
\frac{B_{in,2}}{B_{in,1}}=\frac{2 Z_\pi \exp(i\Lambda_\pi)-w_++w_0}{w_++w_0}.
\end{equation}
The  normalization condition \eqref{normphi} leads now to the following formula for the 
absolute value of the coefficient $B_{in,1}$:
 \begin{equation}\label{absBa1}
|B_{in,1}|^2=\frac{f}{2 (p_a-\pi/2) \epsilon'(p_a)(1+|B_{in,2}/B_{in,1}|^2)}.
\end{equation}
We skip derivation of this formula, since it similar to the described in Appendix \ref{AppB} derivation
of formula \eqref{absB}.

The  analogous to  \eqref{Seq} secular equation following from  \eqref{B12b} reads:
\begin{equation}\label{SeqA}
Z_\pi^2-(-w_{0}+w_{+}) Z_\pi \cos \Lambda_\pi-w_{0}w_{+}=0.
\end{equation}

Using equations \eqref{vare}, \eqref{delOm}, the integral in the right-hand side 
of  \eqref{LamA} can be simplified to the form:
\begin{equation}\label{Lam2aA}
\Lambda_\pi(E,P)=- \frac{2 \sinh \eta }{f}A_- h_1 \int_{p_2}^{p_1} dp \, \frac{\sin p}{\omega(p)},
\end{equation}
where $p_{1,2}$  are the solutions of the  equations
\begin{align}\label{PE2}
p_1+p_2=P-\pi,\quad \omega(p_1)+\omega(p_2)=E,
\end{align}
such that $-\pi/2<p_2<p_1<\pi/2$.
The parameter $\Lambda_\pi$ does not depend on the $x$-component of the transverse magnetic field $h_2$,
as one can see from \eqref{Lam2aA}. It is clear also from \eqref{Lam2aA}, \eqref{PE2}, that 
the parameter $\Lambda_\pi$  vanishes in two cases:
(i) at $h_1=0$, and (ii) at $P=\pi$.
If any of these conditions is satisfied,
equation \eqref{SeqA} has two solutions:
\begin{equation}\label{Zpi}
Z_\pi=-w_0, \quad \text{and   }  Z_\pi=w_+.
\end{equation}
Solutions of the first equation determine the spectrum of the meson modes with $s=\pm1$. At $\mathbf{h}_t=0$ 
and 
$P\in (\pi-P_c(\eta),\pi+P_c(\eta))$, their energies are determined by equation \eqref{EnSem1a}, which can be 
extended in $P$ from the its original interval $(-P_c(\eta),P_c(\eta))$ to the interval $(\pi-P_c(\eta),\pi+P_c(\eta))$  due to the 
periodicity relation \cite{Rut22}:
 $E_{0,n}(P)=E_{0,n}(P+\pi)$.
 
 Solutions of the second equation in \eqref{Zpi} determine at  $\mathbf{h}_t=0$ 
and 
$P\in (\pi-P_c(\eta),\pi+P_c(\eta))$ the energies $E_{+,n}(P)$ of the meson modes with spin 
$s=0$ and parity $\iota=+$. It was shown in \cite{Rut22}, that the energy spectra of mesons 
with $s=0$ and parities $\iota=\pm$ are connected by the relation $E_{+,n}(P)=E_{-,n}(P\pm\pi)$,
see equation (131) in \cite{Rut22}.

The solutions $E_n(P)$ of the secular equation \eqref{SeqA} at ${h}_1=0$ and 
$P\in (\pi-P_c(\eta),\pi+P_c(\eta))$ correspond to transverse ($s=\pm1$) and longitudinal 
($s=0$) meson modes at odd, and even $n$, respectively:
\begin{equation}\label{EnP3}
E_n(P)=
\begin{cases}
E_{0,(n+1)/2}(P-\pi), & n \text{  odd},\\
E_{-,n/2}(P-\pi), & n \text{  even},
\end{cases}
\end{equation}
and the functions $E_{\iota,n}(P-\pi)$ are given by  \eqref{EnSem1a}.  

If $P\ne \pi$, the application of the transverse magnetic field with nonzero $y$-component $h_1$ 
gains a nonzero value to the parameter $\Lambda_\pi$ and induces hybridization of transverse and longitudinal
meson modes due to the term $\sim \cos \Lambda_\pi$ in the secular equation \eqref{SeqA}.

However, at any fixed value of $h_1$, the parameter $\Lambda_\pi$ given by \eqref{Lam2aA} decreases
in absolute value as $P$ approaches to $\pi$, and finally vanishes at $P=\pi$. Therefore, the meson energies $E_n(P)$ at 
$P=\pi$  in presence of the transverse magnetic field $\mathbf{h}_t$ are still described by equation 
\eqref{EnP3}. This means, in particular, that the application of the transverse magnetic field $\mathbf{h}_t$
does not lead to hybridisation of transverse and longitudinal meson modes with $P=\pi$, and the 
energies $E_n(\pi)$ of these mesons do not depend on $\mathbf{h}_t$. Of course, this result holds only
in the adopted approximation corresponding to the leading order in the weak transverse magnetic field $\mathbf{h}_t$. 
\section{Dynamical structure factors of spin operators \label{DStF}}
In this section we study the effect of the mutually orthogonal staggered and uniform 
 transverse  magnetic fields on the 
DSF of the local spin operators in the XXZ spin chain in the weak confinement regime. In the thermodynamic limit, this structure
factor   can be  defined as follows:
\begin{align}\label{Smuab0}
&S^{\mathfrak{a}\mathfrak{b}}(\mathrm{k},\omega)=\frac{1}{8 }\sum_{j=-\infty}^\infty e^{-i\mathrm{k} j}\int_{-\infty}^\infty dt \,e^{i\omega t}\\\nonumber
&\times\Big[\phantom{.}^{(1)}  \langle Vac(\mathbf{h}_t,h_z)|e^{i \mathcal{H}t }\sigma_j^\mathfrak{a}e^{-i \mathcal{H}t } \sigma_{0}^\mathfrak{b} |Vac(\mathbf{h}_t,h_z)\rangle^{(1)}+\\
&\phantom{.}^{(1)} \langle Vac(\mathbf{h}_t,h_z)|e^{i \mathcal{H}t } \sigma_{j+1}^\mathfrak{a}e^{-i \mathcal{H}t }  \sigma_{1}^\mathfrak{b} |
Vac(\mathbf{h}_t,h_z)\rangle^{(1)}\Big] ,\nonumber
\end{align}
where $\mathfrak{a},\mathfrak{b}=x,y,z$,  $\mathcal{H}$ is the  Hamiltonian given by \eqref{Hama}, 
$|Vac(\mathbf{h}_t,h_z)\rangle^{(1)}$ is 
its ground-state. Exploiting equations \eqref{wTs}, \eqref{com}, and \eqref{vacT}, formula \eqref{Smuab0} can be simplified to the form
\begin{align}\label{Smuab}
&S^{\mathfrak{a}\mathfrak{b}}(\mathrm{k},\omega)=\frac{\delta_{d_\mathfrak{a},d_\mathfrak{b}}}{4}
\sum_{j=-\infty}^\infty e^{-i\mathrm{k} j}\int_{-\infty}^\infty dt \,e^{i\omega t}\\\nonumber
&\times\phantom{.}^{(1)}  \langle Vac(\mathbf{h}_t,h_z)|e^{i \mathcal{H}t }\sigma_j^\mathfrak{a}e^{-i \mathcal{H}t } \sigma_{0}^\mathfrak{b} |Vac(\mathbf{h}_t,h_z)\rangle^{(1)},
\end{align}
where $d_x=0$, $d_y=d_z=1$. 
Note, that due to the Kronecker-delta  $\delta_{d_\mathfrak{a},d_\mathfrak{b}}$ in the right-hand of \eqref{Smuab},
only two non-diagonal components $S^{yz}(\mathrm{k},\omega)$, and $S^{zy}(\mathrm{k},\omega)$ of the DSF 
tensor \eqref{Smuab} are  non-zero.

As in \cite{Rut22}, we will use two approximations in calculation of the structure factors \eqref{Smuab}.
First, the analysis will be limited to the leading order in the weak  staggered longitudinal magnetic fields  $h_z$.
This allows us to replace the vacuum state $|Vac(\mathbf{h}_t,h_z)\rangle^{(1)}$  
of the Hamiltonian \eqref{Hama}   in equation
\eqref{Smuab} by its  counterpart  at $h_z=0$:
\[
|Vac(\mathbf{h}_t)\rangle^{(1)}=\lim_{{h}_z\to+0}|Vac(\mathbf{h}_t,h_z)\rangle^{(1)}.
\]
Second, our analysis will be restricted 
solely to the two-spinon contribution $S_{(2)}^{\mathfrak{a}\mathfrak{b}}(\mathrm{k},\omega)$
to the structure factor. To the leading order in  $h_z$, the latter is defined by the equation:
\begin{align}\label{DSF20}
&S_{(2)}^{\mathfrak{a}\mathfrak{b}}(\mathrm{k},\omega)=\frac{\delta_{d_\mathfrak{a},d_\mathfrak{b}}}{4}
\sum_{j=-\infty}^\infty e^{-i\mathrm{k} j}\int_{-\infty}^\infty dt \,
e^{i\omega t}\\\nonumber
&\times \phantom{.}^{(1)} \langle Vac(\mathbf{h}_t)|e^{i \mathcal{H}^{(2)}t}\sigma_j^\mathfrak{a}e^{-i \mathcal{H}^{(2)}t}\mathcal{P}_{11}^{(2)} (\mathbf{h}_t ) \sigma_{0}^\mathfrak{b}|Vac(\mathbf{h}_t)\rangle^{(1)},
\end{align}
where $\mathcal{P}_{11}^{(2)}  (\mathbf{h}_t ) $ is the projector operator \eqref{P112a}
onto the two-spinon subspace 
$\mathcal{L}_{11}^{(2)} (\mathbf{h}_t )$, and $\mathcal{H}^{(2)}$ is given by \eqref{H2}.

After substitution of \eqref{P112a} into \eqref{DSF20} and straightforward calculations following the lines 
described in section III of \cite{Rut22}, the dynamical structure factor \eqref{DSF20} can be represented in the
compact form:
\begin{equation}\label{S2ab}
S_{(2)}^{\mathfrak{a}\mathfrak{b}}(\mathrm{k},\omega)=\delta_{d_\mathfrak{a},d_\mathfrak{b}} \sum_{n=1}^\infty
\delta[\omega-\widetilde{E}_n(\mathrm{k}+\pi d_\mathfrak{a})]\, I_n^{\mathfrak{a}\mathfrak{b}}(\mathrm{k}+\pi d_\mathfrak{a}),
\end{equation}
where 
\begin{align}\label{Int}
I_n^{\mathfrak{a}\mathfrak{b}}(P)=\frac{\pi}{2}
\phantom{.}^{(1)} \langle Vac(\mathbf{h}_t)|\sigma_0^\mathfrak{a}|\tilde{\pi}_n(P)\rangle\\\nonumber
\times
\langle \tilde{\pi}_n(P)|\sigma_0^\mathfrak{b}|Vac(\mathbf{h}_t)\rangle^{(1)}
\end{align}
is the intensity of the $n$th meson mode, and $\widetilde{E}_n(P)$ is the meson energy in the two-kink approximation. 
Following our practice noticed by the end of Section \ref{sec_Mes}, we will neglect the difference \eqref{difE}, and
identify $\widetilde{E}_n(P)$ with ${E}_n(P)$. 

The matrix element of the $\sigma_0^\mathfrak{a}$ 
operator in the right-hand side of \eqref{Int} can be 
expressed in terms of the reduced wave function $\phi_n(p|P,\mathbf{h}_t,h_z)$, which is defined 
by equation \eqref{Phph}  and normalized due to \eqref{normphi}:
\begin{align}\label{mel}
&\phantom{.}^{(1)}\langle  Vac(\mathbf{h}_t) | \sigma_0^\mathfrak{a}|  \tilde\pi_{n}(P) \rangle=
\int_{-\pi}^{\pi}\frac{dp}{4\pi}\,\phi_{n}(p|P,\mathbf{h}_t,h_z)\\\nonumber
&\times\phantom{.}^{(1)} \langle Vac(\mathbf{h}_t)| \sigma_0^\mathfrak{a}
| \mathbb{K}_{10}(P/2+p|\mathbf{h}_t)\mathbb{K}_{01}(P/2-p|\mathbf{h}_t)\rangle.
 \end{align} 
 We replace the matrix element in the second line of the above formula by the 
 zero-order term of its expansion in $\mathbf{h}_t$:
 \begin{align}
\phantom{.}^{(1)} \langle Vac(\mathbf{h}_t)| \sigma_0^\mathfrak{a}
| \mathbb{K}_{10}(p_1|\mathbf{h}_t)\mathbb{K}_{01}(p_2|\mathbf{h}_t)\rangle=\\\nonumber
\phantom{.}^{(1)} \langle vac| \sigma_0^\mathfrak{a}
|{K}_{10}(p_1){K}_{01}(p_2)\rangle+O(|\mathbf{h}_t|),
  \end{align} 
  where $\phantom{.}^{(1)} \langle vac|$ and $|{K}_{10}(p_1){K}_{01}(p_2)\rangle$ are defined by 
  equations \eqref{vac}, \eqref{svac}, and \eqref{K2}, respectively.
The matrix element in the right-hand side can be represented in terms of the two-kink states 
$|\mathcal{K}_{10}(\xi_1)\mathcal{K}_{01}(\xi_2)\rangle$ 
parametrized by the multiplicative  spectral parameters $\xi_{1,2}$:
 \begin{align}\label{2kink}
   \phantom{.}^{(1)} \langle vac| \sigma_0^\mathfrak{a}
|{K}_{10}(p_1){K}_{01}(p_2)\rangle= \frac{ \sinh \eta}{\sqrt{\omega(p_1)\omega(p_2)}}\\\nonumber
\times    \phantom{.}^{(1)} \langle vac| \sigma_0^\mathfrak{a}
|\mathcal{K}_{10}(\xi_1)\mathcal{K}_{01}(\xi_2)\rangle,
  \end{align} 
   where  $\xi_{1,2}=-i e^{i\alpha_{1,2}}$,  
 the  rapidities $\alpha_{1,2}$ are relates with momenta $p_{1,2}$ due to \eqref{pe}, and
 \begin{align}\label{2kink1}
 &|\mathcal{K}_{10}(\xi_1)\mathcal{K}_{01}(\xi_2)\rangle=\sqrt{2}\,  |\mathcal{K}_{10}(\xi_1)\mathcal{K}_{01}(\xi_2)\rangle_{+}\\\nonumber
 &+ |\mathcal{K}_{10}(\xi_1)\mathcal{K}_{01}(\xi_2)\rangle_{1/2,1/2} 
+ |\mathcal{K}_{10}(\xi_1)\mathcal{K}_{01}(\xi_2)\rangle_{-1/2,-1/2}.
 \end{align} 
 The two-kink states in the right-hand side of \eqref{2kink1} are defined by equations \eqref{Kxi} and \eqref{baspm1A} in 
 Appendix \ref{Ap1}.
 
 Using \eqref{2kink1} and \eqref{XXF}, the matrix elements    $\phantom{.}^{(1)} \langle vac| \sigma_0^\mathfrak{a}
|\mathcal{K}_{10}(\xi_1)\mathcal{K}_{01}(\xi_2)\rangle$ in the right-hand side of \eqref{2kink}
can by written in  the explicit form:
  \begin{align}
 & \phantom{.}^{(1)} \langle vac| \sigma_0^x
|\mathcal{K}_{10}(\xi_1)\mathcal{K}_{01}(\xi_2)\rangle \\\nonumber
 &=X^0(\xi_1,\xi_2)+X^1(\xi_1,\xi_2),\\
  &\phantom{.}^{(1)} \langle vac| \sigma_0^y
|\mathcal{K}_{10}(\xi_1)\mathcal{K}_{01}(\xi_2)\rangle \\\nonumber
 &=i[X^0(\xi_1,\xi_2)-X^1(\xi_1,\xi_2)],\\
  &\phantom{.}^{(1)} \langle vac| \sigma_0^z
|\mathcal{K}_{10}(\xi_1)\mathcal{K}_{01}(\xi_2)\rangle =\sqrt{2}\,X_+^z(\xi_1,\xi_2),
    \end{align} 
where the functions   $X^j(\xi_1,\xi_2)$ and $X_\pm^z(\xi_1,\xi_2)$ are determined  by equations \eqref{XXall}.
    
 In order to complete calculation of the intensities \eqref{Int}, it remains to obtain the  explicit expression 
 for the wave function $\phi_{n}(p|P,\mathbf{h}_t,h_z)$, that stays in the integrand in the right-hand side of \eqref{mel}.
 In principle, this can be achievd by the perturbative solution of the Bethe-Salpeter equation \eqref{BS}, as it was 
done in \cite{Rut22} in the case $\mathbf{h}_t=0$. Here we shall use a more simple and less rigorous procedure
 exploiting the heuristic semiclassical solutions obtained above in two cases $P\in (-P_c(\eta),P_c(\eta))$, 
 and  $P\in (\pi-P_c(\eta),\pi+P_c(\eta))$.
 
In the first case  $P\in (-P_c(\eta),P_c(\eta))$, the wave function $\phi_{n}(p|P,\mathbf{h}_t,h_z)$  
in the semiclassical approximation can be written in the form:
  \begin{align}\label{delph}
&\phi_{n}(p|P,\mathbf{h}_t,h_z)=2\pi[
B_{in,1}\delta(p-p_a)\\\nonumber
&+B_{in,2}\delta(p-p_b)
+B_{out,1}\delta(p-p_c)+B_{out,2}\delta(p-p_d)
].
 \end{align}  
 Indeed, the result of substitution of the  right-hand side into the integrand in \eqref{redps} 
reproduces  (to the zero  order in $|\mathbf{h}_t|$)  formula \eqref{ps2ba}.
The coefficients $B_{in,i}$, $B_{out,i}$, with $i=1,2$, are known due to 
\eqref{Bab}, \eqref{Ww}, \eqref{rB}, \eqref{absB}.

Combing\eqref{Int}-\eqref{delph}, we obtain finally:
 \begin{subequations}\label{Iab1}
 \begin{align}\label{Ixx1}
 &I_n^{xx}(P)=\frac{\Delta E_n}{2\epsilon'(p_a)}\frac{\sinh^2\eta}{\omega(p_1)\omega(p_2)}\\\nonumber
& \times|X^0(\xi_1,\xi_2)+X^1(\xi_1,\xi_2)|^2 \frac{|1-B_{in,2}/B_{in,1}|^2}{2(1+|B_{in,2}/B_{in,1}|^2)},\\\label{Iyy}
 & I_n^{yy}(P)=\frac{\Delta E_n}{2\epsilon'(p_a)}\frac{\sinh^2\eta}{\omega(p_1)\omega(p_2)}\\\nonumber
 & \times|X^0(\xi_1,\xi_2)-X^1(\xi_1,\xi_2)|^2 \frac{|1-B_{in,2}/B_{in,1}|^2}{2(1+|B_{in,2}/B_{in,1}|^2)},\\\label{Inzz}
&   I_n^{zz}(P)=\frac{\Delta E_n}{2\epsilon'(p_a)}\frac{\sinh^2\eta}{\omega(p_1)\omega(p_2)}\\\nonumber
 & \times 2 |X_+^z(\xi_1,\xi_2)|^2 \frac{|1+B_{in,2}/B_{in,1}|^2}{2(1+|B_{in,2}/B_{in,1}|^2)},\\
  &   I_n^{xz}(P)=\frac{\sqrt{2}\,\,\Delta E_n}{2 \epsilon'(p_a)}\frac{\sinh^2\eta}{\omega(p_1)\omega(p_2)}\\\nonumber
 &\times[X^0(\xi_1,\xi_2)^*+X^1(\xi_1,\xi_2)^*]\, X_+^z(\xi_1,\xi_2)\\\nonumber
&  \times \frac{(1-B_{in,2}^*/B_{in,1}^*)(1+B_{in,2}/B_{in,1})}{2(1+|B_{in,2}/B_{in,1}|^2)},\\
  &   I_n^{yz}(P)=\frac{-i \sqrt{2}\,\,\Delta E_n}{2 \epsilon'(p_a)}\frac{\sinh^2\eta}{\omega(p_1)\omega(p_2)}\\\nonumber
 &\times[X^0(\xi_1,\xi_2)^*-X^1(\xi_1,\xi_2)^*]\, X_+^z(\xi_1,\xi_2)\\\nonumber
&  \times \frac{(1-B_{in,2}^*/B_{in,1}^*)(1+B_{in,2}/B_{in,1})}{2(1+|B_{in,2}/B_{in,1}|^2)},\\
   &    I_n^{zx}(P) = I_n^{xz}(P)^*, \quad   I_n^{zy}(P) = I_n^{yz}(P)^*.
     \end{align} 
   \end{subequations}   
Here $p_a$ is the solution of equation \eqref{kinEn1} shown in Figure~\ref{eprof1}, 
$p_1=P/2+p_a$, $p_2=P/2-p_a$,  $\alpha_{1,2}$ are the  rapidities corresponding to 
the momenta $p_{1,2}$,
and 
\begin{equation}
\Delta E_n=\frac{\pi f}{p_a}
\end{equation}
is the small interval between  the energies  of $n$th and $(n+2)$th meson  modes at given  $P\in (-P_c,P_c)$ and 
$\mathbf{h}_t=0$. 
\begin{figure}
\centering
\subfloat[]
{\label{fig:Sec1a}\includegraphics[width=.9\linewidth]{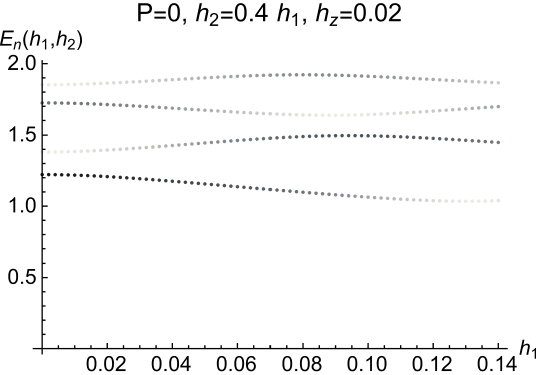}}

\subfloat[]
{\label{fig:Sec2a}\includegraphics[width=.9\linewidth]{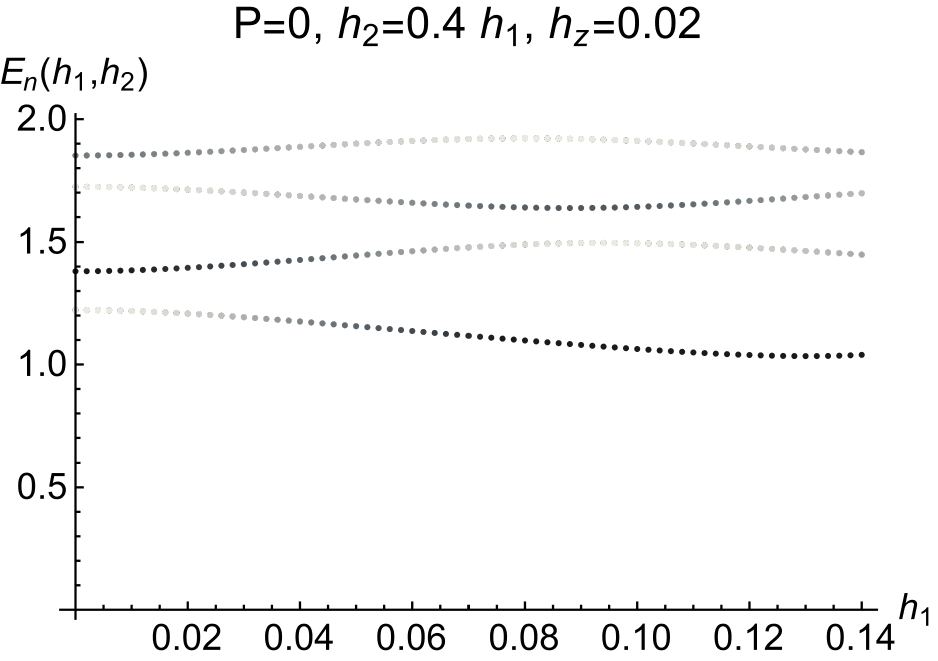}}

\caption{Dependences of energies $E_n$ of four lowest meson modes on the $y$-component $h_1$  of the transverse magnetic field  at 
$\eta=1.25$, $P=0$, $h_z=0.02$, $h_2=0.4 h_1$, according to \eqref{sec}. 
Darkness of the dots in Figures (a), and (b)  characterizes the intensities 
 $I_n^{xx}(0)$, and $I_n^{zz}(0)$, respectively,
which are   given by \eqref{Iab1}. \label{fig:Int1}}
\end{figure}

Obtained results \eqref{Iab1} for the intensities  of the meson modes are
illustrated for the case $P=0$  in Figures \ref{fig:Sec1a}, \ref{fig:Sec2a}. 
 The dotted curves in these figures  are identical with those in Figure \ref{fig:Mag}, and display the 
 dependencies  of the 
energies $E_n(0|\mathbf{h}_t,h_z)$ on the magnetic field $h_1$ at $h_2=0.4 h_1$, and $\eta=1.25$, $h_z=0.02$. 
The darknesses of  dots in Figures \ref{fig:Sec1a},  and Figure \ref{fig:Sec2a}, characterize the intensities  
$I_n^{xx}(P)$, and 
$I_n^{zz}(P)$, respectively, of the presented four lightest meson modes at $P=0$. We did not show the 
intensity  $I_n^{yy}(P)$ determined by equation \eqref{Iyy}, since it is very small for the mesons with zero 
momentum $P=0$
at the   chosen
value  $\eta=1.25$ of the  anisotropy parameter.  As one can see from Figure \ref{fig:Int1}, application of the 
transverse magnetic field causes the effective hybridization of transverse and longitudinal meson
modes at $P=0$. 

In the second case $P\in (\pi-P_c(\eta),\pi+P_c(\eta))$, we can again use formula \eqref{delph}
for the wave function $\phi_{n}(p|P,\mathbf{h}_t,h_z)$. However, the momenta 
$\mathfrak{p}_a,\mathfrak{p}_b,  \mathfrak{p}_c,  \mathfrak{p}_d$  in this formula now are ordered 
in accordance with  \eqref{pp}, and the in- and out-amplitudes in \eqref{delph} are different 
from those in the first case. In particular, the in-amplitudes  solve equations \eqref{B12b}, 
and satisfy equalities \eqref{rBpi} and \eqref{absBa1}. It turns out, that 
 formulas \eqref{Iab1}
can still describe  the intensities $I_n^{\mathfrak{ab}}(P)$  at 
$P\in (\pi-P_c(\eta),\pi+P_c(\eta))$,  upon the following minor modifications: (i) the momentum $p_a$ 
solving equation \eqref{kinEn1}  lies now in the interval $p_a\in (\pi/2,\pi)$, see Figure \ref{eprofPi1},
(ii) the ratio $B_{in,2}/B_{in,1}$ is determined by formula \eqref{rBpi}, instead of \eqref{rB}, (iii) the parameter 
$\Delta E_n$ in \eqref{Iab1} now denotes the ratio:
\begin{equation}
\Delta E_n=\frac{\pi f}{p_a-\pi/2}.
\end{equation}
In particular, the intensities $I_n^{\mathfrak{ab}}(P=\pi)$ do not depend on $h_1, h_2$ in the adopted approximation,
and coincide with their values at $h_1= h_2=0$. 
\section{Comparison with experiment \label{CompExp}}
In previous sections we followed the convention used in the algebraic approach \cite{Jimbo94} by choosing the
$``-"$sign in front of the 
right-hand side of the Hamiltonian \eqref{H0}.  This choice corresponds to the ferromagnetic nearest-neighbor 
exchange coupling  in the $xy$-plane  of the XXZ spin-chain. In the experimental research papers, however, the different and more physically relevant form of the  XXZ model Hamiltonian  is commonly used  with the antiferromagnetic exchange in the $xy$-plane. 
These two forms of the XXZ model Hamiltonian are simply related by a certain unitary transform, which we will recall below in order
to facilitate comparison of our theoretical predictions with the results of  already published and future experiments and computer simulations.

Let us start from the finite-size version $\mathcal{H}_N(\Delta,\mathbf{h}_t,h_z)$ of the  Hamiltonian \eqref{Hama} 
defined on the spin-chain having $N$-sites.  The periodic boundary conditions are implied,
and the number of sites is a multiple of eight, $N\mod 8=0$. 
The appropriate unitary transform $U_N$ is the rotation by $\pi$  around the $z$-axis of all spins at odd sites of the chain:
\begin{equation}\label{Uodd}
U_N=\otimes_{m=-N/4}^{N/4-1} \sigma_{2m+1}^z.
\end{equation}
Operators $\mathcal{H}_N(\Delta,\mathbf{h}_t,h_z)$ and $\widetilde{T}_1=T_1 \mathbb{C}$ modify upon the action
of this transform to:
\begin{align}\label{cH}
&\breve{\mathcal{H}}_N(\breve{\Delta},\mathbf{h}_t,h_z)=U_N\,\mathcal{H}_N(\Delta,\mathbf{h}_t,h_z)\,U_N^{-1},\\\label{T1U}
&\breve{T}_1=U_N \widetilde{T}_1\,U_N^{-1},
\end{align}
where $\breve{\Delta}=-\Delta=\cosh \eta$, and $\breve{T}_1$ is given by \eqref{brTr}. 
The explicit form of the resulting Hamiltonian in the thermodynamic limit 
$N\to\infty$ reads:
\begin{align} \label{Hama2}
&\breve{\mathcal{H}}(\breve{\Delta},\mathbf{h}_t,h_z)=\lim_{N\to\infty} \breve{\mathcal{H}}_N(\breve{\Delta},\mathbf{h}_t,h_z)\\\nonumber
&=\frac{1}{2}\sum_{j=-\infty}^\infty\!\left(\sigma_j^x\sigma_{j+1}^x+\sigma_j^y\sigma_{j+1}^y+
\widetilde{\Delta}\,
\sigma_j^z\sigma_{j+1}^z
\right)\\\nonumber
&-\sum_{j=-\infty}^\infty\left[(-1)^j 
h_{2}\sigma_j^x
+h_{1}\sigma_j^y+(-1)^j h_z \sigma_j^z\right].
\end{align}
We shall use the notation $U$ for the thermodynamic limit of the transform operator \eqref{Uodd}:
\begin{equation}\label{U1}
U=\lim_{N\to\infty} U_N.
\end{equation}
Note, that the unitary transform \eqref{U1}  permutes  the staggered and uniform transverse magnetic fields, 
as one can see by comparison of \eqref{Hama2} with \eqref{Vt}. Note also, that operators 
$\breve{\mathcal{H}}(\breve{\Delta},\mathbf{h}_t,h_z)$ and $\breve{T}_1$  comute:
\begin{equation}\label{comTH}
[\breve{\mathcal{H}}(\breve{\Delta},\mathbf{h}_t,h_z), \breve{T}_1]=0.
\end{equation}

The vacuum and meson states in the new representation are determined by  relations:
\begin{align}\label{VacB}
|\breve{V}ac(\mathbf{h}_t,h_z)\rangle^{(1)}=U  |Vac(\mathbf{h}_t,h_z)\rangle^{(1)},\\\label{mesbr}
|\breve{\pi}_n(P|\mathbf{h}_t,h_z)\rangle=U |{\pi}_n(P|\mathbf{h}_t,h_z)\rangle.
\end{align}
It follows from \eqref{vacT}, \eqref{T1U}, \eqref{U1}, \eqref{VacB} that:
\begin{equation}\label{brT1}
\breve{T}_1 |\breve{V}ac(\mathbf{h}_t,h_z)\rangle^{(1)}=|\breve{V}ac(\mathbf{h}_t,h_z)\rangle^{(1)}.
\end{equation}
 Due to \eqref{cH}-\eqref{U1}, \eqref{mes0}, the meson states \eqref{mesbr}
satisfy equalities:
\begin{subequations}\label{Mes}
\begin{align} \label{mesH1}
&\breve{T}_1|\breve{\pi}_n(P|\mathbf{h}_t,h_z)\rangle=e^{i P}|\breve{\pi}_n(P|\mathbf{h}_t,h_z)\rangle,\\\label{mesP1}
&(\breve{\mathcal{H}}+C)|\breve{\pi}_n(P|\mathbf{h}_t,h_z)\rangle=E_n(P|\mathbf{h}_t,h_z)|\breve{\pi}_n(P|\mathbf{h}_t,h_z)\rangle.
\end{align}
\end{subequations}
The meson dispersion laws $E_n(P|\mathbf{h}_t,h_z)$ and the constant $C$ in equations  \eqref{mesP} and \eqref{mesP1} are the same.

The dynamical structure factor $\breve{S}^{\mathfrak{a}\mathfrak{b}}(\mathrm{k},\omega)$ corresponding to the Hamiltonian \eqref{cH}
is defined by the formula
\begin{align}\label{SmuabA}
&\breve{S}^{\mathfrak{a}\mathfrak{b}}(\mathrm{k},\omega)=\frac{1}{8 }\sum_{j=-\infty}^\infty e^{-i\mathrm{k} j}\int_{-\infty}^\infty dt \,e^{i\omega t}\\\nonumber
&\times\Big[\phantom{.}^{(1)}  \langle \breve{V}ac(\mathbf{h}_t,h_z)|e^{i \breve{\mathcal{H}}t }\sigma_j^\mathfrak{a}e^{-i \breve{\mathcal{H}}t } \sigma_{0}^\mathfrak{b} |\breve{V}ac(\mathbf{h}_t,h_z)\rangle^{(1)}+\\
&\phantom{.}^{(1)} \langle \breve{V}ac(\mathbf{h}_t,h_z)|e^{i \breve{\mathcal{H}}t } \sigma_{j+1}^\mathfrak{a}e^{-i \breve{\mathcal{H}}t }  \sigma_{1}^\mathfrak{b} |
\breve{V}ac(\mathbf{h}_t,h_z)\rangle^{(1)}\Big] ,\nonumber
\end{align}
which is analogous to \eqref{Smuab0}. Exploiting equations \eqref{brTr}, \eqref{comTH},  and   \eqref{brT1}, formula \eqref{SmuabA} 
can be simplified to the form:
\begin{align}\label{Smuab1}
&\breve{S}^{\mathfrak{a}\mathfrak{b}}(\mathrm{k},\omega)=\frac{\delta_{\breve{d}_\mathfrak{a},\breve{d}_\mathfrak{b}}}{4}\sum_{j=-\infty}^\infty e^{-i\mathrm{k} j}\int_{-\infty}^\infty dt \,e^{i\omega t}\\\nonumber
&\times\phantom{.}^{(1)}  \langle \breve{V}ac(\mathbf{h}_t,h_z)|e^{i \breve{\mathcal{H}}t}\sigma_j^\mathfrak{a}
e^{-i \breve{\mathcal{H}}t}\sigma_{0}^\mathfrak{b} |\breve{V}ac(\mathbf{h}_t,h_z)\rangle^{(1)}.
\end{align}
 Due to \eqref{Uodd}, \eqref{cH}, \eqref{U1},  and \eqref{VacB}, this formula can be further rewritten as 
\begin{align}\label{Smuab2}
&\breve{S}^{\mathfrak{a}\mathfrak{b}}(\mathrm{k},\omega)=\frac{\delta_{\breve{d}_\mathfrak{a},\breve{d}_\mathfrak{b}}}{4}\sum_{j=-\infty}^\infty e^{-i\mathrm{k} j}\int_{-\infty}^\infty dt \,e^{i\omega t}\\\nonumber
&\times\phantom{.}^{(1)}  \langle {V}ac(\mathbf{h}_t,h_z)|e^{i {\mathcal{H}}t}\breve{\sigma}_j^\mathfrak{a}
e^{-i {\mathcal{H}}t}\breve{\sigma}_{0}^\mathfrak{b} |{V}ac(\mathbf{h}_t,h_z)\rangle^{(1)},
\end{align}
where the Hamiltonian $\mathcal{H}$ is given by \eqref{Hama}, $ |{V}ac(\mathbf{h}_t,h_z)\rangle^{(1)}$ is its ground state, and
\begin{equation}
\breve{\sigma}_j^\mathfrak{a}= U^{-1}{\sigma}_j^\mathfrak{a}U=\begin{cases}
(-1)^j {\sigma}_j^\mathfrak{a}, & \text{if } \mathfrak{a}=x,y,\\
 {\sigma}_j^z, & \text{if } \mathfrak{a}=z
\end{cases}.
\end{equation}
This leads to the following simple exact relations between the diagonal matrix elements of the DSF  
tensors $\breve{S}^{\mathfrak{ab}}(\mathrm{k},\omega)$ and
${S}^{\mathfrak{ab}}(\mathrm{k},\omega)$:
\begin{subequations}
\begin{align*}
&\breve{S}^{xx}(\mathrm{k},\omega)={S}^{xx}(\mathrm{k}+\pi,\omega), \\
&\breve{S}^{yy}(\mathrm{k},\omega)={S}^{yy}(\mathrm{k}+\pi,\omega),\\
&\breve{S}^{zz}(\mathrm{k},\omega)={S}^{zz}(\mathrm{k},\omega).
\end{align*}
\end{subequations}
On the other hand, the non-diagonal matrix elements of the DSF tensors ${S}^{\mathfrak{ab}}(\mathrm{k},\omega)$
and $\breve{S}^{\mathfrak{ab}}(\mathrm{k}, \omega)$ are different: the only non-vanishing non-diagonal
components of the later are 
$
\breve{S}^{xz}(\mathrm{k}, \omega)
$
and $\breve{S}^{zx}(\mathrm{k}, \omega)$, while the non-vanishing non-diagonal components of $S^{\mathfrak{ab}}(\mathrm{k}, \omega)$
are $S^{yz}(\mathrm{k}, \omega)$ and $S^{zy}(\mathrm{k}, \omega)$, see equation \eqref{Smuab}.

For the non-zero matrix elements $\breve{S}_{(2)}^{\mathfrak{a}\mathfrak{b}}(\mathrm{k},\omega)$ of the DSF \eqref{Smuab1} in the two-kink approximation we get finally:
\begin{subequations}\label{brS}
\begin{align}\label{brSa}
&\breve{S}_{(2)}^{xx}(\mathrm{k},\omega)=\sum_{n=1}^\infty
\delta[\omega-E_n(\mathrm{k}+\pi)]\, I_n^{xx}(\mathrm{k}+\pi),\\\label{brSyy}
&\breve{S}_{(2)}^{yy}(\mathrm{k},\omega)=\sum_{n=1}^\infty
\delta[\omega-E_n(\mathrm{k})]\, I_n^{yy}(\mathrm{k}),\\\label{brSc}
&\breve{S}_{(2)}^{zz}(\mathrm{k},\omega)=\sum_{n=1}^\infty
\delta[\omega-E_n(\mathrm{k}+\pi)]\, I_n^{zz}(\mathrm{k}+\pi),\\
&\breve{S}_{(2)}^{xz}(\mathrm{k},\omega)=\sum_{n=1}^\infty
\delta[\omega-E_n(\mathrm{k}+\pi)]\, I_n^{xz}(\mathrm{k}+\pi),\\
&\breve{S}_{(2)}^{zx}(\mathrm{k},\omega)=\sum_{n=1}^\infty
\delta[\omega-E_n(\mathrm{k}+\pi)]\, I_n^{zx}(\mathrm{k}+\pi).
\end{align}
\end{subequations}
The intensities $I_n^{\mathfrak{a}\mathfrak{b}}(P)$ in the right-hand sides of these relations  are given by \eqref{Iab1}.

\begin{figure}
\centering
\subfloat[Density plot of the longitudinal structure factor $\breve{S}_{(2)}^{zz}(\pi,\omega)$.]
{\label{fig:Szz}\includegraphics[width=.9\linewidth]{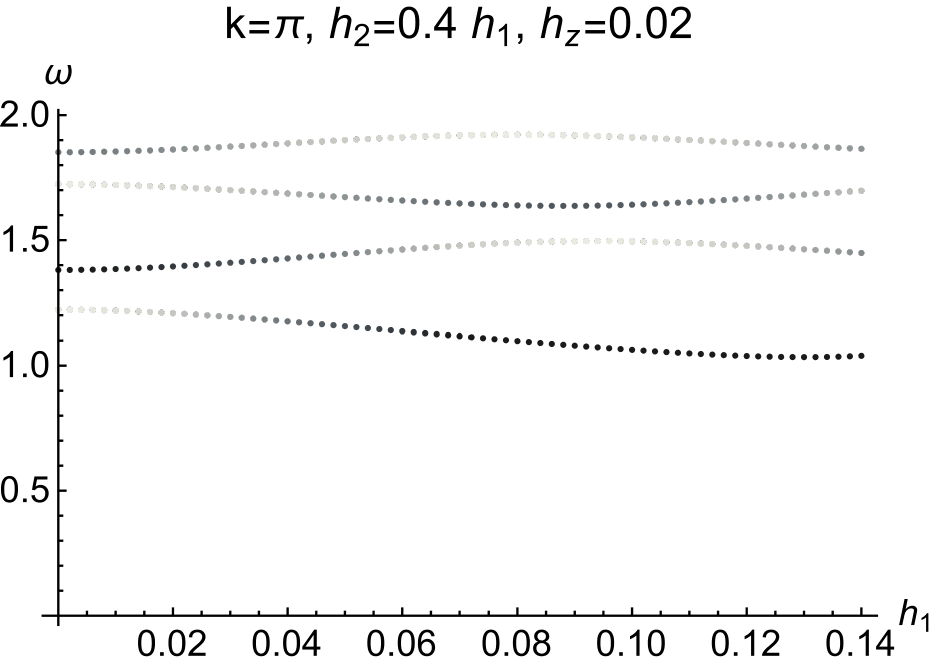}}

\subfloat[Density plots of the transverse DSF. The horizontal straight lines (second and fifth lines from below) display the 
structure factor  $\breve{S}_{(2)}^{yy}(\pi,\omega)$, while the rest curves correspond to the DSF $\breve{S}_{(2)}^{xx}(\pi,\omega)$.]
{\label{fig:Str}\includegraphics[width=.9\linewidth]{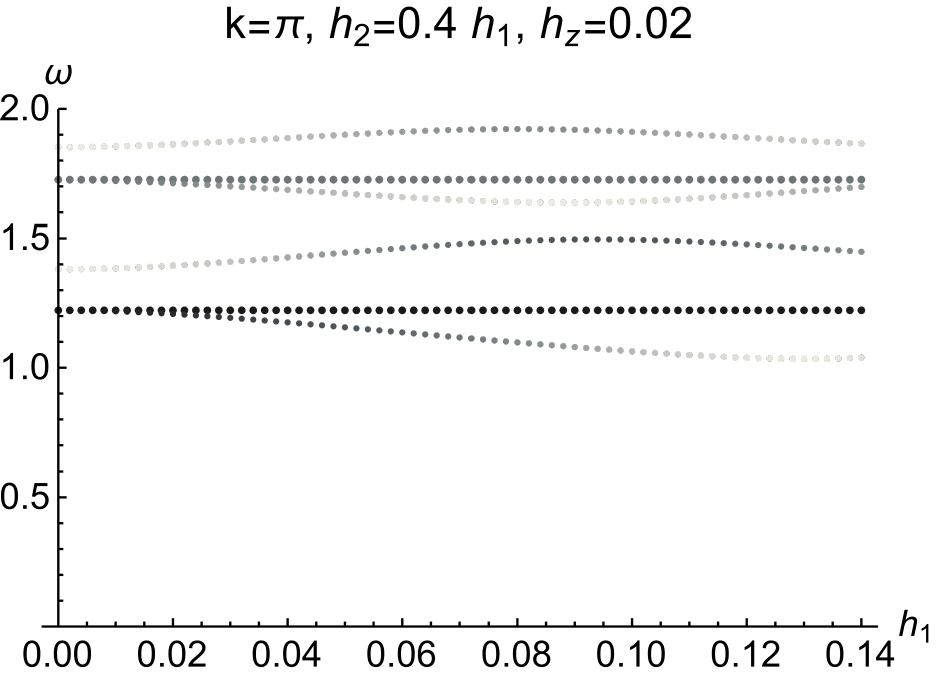}}

\caption{Density plots in the plane $\langle h_1,\omega\rangle$  of the structure factors $\breve{S}_{(2)}^{\mathfrak{a}\mathfrak{a}}(\mathrm{k}=\pi,\omega)$  
due to \eqref{brS} at $h_2=0.4 h_1$, $\eta=1.25$, $h_z=0.02$. Darkness of the 
dots characterizes intensities $I_n^{\mathfrak{a}\mathfrak{a}}(\pi \delta_{y,\mathfrak{a}})$ of the meson modes
with $n=1,2,3,4$, and energies $E_n(\pi \delta_{y,\mathfrak{a}})=\omega$.
\label{fig:St}}
\end{figure}
Figures \ref{fig:Szz} , \ref{fig:Str} display the density plots  in the plane $\langle h_1,\omega\rangle$  of the   DSF $\breve{S}_{(2)}^{\mathfrak{a}\mathfrak{a}}(\mathrm{k}=\pi,\omega)$ with $\mathfrak{a}=x,y,z$, which are   
determined by equations \eqref{brSa}-\eqref{brSc}.  The other parameters are taken at the same values, as in Figure \ref{fig:Int1}:
$h_2=0.4 h_1$, $\eta=1.25$, $h_z=0.02$. 

The density plot of the longitudinal DSF $\breve{S}_{(2)}^{zz}(\pi,\omega)$ is shown in Figure \ref{fig:Szz}. 
Due to the $\delta$-functions in the right-hand side of \eqref{brSc},  it is located along the dispersion curves  
of the mesons with zero quasimomentum:
\begin{equation}\label{ome}
\omega(h_1)=E_n(P=0| h_1,h_2=0.4 h_1),\quad n=1,2,\ldots,
\end{equation}
which were depicted previously in Figure \ref{fig:Int1}. The corresponding intensities 
 $I_n^{zz}(\mathrm{k}=0)$ display substantial variation with increasing $h_1$  in accordance
 with equation \eqref{Inzz}.
 
 The transverse  $\breve{S}_{(2)}^{xx}(\pi,\omega)$ component of the DSF tensor 
 shown in Figure \ref{fig:Str} is located in the 
 plane $\langle h_1,\omega\rangle$  along the same curves \eqref{ome}, and corresponding intensities 
$I_n^{xx}(\mathrm{k}=0)$ also strongly depend on $h_1$ due to \eqref{Ixx1}.
 
 The two remaining horizontal dotted lines in Figure \ref{fig:Str} represent the density plot of the transverse 
component  $\breve{S}_{(2)}^{yy}(\pi,\omega)$  of the DSF tensor, which is given by equation \eqref{brSyy}.
It  is located along the dispersion curves of mesons having the  quasimomentum $P=\pi$:
 \begin{equation}\label{ome2}
\omega(h_1)=E_n(P=\pi| h_1,h_2=0.4 h_1).
\end{equation}
It was shown in Section \ref{Heur}, that 
the energies of such mesons do not 
 depend on $h_1,h_2$:  
 \[
 E_n(P=\pi| h_1,h_2)=  E_n(P=\pi| 0,0),
 \]
to the leading order in $h_1,h_2$. 
 Furthermore, the intensities $I_n^{yy}(\mathrm{k}=\pi)$ that stay in the right-hand side of \eqref{brSyy} 
 also do not depend on $h_1,h_2$: 
  \[
I_n^{yy}(\mathrm{k}=\pi| h_1,h_2)= I_n^{yy}(\mathrm{k}=\pi| 0,0),
 \]
 as it was mentioned by the end of Section \ref{DStF}. These intensities vanish for all even $n=2,4,\ldots$,
 since the meson modes with even $n$ carry zero spin $s=0$ at $\mathbf{h}_t=0$ and do not 
 contribute to the transverse DSF, as it was explained in Section \ref{Heur}, see  equation \eqref{EnP}.
 On the other hand, the following equalities hold for the transverse meson modes with odd $n=1,3,\ldots$ 
 at $\mathbf{h}_t=0$:
 \begin{align*}
&E_n(P=\pi| 0,0)= E_n(P=0| 0,0),\\
&I_n^{yy}(\mathrm{k}=\pi| 0,0)= I_n^{xx}(\mathrm{k}=0| 0,0).
 \end{align*}

 The resulting pattern of the transverse DSF in the $\langle h_1,\omega\rangle$-plane at the antiferromagnetic point $\mathrm{k}=\pi$,
 which is shown in Figure  \ref{fig:Str}, looks  at $h_1=0$ like a set of transverse meson modes.  Each of these modes splits upon 
application of the  transverse magnetic field $h_1>0$ into two branches. The energy of the lower branch has a strong non-linear 
$h_1$-dependence, while the energy of the upper branch does not depend on $h_1$. 
The lower branch of the splited modes becomes polarized in the $\langle x,z\rangle$-plane,
 whereas the upper branch has the linear polarization  along the $y$ direction, that does not change with increase of $h_1$.
 
 The described above rather peculiar evolution of the DSF with increasing transverse magnetic field was indeed observed 
 by  Faure {\it et al.} \cite{Faur17}, who studied in the 
 inelastic neutron scattering experiments the effect of the transverse magnetic field on the magnetic excitation
 energy spectra in the antiferromagnetic crystal $\mathrm{BaCo}_{2}\mathrm{V}_{2}\mathrm{O}_{8}$ at low temperatures. 
 
For interpretation of their   experimental results,  the authors of \cite{Faur17}  have used the
following effective spin-chain Hamiltonian:
\begin{align} \label{HF}
&\mathbb{H}=\frac{J}{4}\sum_{j=-\infty}^\infty\Big[
\varepsilon(\sigma_j^x\sigma_{j+1}^x+\sigma_j^y\sigma_{j+1}^y)+\sigma_j^z\sigma_{j+1}^z
\Big]\\\nonumber
&-\frac{\mu_B}{2}\sum_{j=-\infty}^\infty\Big[
g_{zz} h^{\mathrm{eff}}(-1)^j\sigma_j^z+(-1)^jg_{yx} H \sigma_j^x\\\nonumber
&+g_{yy} H \sigma_j^y+\cos[(2j+1)\pi/4]g_{yz} H \sigma_j^z
\Big],
\end{align}
which was introduced earlier  by Kimura {\it et al.} \cite{Kimura13} for description of the quasi-1D  magnetic 
structure of $\mathrm{BaCo}_{2}\mathrm{V}_{2}\mathrm{O}_{8}$.
 
 In the effective Hamiltonian \eqref{HF}, $J>0$ is the antiferromagnetic intra-chain interaction, 
 $\varepsilon$ is the anisotropy parameter,
 $H$ denotes  the strength of the uniform magnetic field (in Tesla) applied in the 
 $y$-direction, 
 $h^{\mathrm{eff}}$ is the effective staggered longitudinal magnetic field, that mimics in the mean-field approximation the weak
 interchain interaction in the 3D-ordered antiferromagnetic phase, $g_{\mathfrak{a}\mathfrak{b}}$ are the components 
 of the Land{\'e} tensor, and
$\mu_B=5.788\times 10^{-5} \mathrm{eV}/\mathrm{T}$ is the Bohr magneton. 

According to  \cite{Kimura13}, the effective magnetic field $\cos[(2j+1)\pi/4]g_{yz} H$ in the $z$-direction having the 
four-step periodicity arises in \eqref{HF} due to the combination of the anisotropy of the Land{\'e}  $g$-tensor, with the 
specific screw structure of the $\mathrm{Co}^{2+}$
magnetic ion chains in  the compound $\mathrm{BaCo}_{2}\mathrm{V}_{2}\mathrm{O}_{8}$. 
Note also, that the effective spin-chain Hamiltonian
\eqref{HF} can be used to describe the magnetic structure of this crystal only if the external uniform magnetic field
is applied along the $\mathbf{b}$  (or $\mathbf{a}$) crystallographic axis \cite{Kimura13}. 

Comparing \eqref{Hama2} and \eqref{HF}, one can see that the simplified version of the  Hamiltonian $\mathbb{H}$ with $g_{yz}=0$ is proportional to $\breve{\mathcal{H}}(\widetilde{\Delta},\mathbf{h}_t,h_z)$:
\begin{equation}\label{HU}
\mathbb{H}\Big|_{g_{yz}=0}=\frac{J\varepsilon}{2 }\,\breve{\mathcal{H}}(\widetilde{\Delta},\mathbf{h}_t,h_z),
\end{equation}
and parameters of the Hamiltonians $\breve{\mathcal{H}}$ and $\mathbb{H}$ are simply related:
\begin{align}\label{par2}
&\widetilde{\Delta}=\frac{1}{\varepsilon}, &
h_z=\frac{\mu_B}{J\varepsilon} g_{zz}h^{\mathrm{eff}},\\\nonumber
&h_1=\frac{\mu_B}{J\varepsilon}  g_{yy}H,&
h_2=\frac{\mu_B}{J\varepsilon} g_{yx}H.
\end{align}

The following numerical values have been used in  \cite{Faur17} for the parameters in the Hamiltonian  \eqref{HF}:
\begin{align}\label{parN}
J=5.8\, \mathrm{meV}, \;\varepsilon=0.53,\;g_{zz} \mu_B  h^{\mathrm{eff}}=0.06\, \mathrm{meV}, \\\nonumber
g_{zz}=6.07, \;g_{yy} =2.35,\; g_{yx} /g_{yy} =0.4,\;g_{yz} /g_{yy} =0.14.
\end{align}
For these numerical values, we get from \eqref{par2}:
\begin{equation}\label{coef}
h_1(H)\approx0.0443 H, \quad h_2(H)\approx0.0177 H, \quad \breve{\mathcal{H}}\approx 651\mathbb{H},
\end{equation}
where the magnetic field $H$ and the energy $\mathbb{H}$ in the right-hand sides are measured in Tesla and electron-volt, respectively.

Note, that Kimura {\it et al.} \cite{Kimura13}, and  Grenier  {\it et al.} \cite{Gr15,Gr15E} used slightly different from \eqref{parN}
values for the parameters of the effective Hamiltonian \eqref{HF}.

The values of parameters
$\eta=1.25$, $h_z=0.02$, $h_2=0.4 h_1$ of the Hamiltonian \eqref{Hama}, which we have chosen for  illustration of our results in Figures \ref{fig:Sec}, \ref{fig:Mag}, \ref{fig:Int1}, are close to those determined by equations \eqref{par2}, \eqref{parN}.

The evolution of the static magnetic structure of $\mathrm{BaCo}_{2}\mathrm{V}_{2}\mathrm{O}_{8}$ in the low-temperature phase
with the applied in the $y$-direction uniform magnetic field $H$ is displayed in Figure 2 of \cite{Faur17}. It is shown, in particular, 
in Figure 2b, that the staggered magnetization $m_c(H)$ in the $z$-direction monotonically decreases from its maximal value at $H=0$ 
to the zero value at the critical  field $H_c\approx 10\mathrm{T}$. As one can see from this figure, the staggered magnetization 
$m_c(H)$ remains close to its zero-field value $m_c(0)$ at weak enough magnetic fields in the interval  
$0<H\lesssim 3 \mathrm{T}$.
It is natural to expect, that our perturbative treatment of the effect 
of the transverse magnetic field is appropriate  in this  interval.

The dependences of the inelastic neutron scattering intensities on the transverse magnetic field $H$ at three different 
neutron scattering vectors  $\mathbf{Q}=(2,0,1)$,  $\mathbf{Q}=(0,0,2)$, and $\mathbf{Q}=(3,0,1)$ are shown in \cite{Faur17}
in Figures 3 a,b, and c, respectively. Note, that the components of the vector $\mathbf{Q}$ are written here in terms of the crystallographic 
lattice spacings.  This means, in particular, that the 1D momentum transfer $\mathrm{k}$, that stands in equations \eqref{brS},
is proportional to $Q_z$: 
\begin{equation}\label{kQ1}
\mathrm{k}=\frac{2\pi Q_z}{c/c'}, 
 \end{equation}
 where $c$ is the  crystallographic lattice spacing in the $z$-direction, and $c'$ is the spacing  (along the $z$-direction) between the 
$\mathrm{Co}^{2+}$ ions  in the magnetic chains. Since each magnetic chain contains four $\mathrm{Co}^{2+}$ ions in the
crystallographic cell \cite{Kimura13,Faur17}, the denominator in \eqref{kQ1} equals  four, and 
 \begin{equation}\label{kQ2}
\mathrm{k}=\frac{\pi Q_z}{2}.
 \end{equation}
It is well known \cite{Faur17}, further, that neutron scattering experiments probe only the spin fluctuations perpendicular to the scattering 
vector $\mathbf{Q}$.
  
The arguments listed above allow us to expect, that our analytical perturbative  results  
\eqref{brSa}, \eqref{brSyy} for the transverse structure factors 
$\breve{S}_{(2)}^{xx}(\mathrm{k},\omega)$, $\breve{S}_{(2)}^{yy}(\mathrm{k},\omega)$ 
at $\mathrm{k}=\pi$ for the model \eqref{Hama2}, which are illustrated in Figure \ref{fig:Str}, can be directly 
applied to the interpretation of the results of the neutron scattering intensity measurements for $\mathbf{Q}=(0,0,2)$
shown in Figure 3b in \cite{Faur17}.

Figures 3a and 3c in \cite{Faur17} display the measured by Faure {\it et al.} neutron scattering intensities at the scattering vectors 
$\mathbf{Q}$ with $Q_z=1$, which correspond due to \eqref{kQ2}   to  the 1D momentum transfer $\mathrm{k}=\pi/2$.
As one can see, the energies of the meson excitation modes in these two figures  are close to those in Figure 3b,
the only difference is in their intensities. This experimental result cannot be explained  in the frame of the theory based on the 
spin-chain Hamiltonian \eqref{Hama2}. Really, the energy 
of the lightest meson with zero momentum  in the absence of the transverse 
magnetic field is well below the two-spinon threshold $2 \omega(\pi/4)$ at their total momentum $P=\pi/2$:
\begin{align*}
&E_1(P,h_1,h_2)\big|_{P=0,h_1=h_2=0}\\
&<\min_{p}[\omega(P+p/2)+\omega(P-p/2)]\big|_{P=\pi/2}=2\, \omega(\pi/4).
\end{align*}
The natural and well-known \cite{Wang15,Gr15,Bera17,Kimura22,Takay23} way to overcome this difficulty is to perturb the 
Hamiltonian \eqref{Hama2}  with a term $\lambda V_1$
\begin{equation}\label{defH}
\breve{\mathcal{H}}(\breve{\Delta},\mathbf{h}_t,h_z)\to\breve{\mathcal{H}}(\breve{\Delta},\mathbf{h}_t,h_z)+\lambda V_1,
\end{equation}
having the four-site periodicity. 
Note, that such a term proportional to the small parameter $g_{yz}$ is already present in the 
 Hamiltonian \eqref{HF}. Further additional terms of this kind refining  this effective Hamiltonian  were 
introduced in \cite{Kimura22,Takay23}.
 The symmetry-breaking perturbation $\lambda V_1$  reduces the Brillouin zone $P\in(-\pi,\pi)$ by 
the factor of four, and should lead at small $\lambda$ to folding of the meson modes: 
\begin{equation}\label{folding}
E_{n}(P,h_1,h_2)\to E_{n}\left(P+{l \pi}/{2},h_1,h_2\right),\; \text{with  }l=0,1,2,3.
\end{equation}
It is reasonable to expect, that the small deformation \eqref{defH} of the Hamiltonian \eqref{Hama2} would
give rise also to the weighted sum of four ``copies" of the structure factors, which are shifted in 
$\mathrm{k}$ with respect to  one another by $\pi/2$:
\begin{equation}\label{coS}
\breve{S}_{(2)}^{\mathfrak{a}\mathfrak{b}}(\mathrm{k},\omega)\to \breve{S}_{(2)}^{\mathfrak{a}\mathfrak{b}}(\mathrm{k},\omega|\lambda)
\simeq 
\sum_{l=0}^3 A_l(\lambda )\breve{S}_{(2)}^{\mathfrak{a}\mathfrak{b}}(\mathrm{k}+\pi l/2,\omega).
\end{equation}
where the weight coefficients $A_l(\lambda )$ do not depend on $\mathbf{h}_t,h_z$, and $A_l(0 )=\delta_{l,0}$.
In the case of $\mathbf{h}_t=0$, a similar assumption was used by Bera {\it et al.}, see equation (7)
in \cite{Bera17}. Accordingly, we will use the $l=1$ term in the right-hand side of \eqref{coS} at 
$\mathrm{k}=\pi/2$ in order to interpret the  inelastic neutron scattering intensities shown in Figures 3a, and 3c of 
\cite{Faur17}.

So, there are good reasons to expect, that in the region of weak transverse magnetic fields $0<H\lesssim 3 \mathrm{T}$, the experimental inelastic neutron scattering  patterns shown in  Figures~3 a,b,c of \cite{Faur17}, can be, at least, qualitatively described by our 
perturbative results  \eqref{brS}, \eqref{Iab1} for the DSF at $\mathrm{k}=\pi$, which are illustrated in Figure \ref{fig:St}.
We will show below, that this is really the case.
\begin{enumerate}
\item
The  lowest mode in  Figures~3, which is called the {\it spin-flip} (SF) mode in \cite{Faur17}, exhibits a strong 
non-linear monotonic downward variation upon increase of the transverse field in the interval $0<H<10 \mathrm{T}$.
The inelastic neutron scattering measurements using polarized neutrons allow authors of \cite{Faur17} to conclude, 
that this SF mode remains polarized in the $xz$-plane in the whole range of variation of the transverse field $H$. 
At $H=0$, this mode is polarized purely in the $x$-direction. Upon increase of the  magnetic field $H$,
the transverse polarization in the $x$-direction decreases, while longitudinal polarization along the $z$-axis increases.
All listed above qualitative features of the evolution 
 of the  lowest  mode upon increase of the transverse field $H$  are in  agreement with our theoretical predictions
 \eqref{brSa}, \eqref{brSc}
 on the contribution of the meson mode with $n=1$ and $P=0$ into the diagonal DSF  
 components $\breve{S}_{(2)}^{xx}(\pi,\omega)$, 
 $\breve{S}_{(2)}^{zz}(\pi,\omega)$, which are illustrated in Figures \ref{fig:St}a,b.
\item
The second in energy mode in Figures~3  in \cite{Faur17} is called there  the {\it non-spin-flip} (NSF) mode. At $H=0$, the energies 
of the NSF and SF modes are the same. The NSF mode exhibits a small upward variation upon increase 
of the magnetic field in the interval $0<H\lesssim  3\mathrm{T}$. The NSF is polarized along the $y$-direction at
the magnetic fields $H$ inside this interval. The NSF mode observed in the neutron scattering experiments should be
identified with the meson mode with $n=1$ and $P=\pi$, which contributes to the DSF $\breve{S}_{(2)}^{yy}(\pi,\omega)$,
according to  \eqref{brSyy}. In the adopted approximation, the energy and polarization of this mode do not depend 
on the transverse magnetic field.
\item The third and the fourth modes in Figures~3  of \cite{Faur17} are polarized at $H=0$ along the $z$- and $x$-axes, 
respectively. Their energies approach one another upon increase of $H$ up to about $3 \mathrm{T}$, where  
the avoided crossing of  two modes occurs. These two modes can be identified with the meson modes $n=2$ and $n=3$ 
at $P=0$ in our classification. Evolution of energies and polarizations of these two meson modes with increasing magnetic field
$h_1\sim H$, which is illustrated in Figures \ref{fig:St}, is similar to
 those of the third and fourth modes in Figures~3  of \cite{Faur17}.
\item
Contributions of two subsequent $n=3$, $P=\pi$, and $n=4$, $P=0$ meson modes into the DSF, which are  shown in Figures \ref{fig:St}, 
are also clearly seen in Figures~3 a,b,c  of \cite{Faur17}.
\end{enumerate}
To summarise, the observed in the inelastic neutron scattering experiments \cite{Faur17} evolution of the meson energies
and spin polarizations in $\mathrm{BaCo}_{2}\mathrm{V}_{2}\mathrm{O}_{8}$ 
with the increase of the transverse magnetic field are in a good qualitative agreement in the small-field region with our theoretical predictions obtained by  the perturbative analysis of the XXZ spin-chain model 
\eqref{Hama}.  Furthermore, using equations \eqref{coef} one can see, that the evolution of the 
DSF with the transverse magnetic field $h_1$ displayed in Figure \ref{fig:St} is also in a reasonable quantitative
agreement with the experimental data shown in Figure 3 of \cite{Faur17}.

Recently, the evolution of the energies of  magnetic excitations  
upon increase  of the transverse magnetic field was experimentally studied  by Amelin {\it et al.} \cite{Wang_22} 
in the quasi-one-dimensional Ising-like ferromagnet $\mathrm{Co}\mathrm{Nb}_{2}\mathrm{O}_{6}$   and antiferromagnet 
$\mathrm{BaCo}_{2}\mathrm{V}_{2}\mathrm{O}_{8}$ 
by means of the high-resolution  terahertz spectroscopy. The   presented in Figure 2 in \cite{Wang_22}  the  
magnetic-field dependence of the THz absorption spectrum of
$\mathrm{BaCo}_{2}\mathrm{V}_{2}\mathrm{O}_{8}$ has a lot of similarities  with
the inelastic neutron scattering patterns for the same compound displayed in Figure 3 of \cite{Faur17} by Faure  {\it et al.}
However, for the reliable identification of different meson modes contributing into the THz absorption spectrum shown in
Figure 2 in \cite{Wang_22}, the detailed information on the  polarizations of the corresponding spin fluctuations is still required. 

\section{Conclusions \label{Conc}}
In this paper we present a perturbative analysis of the effect of mutually 
orthogonal uniform and staggered  transverse magnetic fields
$h_1,h_2$ on the confinement 
of spinons in the  Heisenberg XXZ spin-1/2 chain \eqref{Hama}, \eqref{H0}. The spinon confinement in this model  
 in the massive antiferromagnetic regime is induced by the staggered longitudinal magnetic field $h_z$. 
The energy spectra of the two-kink bound states in  model \eqref{Hama}, \eqref{H0} are calculated 
perturbatively in two asymptotic regimes: (i) in the extreme anisotropic limit $-\Delta\to \infty$, and (ii) for generic $\Delta<-1$
at weak transverse $h_1$, $h_2$, and staggered longitudinal $h_z$ magnetic fields. In the second regime, the perturbative 
calculations have been performed in two steps. First, in the de-confinement regime at $h_z=0$, 
 the effect of the weak transverse uniform and staggered magnetic fields on the ground states and 
kink excitations of  the XXZ spin chain are calculated by means the Rayleigh-Schr\"odinger perturbation expansion in 
small $h_1,h_2$. Then on top of that, the weak staggered longitudinal magnetic field $h_z>0$ is switched on. It
induces confinement of kinks into the ``mesons" - the two-kink   bound states. 
Their energy spectra and the  dynamical structure factors of local spin operators  for the momenta close to the points $0$ and $\pi$ are
calculated in the second regime by means of the properly modified semiclassical perturbative
technique, which was developed previously in \cite{Rut18,Rut22}.

It is shown, that the superposition of the semiclassical wave-functions corresponding to 
 different classical trajectories of two kinks forming a meson can be very sensitive to 
the strength of the transverse magnetic fields $h_1,h_2$. This leads to  the oscillatory interference patterns
in the transverse magnetic field dependencies of  the energies of mesons, and of the DSF of  local spin operators.  
Our theoretical predictions on the evolution of these quantities with increasing  transverse magnetic field are in 
good qualitative and reasonable quantitative agreement with the results reported by Faure {\it et al.} \cite{Faur17} on the inelastic neutron scattering 
experiments in the quasi-1D antiferromagnetic compound  $\mathrm{BaCo}_{2}\mathrm{V}_{2}\mathrm{O}_{8}$.   
For a detailed quantitative comparison of the obtained theoretical results with experiment, the 
values of the parameters in the effective Hamiltonian  \eqref{HF} should be reliably refined, preferable not
from fitting the experimentally observed meson energy spectra, but from some independent mesurements.   

There are several directions for further study. Our analysis in this paper was based on the XXZ spin-chain Hamiltonian
\eqref{Hama}, which is equivalent to \eqref{Hama2}. The obtained by Kimura {\it et al.} effective Hamiltonian \eqref{HF} 
differs from \eqref{Hama2} by an extra term $\cos[(2j+1)\pi/4]g_{yz} H$. We expect, that this term is responsible, at least
partly, for the  folding of the meson modes and DSF according to equations \eqref{folding}, and \eqref{coS}, respectively. It would be
interesting to account for this term perturbatively in small $g_{yz}$, and to explicitly 
determine the coefficients $A_l$ in equation \eqref{coS}. On the other hand, it would be also interesting to perform 
direct numerical calculations of the  DSF of local spin operators for the  model \eqref{Hama2}, and to compare 
the results with our analytical perturbative formulas \eqref{brS}, \eqref{Iab1} for these DSF. 
On the experimental side, further  detailed measurements  of the meson energy spectra in 
the quasi-1D antiferromagnets $\mathrm{BaCo}_{2}\mathrm{V}_{2}\mathrm{O}_{8}$ and 
$\mathrm{SrCo}_{2}\mathrm{V}_{2}\mathrm{O}_{8}$ in different regimes are desirable.  
\begin{acknowledgments} I am thankful to Frank G\"ohmann for many fruitful discussions.  
This work was supported by Deutsche Forschungsgemeinschaft (DFG) via Grant BO 3401/7-1.
\end{acknowledgments}
\appendix
\section{Two-kink scattering and form factors for the XXZ spin chain \eqref{H0}\label{Ap1}}
In this Appendix we collect some well-known results about the two-kink scattering phases 
and form factors of spin operators for the antiferromagnetic XXZ spin chain 
at zero magnetic field, which Hamiltonian is given by  \eqref{H0}. More details  can be found 
in \cite{Rut22}.

The two-kink Bloch  states $|K_{\mu\nu}(p_1)K_{\nu\mu}(p_2)\rangle_{s_1s_2}$ at zero magnetic field 
are characterized by the quasimomenta $p_1,p_2$, and $z$-projections  $s_1,s_2$ of the spins of particular kinks.
The defining equations for these states read:
\begin{subequations}\label{prop}
\begin{align}
\label{tilT}
&\widetilde{T}_1 |K_{\mu\nu}(p_1)K_{\nu\mu}(p_2)\rangle_{s_1s_2}=e^{i(p_1+p_2)}\\\nonumber
&\times|K_{\mu\nu}(p_1)K_{\nu\mu}(p_2)\rangle_{-s_1,-s_2},\\
&S^z |K_{\mu\nu}(p_1)K_{\nu\mu}(p_2)\rangle_{s_1s_2}=(s_1+s_2)\\\nonumber
&\times|K_{\mu\nu}(p_1)K_{\nu\mu}(p_2)\rangle_{s_1s_2},\\\label{pishift}
&(\mathcal{H}_0-E_{vac}^{(0)})|K_{\mu\nu}(p_1)K_{\nu\mu}(p_2)\rangle_{s_1s_2}\\\nonumber
&= [\omega(p_1)+\omega(p_2)]|K_{\mu\nu}(p_1)K_{\nu\mu}(p_2)\rangle_{s_1s_2},\nonumber
\end{align}
\end{subequations}
where $\mathcal{H}_0$ is the Hamiltonian \eqref{H0},  $E_{vac}^{(0)}$ is its ground-state energy, and $\omega(p)$ is the kink dispersion law \eqref{dl}.
The Bloch states  $|K_{\mu\nu}(p_1)K_{\nu\mu}(p_2)\rangle_{s_1s_2}$ with $0<p_2<p_1<\pi$ and $s_{1,2}=\pm1/2$ 
form the basis in the two-kink subspace $\mathcal{L}^{(2)}$. They are normalized by the 
condition:
\begin{align}
\phantom{.}_{s_2 s_1} \langle K_{\mu\nu}(p_2) K_{\nu\mu}(p_1) ||K_{\mu'\nu'}(p_1')K_{\nu\mu}(p_2')\rangle_{s_1's_2'}\\\nonumber
=
\pi^2 \delta_{\mu \mu'} \delta_{\nu \nu'}\delta_{s_1s_1'}\delta_{s_2s_2'}\delta(p_1-p_1')\delta(p_2-p_2'),
\end{align}
where $0<p_2<p_1<\pi$, $0<p_2'<p_1'<\pi$, and $\mu\ne\nu$.

The following equalities hold \cite{Rut22}:
\begin{align}\label{Kpi}
&|K_{\mu\nu}(p_1)K_{\nu\mu}(p_2)\rangle_{s_1s_2}=\\\nonumber
&\varkappa(\mu,s_1)|K_{\mu\nu}(p_1+\pi)K_{\nu\mu}(p_2)\rangle_{s_1s_2}\\
&=\varkappa(\nu,s_2)|K_{\mu\nu}(p_1)K_{\nu\mu}(p_2+\pi)\rangle_{s_1s_2},\nonumber
\end{align}
where $\varkappa(\nu,s)$ is defined in the in-line formulas below equation \eqref{Kpi1}.

It is useful to define an alternative  basis in the  subspace  of two-kink states with zero total spin $S^z = 0$:
\begin{align}\nonumber
&|K_{\mu\nu}(p_1)K_{\nu\mu}(p_2)\rangle_\pm\equiv
\frac{1}{\sqrt{2}}\Big(
|K_{\mu\nu}(p_1)K_{\nu\mu}(p_2)\rangle_{1/2,-1/2}\\
&\pm
|K_{\mu\nu}(p_1)K_{\nu\mu}(p_2)\rangle_{-1/2,1/2}
\Big).\label{baspm1}
\end{align}

The two-kink scattering can be described by the Faddeev-Zamolodchikov commutation relations:
\begin{subequations}\label{FZC}
\begin{align}\label{FZ2}
|K_{\mu\nu}(p_1)K_{\nu\mu}(p_2)\rangle_{ss}=w_0(p_1,p_2)
 |K_{\mu\nu}(p_2)K_{\nu\mu}(p_1)\rangle_{ss},\\
|K_{\mu\nu}(p_1)K_{\nu\mu}(p_2)\rangle_{\pm}=w_\pm(p_1,p_2)
|K_{\mu\nu}(p_2)K_{\nu\mu}(p_1)\rangle_\pm.\label{FZpm}
\end{align}
\end{subequations}
The three scattering amplitudes $w_\iota(p_1,p_2)$, with $\iota=0,\pm$, can be  parametrized by the 
rapidity variable $\alpha$, 
\begin{subequations}\label{scph}
\begin{align}\label{wio}
&w_\iota(p_1,p_2)=\exp[-i\pi +i \theta_\iota(p_1,p_2)],\\
&\theta_\iota(p_1,p_2)=\Theta_\iota(\alpha_1-\alpha_2),\label{thet}\\
\label{Th0}
&\Theta_0(\alpha)=
 \alpha+\sum_{n=1}^\infty \frac{e^{-n\eta}\sin(2\alpha n)}{n \cosh(n \eta)},\\\label{Thpm}
&\Theta_\pm(\alpha)=\Theta_0(\alpha)+\chi_\pm (\alpha),\\
&\chi_+(\alpha)=-i \ln\left(- \frac{\sin[(\alpha+i \eta)/2]}{\sin[(\alpha-i \eta)/2]} \right),\\
&\chi_-(\alpha)=-i \ln\left( \frac{\cos[(\alpha+i \eta)/2]}{\cos[(\alpha-i \eta)/2]} \right),
\end{align}
\end{subequations}
where $p_j=p(\alpha_j)$, $j=1,2$,  and $\Theta_\iota(\alpha)$ are the scattering phases. 
The  scattering amplitude $w_0(p_1,p_2)$ was found by 
Zabrodin \cite{Zabr92},  and the whole two-kink scattering matrix was determined by 
Davies {\it et al.} \cite{Miwa93}. 

The two-kink states $|\mathcal{K}_{\mu\nu}(\xi_1)\mathcal{K}_{\nu\mu}(\xi_2)\rangle_{s_1s_2}$ parametrized by the multiplicative  spectral parameters $\xi_{1,2}=-i e^{i\alpha_{1,2}}$ are simply related with $|K_{\mu\nu}(p_1)K_{\nu\mu}(p_2)\rangle_{s_1s_2}$:
\begin{align}\label{Kxi}
&|\mathcal{K}_{\mu\nu}(\xi_1)\mathcal{K}_{\nu\mu}(\xi_2)\rangle_{s_1s_2}=\\\nonumber
&\frac{\sqrt{\omega(p_1)\omega(p_2)}}{\sinh\eta}|K_{\mu\nu}(p_1)K_{\nu\mu}(p_2)\rangle_{s_1s_2}.
 \end{align}
 The states $|\mathcal{K}_{\mu\nu}(\xi_1)\mathcal{K}_{\nu\mu}(\xi_2)\rangle_{\pm}$ are related in the same way with 
 $|K_{\mu\nu}(p_1)K_{\nu\mu}(p_2)\rangle_{\pm}$:
\begin{align}\label{baspm1A}
&|\mathcal{K}_{\mu\nu}(\xi_1)\mathcal{K}_{\nu\mu}(\xi_2)\rangle_{\pm}=\\\nonumber
&\frac{\sqrt{\omega(p_1)\omega(p_2)}}{\sinh\eta}|K_{\mu\nu}(p_1)K_{\nu\mu}(p_2)\rangle_{\pm}.
\end{align}

All non-vanishing two-particle form factors of the spin operators $\sigma_0^\pm, \sigma_0^z$ can be expressed in terms of four 
functions $X^{1}(\xi_1,\xi_2)$, $X^{0}(\xi_1,\xi_2)$, and $X_\pm^z(\xi_1,\xi_2)$:
\begin{subequations}\label{XXF}
\begin{align}\nonumber
&X^1(\xi_1,\xi_2)=\phantom{.}^{(1)}\langle vac|\sigma_0^+|\mathcal{K}_{10}(\xi_1)\mathcal{K}_{01}(\xi_2)\rangle_{-1/2,-1/2}=\\
&\phantom{.}^{(0)}\langle vac|\sigma_0^-|\mathcal{K}_{01}(\xi_1)\mathcal{K}_{10}(\xi_2)\rangle_{1/2,1/2},\label{xX}\\\nonumber
&X^0(\xi_1,\xi_2)=\phantom{.}^{(1)}\langle vac|\sigma_0^-|\mathcal{K}_{10}(\xi_1)\mathcal{K}_{01}(\xi_2)\rangle_{1/2,1/2}=\\\label{XX0}
&\phantom{.}^{(0)}\langle vac|\sigma_0^+|\mathcal{K}_{01}(\xi_1)\mathcal{K}_{10}(\xi_2)\rangle_{-1/2,-1/2},\\\nonumber
&X_+^z(\xi_1,\xi_2)=\phantom{.}^{(1)}\langle vac|\sigma_0^z|\mathcal{K}_{10}(\xi_1)\mathcal{K}_{01}(\xi_2)\rangle_+=\\
&-\phantom{.}^{(0)}\langle vac|\sigma_0^z|\mathcal{K}_{01}(\xi_1)\mathcal{K}_{10}(\xi_2)\rangle_+,\\\nonumber
&X_-^z(\xi_1,\xi_2)=\phantom{.}^{(1)}\langle vac|\sigma_0^z|\mathcal{K}_{10}(\xi_1)\mathcal{K}_{01}(\xi_2)\rangle_-=\\
&\phantom{.}^{(0)}\langle vac|\sigma_0^z|\mathcal{K}_{01}(\xi_1)\mathcal{K}_{10}(\xi_2)\rangle_-.
\end{align}
\end{subequations}

The functions $X^{j}(\xi_1,\xi_2)$ and $X_\pm^{z}(\xi_1,\xi_2)$ admit the following explicit representations:
\begin{subequations}\label{XXall}
\begin{align}\label{XX}
&X^{j}(\xi_1,\xi_2)=\rho^2\frac{(q^4;q^4)^2}{(q^2;q^2)^3}\cdot\\\nonumber
&\frac{(-q\xi_1\xi_2)^{1-j}\xi_{\color{black}2}\,\, \gamma(\xi_{\color{black}2}^2/\xi_{\color{black}1}^2)\,
\theta_{q^8}(-\xi_1^{-2}\xi_2^{-2}q^{4j})}{\theta_{q^4}(\xi_1^{-2}q^3)\,\theta_{q^4}(\xi_2^{-2}q^3)},\\\label{Xpl}
&X_+^{z}(\xi_1,\xi_2)=\frac{\sqrt{2}\,e^{-\eta/4}g(\alpha_1+\alpha_2,\eta)}{\sin[(\alpha_1-\alpha_2-i\eta)/2]}\, X^{0}(\xi_1,\xi_2),\\
&X_-^{z}(\xi_1,\xi_2)=-X_+^{z}(-\xi_1,\xi_2),\label{Xmn}
\end{align}
\end{subequations}
where
\begin{align}
&\gamma(\xi)\equiv\frac{(q^4 \xi;q^4;q^4)(\xi^{-1};q^4;q^4)}{(q^6 \xi;q^4;q^4)(q^2\xi^{-1};q^4;q^4)},\\
&\rho\equiv (q^2;q^2)^2\frac{(q^4;q^4;q^4)}{(q^6;q^4;q^4)},\\
&(x;y)\equiv  \prod_{n=0}^\infty (1-x y^n),\\
&(x;y;z)\equiv \prod_{m,n=0}^\infty(1-x\, y^n z^m),\\
&\theta_x(y)=(x;x)(y;x)(x y^{-1};x),\\
&g(\alpha,\eta)=\frac{\vartheta_1(\frac{ \alpha}{2i\eta}|e^{-\pi^2/\eta})}{\vartheta_4(\frac{\alpha}{4i\eta}|e^{-\pi^2/(4\eta)})}.
\end{align}
Here $\vartheta_i(u|p)$  denote the elliptic theta-functions:
\begin{align}\label{theta}
&\vartheta_1(u|p)=2 p^{1/4}\sin(\pi u)\cdot \\\nonumber
&\prod_{n=1}^\infty(1-p^{2n})\left(1-2p^{2n}\cos(2\pi u)+p^{4n}\right),\\\nonumber
&\vartheta_4(u|p)=
\prod_{n=1}^\infty(1-p^{2n})\left(1-2p^{2n-1}\cos(2\pi u)+p^{2(2n-1)}\right),\\\nonumber
&\vartheta_2(u|p)=\vartheta_1(u+1/2|p), \quad \vartheta_3(u|p)=\vartheta_4(u+1/2|p).
\end{align}

The two-kink form factors of the $\sigma_0^\pm $ operators  were determined by means of the vertex-operator formalism
by Jimbo and Miwa \cite{Jimbo94}. The explicit formulas for the form factors of the $\sigma_0^z$ operator in the XYZ spin-1/2 chain were obtained by Lashkevich \cite{Lash02}.
The XXZ limit  of these  formulas used in \eqref{Xpl} and   \eqref{Xmn} can be found in \cite{Dug15}. 
\section{ Derivation of equation \eqref{absB} \label{AppB}}
In this Appendix we describe calculation of the absolute value of the normalization 
constant $B_{in,1}$, which was introduced in equation \eqref{ps2ba}.

Let us start   from the normalization condition \eqref{normphi} for the reduced wave functions 
$\phi_n\!\left(p|P,\mathbf{h}_t,h_z\right)$ describing the relative motion of two kinks forming a meson. 
For $n=n'$, this normalization condition takes the form
\begin{equation}\label{normphi1}
\int_{-\pi}^{\pi}\frac{dp}{2\pi} \,|\phi_n(p|P)|^2=2.
\end{equation}
Here we have skipped parameters $\mathbf{h}_t,h_z$ in the wave function $\phi_n\!\left(p|P,\mathbf{h}_t,h_z\right)$.
The normalization condition \eqref{normphi1} can be rewritten in terms of the Fourier coefficients \eqref{redps} of the reduced wave function:
\begin{equation}\label{NC}
\sum_{j=-\infty}^\infty |\psi_n(j|P)|^2=2.
\end{equation}
One can easily show using \eqref{redps}, \eqref{refl}, that the following equality holds at  $\mathbf{h}_t=0$:
\begin{equation*}
\sum_{j=0}^{-\infty} |\psi_n(j|P)|^2=\sum_{j=0}^\infty |\psi_n(j|P)|^2.
\end{equation*}
Accordingly, the normalization condition \eqref{NC} reduces in this case to the form:
\begin{equation}\label{NC1}
-\frac{1}{2}|\psi_0(j|P)|^2+\sum_{j=0}^{-\infty} |\psi_n(j|P)|^2=1.
\end{equation}
At a small  string tension $f\sim h_z$, the main contribution into the sum in the left-hand side comes from the terms with large 
enough negative
$j$, such that $|j|\gg 2 \xi_c(\eta)$, where $\xi_c(\eta)$ is the correlation length \eqref{corr}.  By this reason, to the leading order 
in $f$, we can neglect the first term in the left-hand side of \eqref{NC1} and replace the functions $\psi_n(j|P)$ in the sum by 
their semiclassical asymptotics \eqref{ps2aa}. This yields:
\begin{align}\label{sumN}
\sum_{j=0}^{-\infty} \Big\{|C_1|^2 |\psi_n^{(1)}(j|P)|^2
+|C_2|^2 |\psi_n^{(2)}(j|P)|^2\\\nonumber
+2\, \mathrm{Re} 
[C_1C_2^* \, \psi_n^{(1)}(j|P)\psi_n^{(2)}(j|P)^*]
\Big\}=1.
\end{align}
For the coefficients $\psi_n^{(1,2)}(j|P)$ in the left-hand side, we obtained in Section \ref{Heur} the asymptotical formulas
\eqref{ps12a}, that hold at small $f>0$.
To the leading order in $f$, these formulas reduce to the form:
\begin{subequations}\label{ps12B}
\begin{align}\label{ps1B}
&\psi^{(1)}(j|P)=\int_{-\pi/2}^{\pi/2} \frac{dp}{2\pi}\,e^{i p j-i F_1(p,E|0)/{f}}[1+O(f)],\\\label{pS2B}
&\psi^{(2)}(j|P)=(-1)^j\psi^{(1)}(j|P)[1+O(f)].
\end{align}
\end{subequations}
The integral in the right-hand side of \eqref{ps1B} is mainly determined at small $f>0$  by the contributions of the vicinity of the saddle points 
$\pm p(j)$, where  $p(j)$ is the solution of the equation:
\begin{equation}\label{Efj}
E+f j =\epsilon(p(j)|P).
\end{equation}
For 
$j\in (j_{min},0)$,
with $j_{min}=[\epsilon(0|P)-E]/f<0$, the solution of saddle-point equation \eqref{Efj} is real and lies 
in the interval $p(j)\in(0,p_a)$ in the left kinematically allowed region shown in Figure \ref{eprof1}.
In this case, the saddle point asymptotics of the integral \eqref{ps1B} reads:
\begin{align}\label{psisin}
\psi^{(i)}(j|P)=2 i\sqrt{\frac{f}{2\pi \epsilon'(p|P)}}\bigg|_{p=p(j)}\\\nonumber
\times \sin\left[
p(j)-\frac{F_1\left(p(j),E|0\right)}{f}-\frac{\pi}{4}
\right][1+O(f)].
\end{align}
After substitution of \eqref{psisin}, \eqref{pS2B} into the left-hand side of equation \eqref{sumN}, one can see, that the 
highly oscillating third term in the curly brackets under the sum sign can be dropped. As the result, the 
normalization condition  \eqref{sumN} reduces to the form:
\begin{align}\label{sumN2}
&1=\frac{|C_1|^2+|C_2|^2 }{\pi}\\\nonumber
&\times \sum_{j=j_{min}}^{0} \Delta p(j)
\left\{1- \sin\left[
2p-{2F_1\left(p(j),E|0\right)}/{f}
\right]\right\},
\end{align}
where
$
\Delta p(j)={f}/{\epsilon'(p|P)}\big|_{p=p(j)}
$.

After dropping the oscillation $\sin$-term in the curly brackets in the second line of \eqref{sumN2} and replacement the 
summation in $j$ by the integration in $p$
\[ 
 \sum_{j=j_{min}}^{0} \Delta p(j)\ldots \to \int_0^{p_a} dp \ldots ,\]
we find from \eqref{sumN2}:
\[
|C_1|^2+|C_2|^2={\pi}/{p_a}.
\]
Combining this result with \eqref{C12AB2a},  \eqref{ht0}, and \eqref{rB}, we arrive to the 
final expression \eqref{absB} for the absolute value of the normalization constant $B_{in,1}$.

\begin{thebibliography}{52}%
\makeatletter
\providecommand \@ifxundefined [1]{%
 \@ifx{#1\undefined}
}%
\providecommand \@ifnum [1]{%
 \ifnum #1\expandafter \@firstoftwo
 \else \expandafter \@secondoftwo
 \fi
}%
\providecommand \@ifx [1]{%
 \ifx #1\expandafter \@firstoftwo
 \else \expandafter \@secondoftwo
 \fi
}%
\providecommand \natexlab [1]{#1}%
\providecommand \enquote  [1]{``#1''}%
\providecommand \bibnamefont  [1]{#1}%
\providecommand \bibfnamefont [1]{#1}%
\providecommand \citenamefont [1]{#1}%
\providecommand \href@noop [0]{\@secondoftwo}%
\providecommand \href [0]{\begingroup \@sanitize@url \@href}%
\providecommand \@href[1]{\@@startlink{#1}\@@href}%
\providecommand \@@href[1]{\endgroup#1\@@endlink}%
\providecommand \@sanitize@url [0]{\catcode `\\12\catcode `\$12\catcode
  `\&12\catcode `\#12\catcode `\^12\catcode `\_12\catcode `\%12\relax}%
\providecommand \@@startlink[1]{}%
\providecommand \@@endlink[0]{}%
\providecommand \url  [0]{\begingroup\@sanitize@url \@url }%
\providecommand \@url [1]{\endgroup\@href {#1}{\urlprefix }}%
\providecommand \urlprefix  [0]{URL }%
\providecommand \Eprint [0]{\href }%
\providecommand \doibase [0]{https://doi.org/}%
\providecommand \selectlanguage [0]{\@gobble}%
\providecommand \bibinfo  [0]{\@secondoftwo}%
\providecommand \bibfield  [0]{\@secondoftwo}%
\providecommand \translation [1]{[#1]}%
\providecommand \BibitemOpen [0]{}%
\providecommand \bibitemStop [0]{}%
\providecommand \bibitemNoStop [0]{.\EOS\space}%
\providecommand \EOS [0]{\spacefactor3000\relax}%
\providecommand \BibitemShut  [1]{\csname bibitem#1\endcsname}%
\let\auto@bib@innerbib\@empty
\bibitem [{\citenamefont {Narison}(2004)}]{Nar04}%
  \BibitemOpen
  \bibfield  {author} {\bibinfo {author} {\bibfnamefont {S.}~\bibnamefont
  {Narison}},\ }\href@noop {} {\emph {\bibinfo {title} {QCD as a Theory of
  Hadrons}}}\ (\bibinfo  {publisher} {Cambridge University Press},\ \bibinfo
  {address} {Cambridge},\ \bibinfo {year} {2004})\BibitemShut {NoStop}%
\bibitem [{\citenamefont {Wilson}(1974)}]{Wilson74}%
  \BibitemOpen
  \bibfield  {author} {\bibinfo {author} {\bibfnamefont {K.~G.}\ \bibnamefont
  {Wilson}},\ }\bibfield  {title} {\bibinfo {title} {Confinement of quarks},\
  }\href {https://doi.org/10.1103/PhysRevD.10.2445} {\bibfield  {journal}
  {\bibinfo  {journal} {Phys. Rev. D}\ }\textbf {\bibinfo {volume} {10}},\
  \bibinfo {pages} {2445} (\bibinfo {year} {1974})}\BibitemShut {NoStop}%
\bibitem [{\citenamefont {{}'t Hooft}(1974)}]{Hooft74}%
  \BibitemOpen
  \bibfield  {author} {\bibinfo {author} {\bibfnamefont {G.}~\bibnamefont {{}'t
  Hooft}},\ }\bibfield  {title} {\bibinfo {title} {A two-dimensional model for
  mesons},\ }\href@noop {} {\bibfield  {journal} {\bibinfo  {journal} {Nucl.
  Phys. B}\ }\textbf {\bibinfo {volume} {75}},\ \bibinfo {pages} {461}
  (\bibinfo {year} {1974})}\BibitemShut {NoStop}%
\bibitem [{\citenamefont {McCoy}\ and\ \citenamefont {Wu}(1978)}]{McCoy78}%
  \BibitemOpen
  \bibfield  {author} {\bibinfo {author} {\bibfnamefont {B.~M.}\ \bibnamefont
  {McCoy}}\ and\ \bibinfo {author} {\bibfnamefont {T.~T.}\ \bibnamefont {Wu}},\
  }\bibfield  {title} {\bibinfo {title} {Two dimensional {I}sing field theory
  in a magnetic field: Breakup of the cut in the two-point function},\ }\href
  {https://doi.org/10.1103/PhysRevD.18.1259} {\bibfield  {journal} {\bibinfo
  {journal} {Phys. Rev. D}\ }\textbf {\bibinfo {volume} {18}},\ \bibinfo
  {pages} {1259} (\bibinfo {year} {1978})}\BibitemShut {NoStop}%
\bibitem [{\citenamefont {Shiba}(1980)}]{Shiba80}%
  \BibitemOpen
  \bibfield  {author} {\bibinfo {author} {\bibfnamefont {H.}~\bibnamefont
  {Shiba}},\ }\bibfield  {title} {\bibinfo {title} {{Quantization of magnetic
  excitation continuum due to interchain coupling in nearly one-dimensional
  {I}sing-like antiferromagnets}},\ }\href {https://doi.org/10.1143/PTP.64.466}
  {\bibfield  {journal} {\bibinfo  {journal} {Progress of Theoretical Physics}\
  }\textbf {\bibinfo {volume} {64}},\ \bibinfo {pages} {466} (\bibinfo {year}
  {1980})}\BibitemShut {NoStop}%
\bibitem [{\citenamefont {Grenier}\ \emph
  {et~al.}(2015{\natexlab{a}})\citenamefont {Grenier}, \citenamefont {Petit},
  \citenamefont {Simonet}, \citenamefont {Can\'evet}, \citenamefont {Regnault},
  \citenamefont {Raymond}, \citenamefont {Canals}, \citenamefont {Berthier},\
  and\ \citenamefont {Lejay}}]{Gr15}%
  \BibitemOpen
  \bibfield  {author} {\bibinfo {author} {\bibfnamefont {B.}~\bibnamefont
  {Grenier}}, \bibinfo {author} {\bibfnamefont {S.}~\bibnamefont {Petit}},
  \bibinfo {author} {\bibfnamefont {V.}~\bibnamefont {Simonet}}, \bibinfo
  {author} {\bibfnamefont {E.}~\bibnamefont {Can\'evet}}, \bibinfo {author}
  {\bibfnamefont {L.-P.}\ \bibnamefont {Regnault}}, \bibinfo {author}
  {\bibfnamefont {S.}~\bibnamefont {Raymond}}, \bibinfo {author} {\bibfnamefont
  {B.}~\bibnamefont {Canals}}, \bibinfo {author} {\bibfnamefont
  {C.}~\bibnamefont {Berthier}},\ and\ \bibinfo {author} {\bibfnamefont
  {P.}~\bibnamefont {Lejay}},\ }\bibfield  {title} {\bibinfo {title}
  {Longitudinal and transverse {Z}eeman ladders in the {I}sing-like chain
  antiferromagnet $\mathrm{BaCo}_{2}\mathrm{V}_{2}\mathrm{O}_{8}$},\ }\href
  {https://doi.org/10.1103/PhysRevLett.114.017201} {\bibfield  {journal}
  {\bibinfo  {journal} {Phys. Rev. Lett.}\ }\textbf {\bibinfo {volume} {114}},\
  \bibinfo {pages} {017201} (\bibinfo {year} {2015}{\natexlab{a}})}\BibitemShut
  {NoStop}%
\bibitem [{\citenamefont {Bera}\ \emph {et~al.}(2017)\citenamefont {Bera},
  \citenamefont {Lake}, \citenamefont {Essler}, \citenamefont {Vanderstraeten},
  \citenamefont {Hubig}, \citenamefont {Schollw\"ock}, \citenamefont {Islam},
  \citenamefont {Schneidewind},\ and\ \citenamefont
  {Quintero-Castro}}]{Bera17}%
  \BibitemOpen
  \bibfield  {author} {\bibinfo {author} {\bibfnamefont {A.~K.}\ \bibnamefont
  {Bera}}, \bibinfo {author} {\bibfnamefont {B.}~\bibnamefont {Lake}}, \bibinfo
  {author} {\bibfnamefont {F.~H.~L.}\ \bibnamefont {Essler}}, \bibinfo {author}
  {\bibfnamefont {L.}~\bibnamefont {Vanderstraeten}}, \bibinfo {author}
  {\bibfnamefont {C.}~\bibnamefont {Hubig}}, \bibinfo {author} {\bibfnamefont
  {U.}~\bibnamefont {Schollw\"ock}}, \bibinfo {author} {\bibfnamefont {A.~T.
  M.~N.}\ \bibnamefont {Islam}}, \bibinfo {author} {\bibfnamefont
  {A.}~\bibnamefont {Schneidewind}},\ and\ \bibinfo {author} {\bibfnamefont
  {D.~L.}\ \bibnamefont {Quintero-Castro}},\ }\bibfield  {title} {\bibinfo
  {title} {Spinon confinement in a quasi-one-dimensional anisotropic
  {H}eisenberg magnet},\ }\href {https://doi.org/10.1103/PhysRevB.96.054423}
  {\bibfield  {journal} {\bibinfo  {journal} {Phys. Rev. B}\ }\textbf {\bibinfo
  {volume} {96}},\ \bibinfo {pages} {054423} (\bibinfo {year}
  {2017})}\BibitemShut {NoStop}%
\bibitem [{\citenamefont {Wang}\ \emph {et~al.}(2015)\citenamefont {Wang},
  \citenamefont {Schmidt}, \citenamefont {Bera}, \citenamefont {Islam},
  \citenamefont {Lake}, \citenamefont {Loidl},\ and\ \citenamefont
  {Deisenhofer}}]{Wang15}%
  \BibitemOpen
  \bibfield  {author} {\bibinfo {author} {\bibfnamefont {Z.}~\bibnamefont
  {Wang}}, \bibinfo {author} {\bibfnamefont {M.}~\bibnamefont {Schmidt}},
  \bibinfo {author} {\bibfnamefont {A.~K.}\ \bibnamefont {Bera}}, \bibinfo
  {author} {\bibfnamefont {A.~T. M.~N.}\ \bibnamefont {Islam}}, \bibinfo
  {author} {\bibfnamefont {B.}~\bibnamefont {Lake}}, \bibinfo {author}
  {\bibfnamefont {A.}~\bibnamefont {Loidl}},\ and\ \bibinfo {author}
  {\bibfnamefont {J.}~\bibnamefont {Deisenhofer}},\ }\bibfield  {title}
  {\bibinfo {title} {Spinon confinement in the one-dimensional {I}sing-like
  antiferromagnet $\mathrm{SrCo}_2\mathrm{V}_{2}\mathrm{O}_8$},\ }\href
  {https://doi.org/10.1103/PhysRevB.91.140404} {\bibfield  {journal} {\bibinfo
  {journal} {Phys. Rev. B}\ }\textbf {\bibinfo {volume} {91}},\ \bibinfo
  {pages} {140404} (\bibinfo {year} {2015})}\BibitemShut {NoStop}%
\bibitem [{\citenamefont {Wang}\ \emph {et~al.}(2019)\citenamefont {Wang},
  \citenamefont {Schmidt}, \citenamefont {Loidl}, \citenamefont {Wu},
  \citenamefont {Zou}, \citenamefont {Yang}, \citenamefont {Dong},
  \citenamefont {Kohama}, \citenamefont {Kindo}, \citenamefont {Gorbunov},
  \citenamefont {Niesen}, \citenamefont {Breunig}, \citenamefont {Engelmayer},\
  and\ \citenamefont {Lorenz}}]{Wang19}%
  \BibitemOpen
  \bibfield  {author} {\bibinfo {author} {\bibfnamefont {Z.}~\bibnamefont
  {Wang}}, \bibinfo {author} {\bibfnamefont {M.}~\bibnamefont {Schmidt}},
  \bibinfo {author} {\bibfnamefont {A.}~\bibnamefont {Loidl}}, \bibinfo
  {author} {\bibfnamefont {J.}~\bibnamefont {Wu}}, \bibinfo {author}
  {\bibfnamefont {H.}~\bibnamefont {Zou}}, \bibinfo {author} {\bibfnamefont
  {W.}~\bibnamefont {Yang}}, \bibinfo {author} {\bibfnamefont {C.}~\bibnamefont
  {Dong}}, \bibinfo {author} {\bibfnamefont {Y.}~\bibnamefont {Kohama}},
  \bibinfo {author} {\bibfnamefont {K.}~\bibnamefont {Kindo}}, \bibinfo
  {author} {\bibfnamefont {D.~I.}\ \bibnamefont {Gorbunov}}, \bibinfo {author}
  {\bibfnamefont {S.}~\bibnamefont {Niesen}}, \bibinfo {author} {\bibfnamefont
  {O.}~\bibnamefont {Breunig}}, \bibinfo {author} {\bibfnamefont
  {J.}~\bibnamefont {Engelmayer}},\ and\ \bibinfo {author} {\bibfnamefont
  {T.}~\bibnamefont {Lorenz}},\ }\bibfield  {title} {\bibinfo {title} {Quantum
  critical dynamics of a {H}eisenberg-{I}sing chain in a longitudinal field:
  {M}any-body strings versus fractional excitations},\ }\href
  {https://doi.org/10.1103/PhysRevLett.123.067202} {\bibfield  {journal}
  {\bibinfo  {journal} {Phys. Rev. Lett.}\ }\textbf {\bibinfo {volume} {123}},\
  \bibinfo {pages} {067202} (\bibinfo {year} {2019})}\BibitemShut {NoStop}%
\bibitem [{\citenamefont {{Faure}}\ \emph {et~al.}(2018)\citenamefont
  {{Faure}}, \citenamefont {{Takayoshi}}, \citenamefont {{Petit}},
  \citenamefont {{Simonet}}, \citenamefont {{Raymond}}, \citenamefont
  {{Regnault}}, \citenamefont {{Boehm}}, \citenamefont {{White}}, \citenamefont
  {{M{\aa}nsson}}, \citenamefont {{R{\"u}egg}}, \citenamefont {{Lejay}},
  \citenamefont {{Canals}}, \citenamefont {{Lorenz}}, \citenamefont {{Furuya}},
  \citenamefont {{Giamarchi}},\ and\ \citenamefont {{Grenier}}}]{Faur17}%
  \BibitemOpen
  \bibfield  {author} {\bibinfo {author} {\bibfnamefont {Q.}~\bibnamefont
  {{Faure}}}, \bibinfo {author} {\bibfnamefont {S.}~\bibnamefont
  {{Takayoshi}}}, \bibinfo {author} {\bibfnamefont {S.}~\bibnamefont
  {{Petit}}}, \bibinfo {author} {\bibfnamefont {V.}~\bibnamefont {{Simonet}}},
  \bibinfo {author} {\bibfnamefont {S.}~\bibnamefont {{Raymond}}}, \bibinfo
  {author} {\bibfnamefont {L.-P.}\ \bibnamefont {{Regnault}}}, \bibinfo
  {author} {\bibfnamefont {M.}~\bibnamefont {{Boehm}}}, \bibinfo {author}
  {\bibfnamefont {J.~S.}\ \bibnamefont {{White}}}, \bibinfo {author}
  {\bibfnamefont {M.}~\bibnamefont {{M{\aa}nsson}}}, \bibinfo {author}
  {\bibfnamefont {C.}~\bibnamefont {{R{\"u}egg}}}, \bibinfo {author}
  {\bibfnamefont {P.}~\bibnamefont {{Lejay}}}, \bibinfo {author} {\bibfnamefont
  {B.}~\bibnamefont {{Canals}}}, \bibinfo {author} {\bibfnamefont
  {T.}~\bibnamefont {{Lorenz}}}, \bibinfo {author} {\bibfnamefont {S.~C.}\
  \bibnamefont {{Furuya}}}, \bibinfo {author} {\bibfnamefont {T.}~\bibnamefont
  {{Giamarchi}}},\ and\ \bibinfo {author} {\bibfnamefont {B.}~\bibnamefont
  {{Grenier}}},\ }\bibfield  {title} {\bibinfo {title} {Topological quantum
  phase transition in the {I}sing-like antiferromagnetic spin chain
  $\mathrm{BaCo}_{2}\mathrm{V}_{2}\mathrm{O}_{8}$},\ }\href
  {https://doi.org/10.1038/s41567-018-0126-8} {\bibfield  {journal} {\bibinfo
  {journal} {Nature Physics}\ }\textbf {\bibinfo {volume} {14}},\ \bibinfo
  {pages} {716} (\bibinfo {year} {2018})}\BibitemShut {NoStop}%
\bibitem [{\citenamefont {Rutkevich}(2022)}]{Rut22}%
  \BibitemOpen
  \bibfield  {author} {\bibinfo {author} {\bibfnamefont {S.~B.}\ \bibnamefont
  {Rutkevich}},\ }\bibfield  {title} {\bibinfo {title} {Spinon confinement in
  the gapped antiferromagnetic {XXZ} spin-$\frac{1}{2}$ chain},\ }\href
  {https://doi.org/10.1103/PhysRevB.106.134405} {\bibfield  {journal} {\bibinfo
   {journal} {Phys. Rev. B}\ }\textbf {\bibinfo {volume} {106}},\ \bibinfo
  {pages} {134405} (\bibinfo {year} {2022})}\BibitemShut {NoStop}%
\bibitem [{\citenamefont {Rutkevich}(2018)}]{Rut18}%
  \BibitemOpen
  \bibfield  {author} {\bibinfo {author} {\bibfnamefont {S.~B.}\ \bibnamefont
  {Rutkevich}},\ }\bibfield  {title} {\bibinfo {title} {Kink confinement in the
  antiferromagnetic {XXZ} spin-(1/2) chain in a weak staggered magnetic
  field},\ }\href {https://doi.org/10.1209/0295-5075/121/37001} {\bibfield
  {journal} {\bibinfo  {journal} {{EPL} (Europhysics Letters)}\ }\textbf
  {\bibinfo {volume} {121}},\ \bibinfo {pages} {37001} (\bibinfo {year}
  {2018})}\BibitemShut {NoStop}%
\bibitem [{\citenamefont {Fonseca}\ and\ \citenamefont
  {Zamolodchikov}(2003)}]{FonZam2003}%
  \BibitemOpen
  \bibfield  {author} {\bibinfo {author} {\bibfnamefont {P.}~\bibnamefont
  {Fonseca}}\ and\ \bibinfo {author} {\bibfnamefont {A.~B.}\ \bibnamefont
  {Zamolodchikov}},\ }\bibfield  {title} {\bibinfo {title} {Ising field theory
  in a magnetic field: Analytic properties of the free energy},\ }\href
  {https://doi.org/https://doi.org/10.1023/A:1022147532606} {\bibfield
  {journal} {\bibinfo  {journal} {J. Stat. Phys.}\ }\textbf {\bibinfo {volume}
  {110}},\ \bibinfo {pages} {527} (\bibinfo {year} {2003})}\BibitemShut
  {NoStop}%
\bibitem [{\citenamefont {Fonseca}\ and\ \citenamefont
  {Zamolodchikov}(2006)}]{FZ06}%
  \BibitemOpen
  \bibfield  {author} {\bibinfo {author} {\bibfnamefont {P.}~\bibnamefont
  {Fonseca}}\ and\ \bibinfo {author} {\bibfnamefont {A.~B.}\ \bibnamefont
  {Zamolodchikov}},\ }\href
  {https://doi.org/https://doi.org/10.48550/arXiv.hep-th/0612304} {\bibinfo
  {title} {{I}sing spectroscopy ${{\rm I}} $: Mesons at ${T}<{T}_c$}} (\bibinfo
  {year} {2006}),\ \Eprint {https://arxiv.org/abs/0612304} {arXiv:0612304
  [hep-th]} \BibitemShut {NoStop}%
\bibitem [{\citenamefont {Rutkevich}(2005)}]{Rut05}%
  \BibitemOpen
  \bibfield  {author} {\bibinfo {author} {\bibfnamefont {S.~B.}\ \bibnamefont
  {Rutkevich}},\ }\bibfield  {title} {\bibinfo {title} {Large-$n$ excitations
  in the ferromagnetic {I}sing field theory in a weak magnetic field: Mass
  spectrum and decay widths},\ }\href
  {https://doi.org/10.1103/PhysRevLett.95.250601} {\bibfield  {journal}
  {\bibinfo  {journal} {Phys. Rev. Lett.}\ }\textbf {\bibinfo {volume} {95}},\
  \bibinfo {pages} {250601} (\bibinfo {year} {2005})}\BibitemShut {NoStop}%
\bibitem [{\citenamefont {{Rutkevich}}(2010)}]{RutP09}%
  \BibitemOpen
  \bibfield  {author} {\bibinfo {author} {\bibfnamefont {S.~B.}\ \bibnamefont
  {{Rutkevich}}},\ }\bibfield  {title} {\bibinfo {title} {Two-kink bound states
  in the magnetically perturbed {P}otts field theory at ${T}<{T}_c$},\ }\href
  {https://doi.org/10.1088/1751-8113/43/23/235004} {\bibfield  {journal}
  {\bibinfo  {journal} {J. Phys. A}\ }\textbf {\bibinfo {volume} {43}},\
  \bibinfo {pages} {235004} (\bibinfo {year} {2010})}\BibitemShut {NoStop}%
\bibitem [{\citenamefont {Lencs\'es}\ \emph {et~al.}(2022)\citenamefont
  {Lencs\'es}, \citenamefont {Mussardo},\ and\ \citenamefont
  {Tak\'acs}}]{Mus22}%
  \BibitemOpen
  \bibfield  {author} {\bibinfo {author} {\bibfnamefont {M.}~\bibnamefont
  {Lencs\'es}}, \bibinfo {author} {\bibfnamefont {G.}~\bibnamefont
  {Mussardo}},\ and\ \bibinfo {author} {\bibfnamefont {G.}~\bibnamefont
  {Tak\'acs}},\ }\bibfield  {title} {\bibinfo {title} {Confinement in the
  tricritical {I}sing model},\ }\href
  {https://doi.org/10.1016/j.physletb.2022.137008} {\bibfield  {journal}
  {\bibinfo  {journal} {Physics Letters B}\ }\textbf {\bibinfo {volume}
  {828}},\ \bibinfo {pages} {137008} (\bibinfo {year} {2022})}\BibitemShut
  {NoStop}%
\bibitem [{\citenamefont {{Rutkevich}}(2008)}]{Rut08a}%
  \BibitemOpen
  \bibfield  {author} {\bibinfo {author} {\bibfnamefont {S.~B.}\ \bibnamefont
  {{Rutkevich}}},\ }\bibfield  {title} {\bibinfo {title} {Energy spectrum of
  bound-spinons in the quantum {I}sing spin-chain ferromagnet},\ }\href
  {https://doi.org/https://doi.org/10.1007/s10955-008-9495-1} {\bibfield
  {journal} {\bibinfo  {journal} {J. Stat. Phys.}\ }\textbf {\bibinfo {volume}
  {131}},\ \bibinfo {pages} {917} (\bibinfo {year} {2008})}\BibitemShut
  {NoStop}%
\bibitem [{\citenamefont {Lagnese}\ \emph {et~al.}(2022)\citenamefont
  {Lagnese}, \citenamefont {Surace}, \citenamefont {Kormos},\ and\
  \citenamefont {Calabrese}}]{Lagnese_2022}%
  \BibitemOpen
  \bibfield  {author} {\bibinfo {author} {\bibfnamefont {G.}~\bibnamefont
  {Lagnese}}, \bibinfo {author} {\bibfnamefont {F.~M.}\ \bibnamefont {Surace}},
  \bibinfo {author} {\bibfnamefont {M.}~\bibnamefont {Kormos}},\ and\ \bibinfo
  {author} {\bibfnamefont {P.}~\bibnamefont {Calabrese}},\ }\bibfield  {title}
  {\bibinfo {title} {Quenches and confinement in a {H}eisenberg-{I}sing spin
  ladder},\ }\href {https://doi.org/10.1088/1751-8121/ac5215} {\bibfield
  {journal} {\bibinfo  {journal} {Journal of Physics A: Mathematical and
  Theoretical}\ }\textbf {\bibinfo {volume} {55}},\ \bibinfo {pages} {124003}
  (\bibinfo {year} {2022})}\BibitemShut {NoStop}%
\bibitem [{\citenamefont {Ramos}\ \emph {et~al.}(2020)\citenamefont {Ramos},
  \citenamefont {Lencs\'es}, \citenamefont {Xavier},\ and\ \citenamefont
  {Pereira}}]{Ram20}%
  \BibitemOpen
  \bibfield  {author} {\bibinfo {author} {\bibfnamefont {F.~B.}\ \bibnamefont
  {Ramos}}, \bibinfo {author} {\bibfnamefont {M.}~\bibnamefont {Lencs\'es}},
  \bibinfo {author} {\bibfnamefont {J.~C.}\ \bibnamefont {Xavier}},\ and\
  \bibinfo {author} {\bibfnamefont {R.~G.}\ \bibnamefont {Pereira}},\
  }\bibfield  {title} {\bibinfo {title} {Confinement and bound states of bound
  states in a transverse-field two-leg {I}sing ladder},\ }\href
  {https://doi.org/10.1103/PhysRevB.102.014426} {\bibfield  {journal} {\bibinfo
   {journal} {Phys. Rev. B}\ }\textbf {\bibinfo {volume} {102}},\ \bibinfo
  {pages} {014426} (\bibinfo {year} {2020})}\BibitemShut {NoStop}%
\bibitem [{\citenamefont {Rutkevich}(2023)}]{rutkevich2023soliton}%
  \BibitemOpen
  \bibfield  {author} {\bibinfo {author} {\bibfnamefont {S.~B.}\ \bibnamefont
  {Rutkevich}},\ }\href@noop {} {\bibinfo {title} {Soliton confinement in the
  double sine-{G}ordon model}} (\bibinfo {year} {2023}),\ \Eprint
  {https://arxiv.org/abs/2311.07303} {arXiv:2311.07303 [hep-th]} \BibitemShut
  {NoStop}%
\bibitem [{\citenamefont {Kimura}\ \emph {et~al.}(2013)\citenamefont {Kimura},
  \citenamefont {Okunishi}, \citenamefont {Hagiwara}, \citenamefont {Kindo},
  \citenamefont {He}, \citenamefont {Taniyama}, \citenamefont {Itoh},
  \citenamefont {Koyama},\ and\ \citenamefont {Watanabe}}]{Kimura13}%
  \BibitemOpen
  \bibfield  {author} {\bibinfo {author} {\bibfnamefont {S.}~\bibnamefont
  {Kimura}}, \bibinfo {author} {\bibfnamefont {K.}~\bibnamefont {Okunishi}},
  \bibinfo {author} {\bibfnamefont {M.}~\bibnamefont {Hagiwara}}, \bibinfo
  {author} {\bibfnamefont {K.}~\bibnamefont {Kindo}}, \bibinfo {author}
  {\bibfnamefont {Z.}~\bibnamefont {He}}, \bibinfo {author} {\bibfnamefont
  {T.}~\bibnamefont {Taniyama}}, \bibinfo {author} {\bibfnamefont
  {M.}~\bibnamefont {Itoh}}, \bibinfo {author} {\bibfnamefont {K.}~\bibnamefont
  {Koyama}},\ and\ \bibinfo {author} {\bibfnamefont {K.}~\bibnamefont
  {Watanabe}},\ }\bibfield  {title} {\bibinfo {title} {Collapse of magnetic
  order of the quasi one-dimensional {I}sing-like antiferromagnet
  {$\mathrm{BaCo}_{2}\mathrm{V}_{2}\mathrm{O}_{8}$} in transverse fields},\
  }\href {https://doi.org/10.7566/JPSJ.82.033706} {\bibfield  {journal}
  {\bibinfo  {journal} {Journal of the Physical Society of Japan}\ }\textbf
  {\bibinfo {volume} {82}},\ \bibinfo {pages} {033706} (\bibinfo {year}
  {2013})}\BibitemShut {NoStop}%
\bibitem [{\citenamefont {Kimura}\ \emph {et~al.}(2008)\citenamefont {Kimura},
  \citenamefont {Takeuchi}, \citenamefont {Okunishi}, \citenamefont {Hagiwara},
  \citenamefont {He}, \citenamefont {Kindo}, \citenamefont {Taniyama},\ and\
  \citenamefont {Itoh}}]{Kimura08}%
  \BibitemOpen
  \bibfield  {author} {\bibinfo {author} {\bibfnamefont {S.}~\bibnamefont
  {Kimura}}, \bibinfo {author} {\bibfnamefont {T.}~\bibnamefont {Takeuchi}},
  \bibinfo {author} {\bibfnamefont {K.}~\bibnamefont {Okunishi}}, \bibinfo
  {author} {\bibfnamefont {M.}~\bibnamefont {Hagiwara}}, \bibinfo {author}
  {\bibfnamefont {Z.}~\bibnamefont {He}}, \bibinfo {author} {\bibfnamefont
  {K.}~\bibnamefont {Kindo}}, \bibinfo {author} {\bibfnamefont
  {T.}~\bibnamefont {Taniyama}},\ and\ \bibinfo {author} {\bibfnamefont
  {M.}~\bibnamefont {Itoh}},\ }\bibfield  {title} {\bibinfo {title} {Novel
  ordering of an $s=1/2$ quasi-1d {I}sing-like antiferromagnet in magnetic
  field},\ }\href {https://doi.org/10.1103/PhysRevLett.100.057202} {\bibfield
  {journal} {\bibinfo  {journal} {Phys. Rev. Lett.}\ }\textbf {\bibinfo
  {volume} {100}},\ \bibinfo {pages} {057202} (\bibinfo {year}
  {2008})}\BibitemShut {NoStop}%
\bibitem [{\citenamefont {Faure}\ \emph {et~al.}(2021)\citenamefont {Faure},
  \citenamefont {Takayoshi}, \citenamefont {Grenier}, \citenamefont {Petit},
  \citenamefont {Raymond}, \citenamefont {Boehm}, \citenamefont {Lejay},
  \citenamefont {Giamarchi},\ and\ \citenamefont {Simonet}}]{Faure_21}%
  \BibitemOpen
  \bibfield  {author} {\bibinfo {author} {\bibfnamefont {Q.}~\bibnamefont
  {Faure}}, \bibinfo {author} {\bibfnamefont {S.}~\bibnamefont {Takayoshi}},
  \bibinfo {author} {\bibfnamefont {B.}~\bibnamefont {Grenier}}, \bibinfo
  {author} {\bibfnamefont {S.}~\bibnamefont {Petit}}, \bibinfo {author}
  {\bibfnamefont {S.}~\bibnamefont {Raymond}}, \bibinfo {author} {\bibfnamefont
  {M.}~\bibnamefont {Boehm}}, \bibinfo {author} {\bibfnamefont
  {P.}~\bibnamefont {Lejay}}, \bibinfo {author} {\bibfnamefont
  {T.}~\bibnamefont {Giamarchi}},\ and\ \bibinfo {author} {\bibfnamefont
  {V.}~\bibnamefont {Simonet}},\ }\bibfield  {title} {\bibinfo {title}
  {Solitonic excitations in the {I}sing anisotropic chain
  {$\mathrm{BaCo}_{2}\mathrm{V}_{2}\mathrm{O}_{8}$} under large transverse
  magnetic field},\ }\href {https://doi.org/10.1103/PhysRevResearch.3.043227}
  {\bibfield  {journal} {\bibinfo  {journal} {Phys. Rev. Research}\ }\textbf
  {\bibinfo {volume} {3}},\ \bibinfo {pages} {043227} (\bibinfo {year}
  {2021})}\BibitemShut {NoStop}%
\bibitem [{\citenamefont {Wang}\ \emph {et~al.}(2016)\citenamefont {Wang},
  \citenamefont {Wu}, \citenamefont {Xu}, \citenamefont {Yang}, \citenamefont
  {Wu}, \citenamefont {Bera}, \citenamefont {Islam}, \citenamefont {Lake},
  \citenamefont {Kamenskyi}, \citenamefont {Gogoi}, \citenamefont {Engelkamp},
  \citenamefont {Wang}, \citenamefont {Deisenhofer},\ and\ \citenamefont
  {Loidl}}]{Wang16}%
  \BibitemOpen
  \bibfield  {author} {\bibinfo {author} {\bibfnamefont {Z.}~\bibnamefont
  {Wang}}, \bibinfo {author} {\bibfnamefont {J.}~\bibnamefont {Wu}}, \bibinfo
  {author} {\bibfnamefont {S.}~\bibnamefont {Xu}}, \bibinfo {author}
  {\bibfnamefont {W.}~\bibnamefont {Yang}}, \bibinfo {author} {\bibfnamefont
  {C.}~\bibnamefont {Wu}}, \bibinfo {author} {\bibfnamefont {A.~K.}\
  \bibnamefont {Bera}}, \bibinfo {author} {\bibfnamefont {A.~T. M.~N.}\
  \bibnamefont {Islam}}, \bibinfo {author} {\bibfnamefont {B.}~\bibnamefont
  {Lake}}, \bibinfo {author} {\bibfnamefont {D.}~\bibnamefont {Kamenskyi}},
  \bibinfo {author} {\bibfnamefont {P.}~\bibnamefont {Gogoi}}, \bibinfo
  {author} {\bibfnamefont {H.}~\bibnamefont {Engelkamp}}, \bibinfo {author}
  {\bibfnamefont {N.}~\bibnamefont {Wang}}, \bibinfo {author} {\bibfnamefont
  {J.}~\bibnamefont {Deisenhofer}},\ and\ \bibinfo {author} {\bibfnamefont
  {A.}~\bibnamefont {Loidl}},\ }\bibfield  {title} {\bibinfo {title} {From
  confined spinons to emergent fermions: Observation of elementary magnetic
  excitations in a transverse-field {I}sing chain},\ }\href
  {https://doi.org/10.1103/PhysRevB.94.125130} {\bibfield  {journal} {\bibinfo
  {journal} {Phys. Rev. B}\ }\textbf {\bibinfo {volume} {94}},\ \bibinfo
  {pages} {125130} (\bibinfo {year} {2016})}\BibitemShut {NoStop}%
\bibitem [{\citenamefont {Zou}\ \emph {et~al.}(2021)\citenamefont {Zou},
  \citenamefont {Cui}, \citenamefont {Wang}, \citenamefont {Zhang},
  \citenamefont {Yang}, \citenamefont {Xu}, \citenamefont {Okutani},
  \citenamefont {Hagiwara}, \citenamefont {Matsuda}, \citenamefont {Wang},
  \citenamefont {Mussardo}, \citenamefont {H\'ods\'agi}, \citenamefont
  {Kormos}, \citenamefont {He}, \citenamefont {Kimura}, \citenamefont {Yu},
  \citenamefont {Yu}, \citenamefont {Ma},\ and\ \citenamefont {Wu}}]{Muss21}%
  \BibitemOpen
  \bibfield  {author} {\bibinfo {author} {\bibfnamefont {H.}~\bibnamefont
  {Zou}}, \bibinfo {author} {\bibfnamefont {Y.}~\bibnamefont {Cui}}, \bibinfo
  {author} {\bibfnamefont {X.}~\bibnamefont {Wang}}, \bibinfo {author}
  {\bibfnamefont {Z.}~\bibnamefont {Zhang}}, \bibinfo {author} {\bibfnamefont
  {J.}~\bibnamefont {Yang}}, \bibinfo {author} {\bibfnamefont {G.}~\bibnamefont
  {Xu}}, \bibinfo {author} {\bibfnamefont {A.}~\bibnamefont {Okutani}},
  \bibinfo {author} {\bibfnamefont {M.}~\bibnamefont {Hagiwara}}, \bibinfo
  {author} {\bibfnamefont {M.}~\bibnamefont {Matsuda}}, \bibinfo {author}
  {\bibfnamefont {G.}~\bibnamefont {Wang}}, \bibinfo {author} {\bibfnamefont
  {G.}~\bibnamefont {Mussardo}}, \bibinfo {author} {\bibfnamefont
  {K.}~\bibnamefont {H\'ods\'agi}}, \bibinfo {author} {\bibfnamefont
  {M.}~\bibnamefont {Kormos}}, \bibinfo {author} {\bibfnamefont
  {Z.}~\bibnamefont {He}}, \bibinfo {author} {\bibfnamefont {S.}~\bibnamefont
  {Kimura}}, \bibinfo {author} {\bibfnamefont {R.}~\bibnamefont {Yu}}, \bibinfo
  {author} {\bibfnamefont {W.}~\bibnamefont {Yu}}, \bibinfo {author}
  {\bibfnamefont {J.}~\bibnamefont {Ma}},\ and\ \bibinfo {author}
  {\bibfnamefont {J.}~\bibnamefont {Wu}},\ }\bibfield  {title} {\bibinfo
  {title} {${E}_{8}$ spectra of quasi-one-dimensional antiferromagnet
  {$\mathrm{BaCo}_{2}\mathrm{V}_{2}\mathrm{O}_{8}$} under transverse field},\
  }\href {https://doi.org/10.1103/PhysRevLett.127.077201} {\bibfield  {journal}
  {\bibinfo  {journal} {Phys. Rev. Lett.}\ }\textbf {\bibinfo {volume} {127}},\
  \bibinfo {pages} {077201} (\bibinfo {year} {2021})}\BibitemShut {NoStop}%
\bibitem [{\citenamefont {Amelin}\ \emph {et~al.}(2022)\citenamefont {Amelin},
  \citenamefont {Viirok}, \citenamefont {Nagel}, \citenamefont
  {R$\tilde{\rm{o}}\tilde{\rm{o}}$m}, \citenamefont {Engelmayer}, \citenamefont
  {Dey}, \citenamefont {Nugroho}, \citenamefont {Lorenz},\ and\ \citenamefont
  {Wang}}]{Wang_22}%
  \BibitemOpen
  \bibfield  {author} {\bibinfo {author} {\bibfnamefont {K.}~\bibnamefont
  {Amelin}}, \bibinfo {author} {\bibfnamefont {J.}~\bibnamefont {Viirok}},
  \bibinfo {author} {\bibfnamefont {U.}~\bibnamefont {Nagel}}, \bibinfo
  {author} {\bibfnamefont {T.}~\bibnamefont
  {R$\tilde{\rm{o}}\tilde{\rm{o}}$m}}, \bibinfo {author} {\bibfnamefont
  {J.}~\bibnamefont {Engelmayer}}, \bibinfo {author} {\bibfnamefont
  {T.}~\bibnamefont {Dey}}, \bibinfo {author} {\bibfnamefont {A.~A.}\
  \bibnamefont {Nugroho}}, \bibinfo {author} {\bibfnamefont {T.}~\bibnamefont
  {Lorenz}},\ and\ \bibinfo {author} {\bibfnamefont {Z.}~\bibnamefont {Wang}},\
  }\bibfield  {title} {\bibinfo {title} {Quantum spin dynamics of
  quasi-one-dimensional {H}eisenberg-{I}sing magnets in a transverse field:
  confined spinons, {E}8 spectrum, and quantum phase transitions},\ }\href
  {https://doi.org/10.1088/1751-8121/aca6b8} {\bibfield  {journal} {\bibinfo
  {journal} {Journal of Physics A: Mathematical and Theoretical}\ }\textbf
  {\bibinfo {volume} {55}},\ \bibinfo {pages} {484005} (\bibinfo {year}
  {2022})}\BibitemShut {NoStop}%
\bibitem [{\citenamefont {Takayoshi}\ \emph {et~al.}(2023)\citenamefont
  {Takayoshi}, \citenamefont {Faure}, \citenamefont {Simonet}, \citenamefont
  {Grenier}, \citenamefont {Petit}, \citenamefont {Ollivier}, \citenamefont
  {Lejay},\ and\ \citenamefont {Giamarchi}}]{Takay23}%
  \BibitemOpen
  \bibfield  {author} {\bibinfo {author} {\bibfnamefont {S.}~\bibnamefont
  {Takayoshi}}, \bibinfo {author} {\bibfnamefont {Q.}~\bibnamefont {Faure}},
  \bibinfo {author} {\bibfnamefont {V.}~\bibnamefont {Simonet}}, \bibinfo
  {author} {\bibfnamefont {B.}~\bibnamefont {Grenier}}, \bibinfo {author}
  {\bibfnamefont {S.}~\bibnamefont {Petit}}, \bibinfo {author} {\bibfnamefont
  {J.}~\bibnamefont {Ollivier}}, \bibinfo {author} {\bibfnamefont
  {P.}~\bibnamefont {Lejay}},\ and\ \bibinfo {author} {\bibfnamefont
  {T.}~\bibnamefont {Giamarchi}},\ }\bibfield  {title} {\bibinfo {title} {Phase
  transitions and spin dynamics of the quasi-one dimensional {I}sing-like
  antiferromagnet {$\mathrm{BaCo}_{2}\mathrm{V}_{2}\mathrm{O}_{8}$} in a
  longitudinal magnetic field},\ }\href
  {https://doi.org/10.1103/PhysRevResearch.5.023205} {\bibfield  {journal}
  {\bibinfo  {journal} {Phys. Rev. Res.}\ }\textbf {\bibinfo {volume} {5}},\
  \bibinfo {pages} {023205} (\bibinfo {year} {2023})}\BibitemShut {NoStop}%
\bibitem [{\citenamefont {Dmitriev}\ \emph {et~al.}(2002)\citenamefont
  {Dmitriev}, \citenamefont {Krivnov}, \citenamefont {Ovchinnikov},\ and\
  \citenamefont {Langari}}]{Dmitr02}%
  \BibitemOpen
  \bibfield  {author} {\bibinfo {author} {\bibfnamefont {D.~V.}\ \bibnamefont
  {Dmitriev}}, \bibinfo {author} {\bibfnamefont {V.~Y.}\ \bibnamefont
  {Krivnov}}, \bibinfo {author} {\bibfnamefont {A.~A.}\ \bibnamefont
  {Ovchinnikov}},\ and\ \bibinfo {author} {\bibfnamefont {A.}~\bibnamefont
  {Langari}},\ }\bibfield  {title} {\bibinfo {title} {One-dimensional
  anisotropic {H}eisenberg model in the transverse magnetic field},\
  }\href@noop {} {\bibfield  {journal} {\bibinfo  {journal} {JETP}\ }\textbf
  {\bibinfo {volume} {95}},\ \bibinfo {pages} {538} (\bibinfo {year}
  {2002})}\BibitemShut {NoStop}%
\bibitem [{\citenamefont {Takayoshi}\ \emph {et~al.}(2018)\citenamefont
  {Takayoshi}, \citenamefont {Furuya},\ and\ \citenamefont
  {Giamarchi}}]{Tako18}%
  \BibitemOpen
  \bibfield  {author} {\bibinfo {author} {\bibfnamefont {S.}~\bibnamefont
  {Takayoshi}}, \bibinfo {author} {\bibfnamefont {S.~C.}\ \bibnamefont
  {Furuya}},\ and\ \bibinfo {author} {\bibfnamefont {T.}~\bibnamefont
  {Giamarchi}},\ }\bibfield  {title} {\bibinfo {title} {Topological transition
  between competing orders in quantum spin chains},\ }\href
  {https://doi.org/10.1103/PhysRevB.98.184429} {\bibfield  {journal} {\bibinfo
  {journal} {Phys. Rev. B}\ }\textbf {\bibinfo {volume} {98}},\ \bibinfo
  {pages} {184429} (\bibinfo {year} {2018})}\BibitemShut {NoStop}%
\bibitem [{\citenamefont {Okutani}\ \emph {et~al.}(2021)\citenamefont
  {Okutani}, \citenamefont {Onishi}, \citenamefont {Kimura}, \citenamefont
  {Takeuchi}, \citenamefont {Kida}, \citenamefont {Mori}, \citenamefont
  {Miyake}, \citenamefont {Tokunaga}, \citenamefont {Kindo},\ and\
  \citenamefont {Hagiwara}}]{Kimura_21}%
  \BibitemOpen
  \bibfield  {author} {\bibinfo {author} {\bibfnamefont {A.}~\bibnamefont
  {Okutani}}, \bibinfo {author} {\bibfnamefont {H.}~\bibnamefont {Onishi}},
  \bibinfo {author} {\bibfnamefont {S.}~\bibnamefont {Kimura}}, \bibinfo
  {author} {\bibfnamefont {T.}~\bibnamefont {Takeuchi}}, \bibinfo {author}
  {\bibfnamefont {T.}~\bibnamefont {Kida}}, \bibinfo {author} {\bibfnamefont
  {M.}~\bibnamefont {Mori}}, \bibinfo {author} {\bibfnamefont {A.}~\bibnamefont
  {Miyake}}, \bibinfo {author} {\bibfnamefont {M.}~\bibnamefont {Tokunaga}},
  \bibinfo {author} {\bibfnamefont {K.}~\bibnamefont {Kindo}},\ and\ \bibinfo
  {author} {\bibfnamefont {M.}~\bibnamefont {Hagiwara}},\ }\bibfield  {title}
  {\bibinfo {title} {Spin excitations of the s = 1/2 one-dimensional
  {I}sing-like antiferromagnet
  {$\mathrm{BaCo}_{2}\mathrm{V}_{2}\mathrm{O}_{8}$} in transverse magnetic
  fields},\ }\href {https://doi.org/10.7566/JPSJ.90.044704} {\bibfield
  {journal} {\bibinfo  {journal} {Journal of the Physical Society of Japan}\
  }\textbf {\bibinfo {volume} {90}},\ \bibinfo {pages} {044704} (\bibinfo
  {year} {2021})}\BibitemShut {NoStop}%
\bibitem [{\citenamefont {Halati}\ \emph {et~al.}(2023)\citenamefont {Halati},
  \citenamefont {Wang}, \citenamefont {Lorenz}, \citenamefont {Kollath},\ and\
  \citenamefont {Bernier}}]{Wang23a}%
  \BibitemOpen
  \bibfield  {author} {\bibinfo {author} {\bibfnamefont {C.-M.}\ \bibnamefont
  {Halati}}, \bibinfo {author} {\bibfnamefont {Z.}~\bibnamefont {Wang}},
  \bibinfo {author} {\bibfnamefont {T.}~\bibnamefont {Lorenz}}, \bibinfo
  {author} {\bibfnamefont {C.}~\bibnamefont {Kollath}},\ and\ \bibinfo {author}
  {\bibfnamefont {J.-S.}\ \bibnamefont {Bernier}},\ }\bibfield  {title}
  {\bibinfo {title} {Repulsively bound magnon excitations of a
  spin-$\frac{1}{2}$ xxz chain in a staggered transverse field},\ }\href
  {https://doi.org/10.1103/PhysRevB.108.224429} {\bibfield  {journal} {\bibinfo
   {journal} {Phys. Rev. B}\ }\textbf {\bibinfo {volume} {108}},\ \bibinfo
  {pages} {224429} (\bibinfo {year} {2023})}\BibitemShut {NoStop}%
\bibitem [{\citenamefont {Jimbo}\ and\ \citenamefont {Miwa}(1995)}]{Jimbo94}%
  \BibitemOpen
  \bibfield  {author} {\bibinfo {author} {\bibfnamefont {M.}~\bibnamefont
  {Jimbo}}\ and\ \bibinfo {author} {\bibfnamefont {T.}~\bibnamefont {Miwa}},\
  }\href@noop {} {\emph {\bibinfo {title} {Algebraic {A}nalysis of {S}olvable
  {L}attice {M}odels}}},\ \bibinfo {series} {Conference Board of the
  Mathematical Sciences}\ No.~\bibinfo {number} {85}\ (\bibinfo  {publisher}
  {American Mathematical Soc.},\ \bibinfo {year} {1995})\BibitemShut {NoStop}%
\bibitem [{\citenamefont {Lukyanov}\ and\ \citenamefont
  {Terras}(2003)}]{LukTer03}%
  \BibitemOpen
  \bibfield  {author} {\bibinfo {author} {\bibfnamefont {S.}~\bibnamefont
  {Lukyanov}}\ and\ \bibinfo {author} {\bibfnamefont {V.}~\bibnamefont
  {Terras}},\ }\bibfield  {title} {\bibinfo {title} {Long-distance asymptotics
  of spin-spin correlation functions for the {XXZ} spin chain},\ }\href
  {https://doi.org/10.1016/S0550-3213(02)01141-0} {\bibfield  {journal}
  {\bibinfo  {journal} {Nuclear Physics B}\ }\textbf {\bibinfo {volume}
  {654}},\ \bibinfo {pages} {323 } (\bibinfo {year} {2003})}\BibitemShut
  {NoStop}%
\bibitem [{\citenamefont {Baxter}(1973)}]{Baxter1973}%
  \BibitemOpen
  \bibfield  {author} {\bibinfo {author} {\bibfnamefont {R.~J.}\ \bibnamefont
  {Baxter}},\ }\bibfield  {title} {\bibinfo {title} {Spontaneous staggered
  polarization of the {F}-model},\ }\href {https://doi.org/10.1007/BF01016845}
  {\bibfield  {journal} {\bibinfo  {journal} {Journal of Statistical Physics}\
  }\textbf {\bibinfo {volume} {9}},\ \bibinfo {pages} {145} (\bibinfo {year}
  {1973})}\BibitemShut {NoStop}%
\bibitem [{\citenamefont {Baxter}(1976)}]{Baxter1976}%
  \BibitemOpen
  \bibfield  {author} {\bibinfo {author} {\bibfnamefont {R.~J.}\ \bibnamefont
  {Baxter}},\ }\bibfield  {title} {\bibinfo {title} {Corner transfer matrices
  of the eight-vertex model. {I}. {L}ow-temperature expansions and conjectured
  properties},\ }\href {https://doi.org/10.1007/BF01020802} {\bibfield
  {journal} {\bibinfo  {journal} {Journal of Statistical Physics}\ }\textbf
  {\bibinfo {volume} {15}},\ \bibinfo {pages} {485} (\bibinfo {year}
  {1976})}\BibitemShut {NoStop}%
\bibitem [{\citenamefont {Izergin}\ \emph {et~al.}(1999)\citenamefont
  {Izergin}, \citenamefont {Kitanine}, \citenamefont {Maillet},\ and\
  \citenamefont {Terras}}]{Iz99}%
  \BibitemOpen
  \bibfield  {author} {\bibinfo {author} {\bibfnamefont {A.~G.}\ \bibnamefont
  {Izergin}}, \bibinfo {author} {\bibfnamefont {N.}~\bibnamefont {Kitanine}},
  \bibinfo {author} {\bibfnamefont {J.~M.}\ \bibnamefont {Maillet}},\ and\
  \bibinfo {author} {\bibfnamefont {V.}~\bibnamefont {Terras}},\ }\bibfield
  {title} {\bibinfo {title} {Spontaneous magnetization of the {XXZ H}eisenberg
  spin-1/2 chain},\ }\href {https://doi.org/10.1016/S0550-3213(99)00273-4}
  {\bibfield  {journal} {\bibinfo  {journal} {Nuclear Physics B}\ }\textbf
  {\bibinfo {volume} {554}},\ \bibinfo {pages} {679 } (\bibinfo {year}
  {1999})}\BibitemShut {NoStop}%
\bibitem [{\citenamefont {Ishimura}\ and\ \citenamefont
  {Shiba}(1980)}]{Shiba_80}%
  \BibitemOpen
  \bibfield  {author} {\bibinfo {author} {\bibfnamefont {N.}~\bibnamefont
  {Ishimura}}\ and\ \bibinfo {author} {\bibfnamefont {H.}~\bibnamefont
  {Shiba}},\ }\bibfield  {title} {\bibinfo {title} {Dynamical correlation
  functions of one-dimensional anisotropic {H}eisenberg model with spin 1/2.
  {I}: {I}sing-like antiferromagnets},\ }\href
  {https://doi.org/10.1143/PTP.63.743} {\bibfield  {journal} {\bibinfo
  {journal} {Progress of Theoretical Physics}\ }\textbf {\bibinfo {volume}
  {63}},\ \bibinfo {pages} {743} (\bibinfo {year} {1980})}\BibitemShut
  {NoStop}%
\bibitem [{\citenamefont {Teschl}(2000)}]{teschl2000jacobi}%
  \BibitemOpen
  \bibfield  {author} {\bibinfo {author} {\bibfnamefont {G.}~\bibnamefont
  {Teschl}},\ }\href {https://books.google.de/books?id=xaHyBwAAQBAJ} {\emph
  {\bibinfo {title} {Jacobi Operators and Completely Integrable Nonlinear
  Lattices}}},\ Mathematical surveys and monographs\ (\bibinfo  {publisher}
  {American Mathematical Society},\ \bibinfo {year} {2000})\BibitemShut
  {NoStop}%
\bibitem [{\citenamefont {Landau}\ and\ \citenamefont {Lifshitz}(1981)}]{LL3}%
  \BibitemOpen
  \bibfield  {author} {\bibinfo {author} {\bibfnamefont {L.~D.}\ \bibnamefont
  {Landau}}\ and\ \bibinfo {author} {\bibfnamefont {E.~M.}\ \bibnamefont
  {Lifshitz}},\ }\href@noop {} {\emph {\bibinfo {title} {Quantum Mechanics:
  Non-Relativistic Theory}}},\ Course of Theoretical Physics\ (\bibinfo
  {publisher} {Elsevier Science},\ \bibinfo {year} {1981})\BibitemShut
  {NoStop}%
\bibitem [{\citenamefont {Johnson}\ \emph {et~al.}(1973)\citenamefont
  {Johnson}, \citenamefont {Krinsky},\ and\ \citenamefont {McCoy}}]{McCoy73}%
  \BibitemOpen
  \bibfield  {author} {\bibinfo {author} {\bibfnamefont {J.~D.}\ \bibnamefont
  {Johnson}}, \bibinfo {author} {\bibfnamefont {S.}~\bibnamefont {Krinsky}},\
  and\ \bibinfo {author} {\bibfnamefont {B.~M.}\ \bibnamefont {McCoy}},\
  }\bibfield  {title} {\bibinfo {title} {Vertical-arrow correlation length in
  the eight-vertex model and the low-lying excitations of the ${X}{-}{Y}{-}{Z}$
  {H}amiltonian},\ }\href {https://doi.org/10.1103/PhysRevA.8.2526} {\bibfield
  {journal} {\bibinfo  {journal} {Phys. Rev. A}\ }\textbf {\bibinfo {volume}
  {8}},\ \bibinfo {pages} {2526} (\bibinfo {year} {1973})}\BibitemShut
  {NoStop}%
\bibitem [{\citenamefont {Lencs{\'e}s}\ and\ \citenamefont
  {Tak{\'a}cs}(2014)}]{Tak14}%
  \BibitemOpen
  \bibfield  {author} {\bibinfo {author} {\bibfnamefont {M.}~\bibnamefont
  {Lencs{\'e}s}}\ and\ \bibinfo {author} {\bibfnamefont {G.}~\bibnamefont
  {Tak{\'a}cs}},\ }\bibfield  {title} {\bibinfo {title} {Excited state {TBA}
  and renormalized {TCSA} in the scaling {P}otts model},\ }\href
  {https://doi.org/10.1007/JHEP09(2014)052} {\bibfield  {journal} {\bibinfo
  {journal} {Journal of High Energy Physics}\ }\textbf {\bibinfo {volume}
  {2014}},\ \bibinfo {pages} {52} (\bibinfo {year} {2014})}\BibitemShut
  {NoStop}%
\bibitem [{\citenamefont {Kormos}\ \emph {et~al.}(2017)\citenamefont {Kormos},
  \citenamefont {Collura}, \citenamefont {Tak{\'a}cs},\ and\ \citenamefont
  {Calabrese}}]{Kor16}%
  \BibitemOpen
  \bibfield  {author} {\bibinfo {author} {\bibfnamefont {M.}~\bibnamefont
  {Kormos}}, \bibinfo {author} {\bibfnamefont {M.}~\bibnamefont {Collura}},
  \bibinfo {author} {\bibfnamefont {G.}~\bibnamefont {Tak{\'a}cs}},\ and\
  \bibinfo {author} {\bibfnamefont {P.}~\bibnamefont {Calabrese}},\ }\bibfield
  {title} {\bibinfo {title} {Real-time confinement following a quantum quench
  to a non-integrable model},\ }\href {https://doi.org/10.1038/nphys3934}
  {\bibfield  {journal} {\bibinfo  {journal} {Nat. Phys.}\ }\textbf {\bibinfo
  {volume} {13}},\ \bibinfo {pages} {246} (\bibinfo {year} {2017})}\BibitemShut
  {NoStop}%
\bibitem [{\citenamefont {Lagnese}\ \emph {et~al.}(2020)\citenamefont
  {Lagnese}, \citenamefont {Surace}, \citenamefont {Kormos},\ and\
  \citenamefont {Calabrese}}]{Lagnese_2020}%
  \BibitemOpen
  \bibfield  {author} {\bibinfo {author} {\bibfnamefont {G.}~\bibnamefont
  {Lagnese}}, \bibinfo {author} {\bibfnamefont {F.~M.}\ \bibnamefont {Surace}},
  \bibinfo {author} {\bibfnamefont {M.}~\bibnamefont {Kormos}},\ and\ \bibinfo
  {author} {\bibfnamefont {P.}~\bibnamefont {Calabrese}},\ }\bibfield  {title}
  {\bibinfo {title} {Confinement in the spectrum of a
  {H}eisenberg{\textendash}{I}sing spin ladder},\ }\href
  {https://doi.org/10.1088/1742-5468/abb368} {\bibfield  {journal} {\bibinfo
  {journal} {Journal of Statistical Mechanics: Theory and Experiment}\ }\textbf
  {\bibinfo {volume} {2020}},\ \bibinfo {pages} {093106} (\bibinfo {year}
  {2020})}\BibitemShut {NoStop}%
\bibitem [{\citenamefont {Kadigrobov}\ and\ \citenamefont
  {Slutskin}(1972)}]{KadSl72}%
  \BibitemOpen
  \bibfield  {author} {\bibinfo {author} {\bibfnamefont {A.~M.}\ \bibnamefont
  {Kadigrobov}}\ and\ \bibinfo {author} {\bibfnamefont {A.~A.}\ \bibnamefont
  {Slutskin}},\ }\bibfield  {title} {\bibinfo {title} {Influence of magnetic
  breakdown on low-frequency conductivity and weak damping electromagnetic
  waves in metals},\ }\href {https://doi.org/10.1007/BF00630910} {\bibfield
  {journal} {\bibinfo  {journal} {Journal of Low Temp. Phys.}\ }\textbf
  {\bibinfo {volume} {6}},\ \bibinfo {pages} {69} (\bibinfo {year}
  {1972})}\BibitemShut {NoStop}%
\bibitem [{\citenamefont {Abrikosov}(2017)}]{Abr17}%
  \BibitemOpen
  \bibfield  {author} {\bibinfo {author} {\bibfnamefont {A.~A.}\ \bibnamefont
  {Abrikosov}},\ }\href@noop {} {\emph {\bibinfo {title} {Fundamentals of the
  Theory of Metals}}}\ (\bibinfo  {publisher} {Dover Publications Inc.},\
  \bibinfo {year} {2017})\BibitemShut {NoStop}%
\bibitem [{\citenamefont {Grenier}\ \emph
  {et~al.}(2015{\natexlab{b}})\citenamefont {Grenier}, \citenamefont {Petit},
  \citenamefont {Simonet}, \citenamefont {Can\'evet}, \citenamefont {Regnault},
  \citenamefont {Raymond}, \citenamefont {Canals}, \citenamefont {Berthier},\
  and\ \citenamefont {Lejay}}]{Gr15E}%
  \BibitemOpen
  \bibfield  {author} {\bibinfo {author} {\bibfnamefont {B.}~\bibnamefont
  {Grenier}}, \bibinfo {author} {\bibfnamefont {S.}~\bibnamefont {Petit}},
  \bibinfo {author} {\bibfnamefont {V.}~\bibnamefont {Simonet}}, \bibinfo
  {author} {\bibfnamefont {E.}~\bibnamefont {Can\'evet}}, \bibinfo {author}
  {\bibfnamefont {L.-P.}\ \bibnamefont {Regnault}}, \bibinfo {author}
  {\bibfnamefont {S.}~\bibnamefont {Raymond}}, \bibinfo {author} {\bibfnamefont
  {B.}~\bibnamefont {Canals}}, \bibinfo {author} {\bibfnamefont
  {C.}~\bibnamefont {Berthier}},\ and\ \bibinfo {author} {\bibfnamefont
  {P.}~\bibnamefont {Lejay}},\ }\bibfield  {title} {\bibinfo {title} {Erratum:
  Longitudinal and transverse {Z}eeman ladders in the {I}sing-like chain
  antiferromagnet {$\mathrm{BaCo}_{2}\mathrm{V}_{2}\mathrm{O}_{8}$} [{P}hys.
  {R}ev. {L}ett. {\bf{114}}, 017201 (2015)]},\ }\href
  {https://doi.org/doi.org/10.1103/PhysRevLett.115.119902} {\bibfield
  {journal} {\bibinfo  {journal} {Phys. Rev. Lett.}\ }\textbf {\bibinfo
  {volume} {115}},\ \bibinfo {pages} {119902} (\bibinfo {year}
  {2015}{\natexlab{b}})}\BibitemShut {NoStop}%
\bibitem [{\citenamefont {Kimura}\ \emph {et~al.}(2022)\citenamefont {Kimura},
  \citenamefont {Onishi}, \citenamefont {Okutani}, \citenamefont {Akaki},
  \citenamefont {Narumi}, \citenamefont {Hagiwara}, \citenamefont {Okunishi},
  \citenamefont {Kindo}, \citenamefont {He}, \citenamefont {Taniyama},\ and\
  \citenamefont {Itoh}}]{Kimura22}%
  \BibitemOpen
  \bibfield  {author} {\bibinfo {author} {\bibfnamefont {S.}~\bibnamefont
  {Kimura}}, \bibinfo {author} {\bibfnamefont {H.}~\bibnamefont {Onishi}},
  \bibinfo {author} {\bibfnamefont {A.}~\bibnamefont {Okutani}}, \bibinfo
  {author} {\bibfnamefont {M.}~\bibnamefont {Akaki}}, \bibinfo {author}
  {\bibfnamefont {Y.}~\bibnamefont {Narumi}}, \bibinfo {author} {\bibfnamefont
  {M.}~\bibnamefont {Hagiwara}}, \bibinfo {author} {\bibfnamefont
  {K.}~\bibnamefont {Okunishi}}, \bibinfo {author} {\bibfnamefont
  {K.}~\bibnamefont {Kindo}}, \bibinfo {author} {\bibfnamefont
  {Z.}~\bibnamefont {He}}, \bibinfo {author} {\bibfnamefont {T.}~\bibnamefont
  {Taniyama}},\ and\ \bibinfo {author} {\bibfnamefont {M.}~\bibnamefont
  {Itoh}},\ }\bibfield  {title} {\bibinfo {title} {Optical selection rules of
  the magnetic excitation in the ${S}=\frac{1}{2}$ one-dimensional {I}sing-like
  antiferromagnet {$\mathrm{BaCo}_{2}\mathrm{V}_{2}\mathrm{O}_{8}$}},\ }\href
  {https://doi.org/10.1103/PhysRevB.105.014417} {\bibfield  {journal} {\bibinfo
   {journal} {Phys. Rev. B}\ }\textbf {\bibinfo {volume} {105}},\ \bibinfo
  {pages} {014417} (\bibinfo {year} {2022})}\BibitemShut {NoStop}%
\bibitem [{\citenamefont {Zabrodin}(1992)}]{Zabr92}%
  \BibitemOpen
  \bibfield  {author} {\bibinfo {author} {\bibfnamefont {A.}~\bibnamefont
  {Zabrodin}},\ }\bibfield  {title} {\bibinfo {title} {Integrable models of
  field theory and scattering on quantum hyperboloids},\ }\href
  {https://doi.org/10.1142/S0217732392000392} {\bibfield  {journal} {\bibinfo
  {journal} {Modern Physics Letters A}\ }\textbf {\bibinfo {volume} {07}},\
  \bibinfo {pages} {441} (\bibinfo {year} {1992})}\BibitemShut {NoStop}%
\bibitem [{\citenamefont {Davies}\ \emph {et~al.}(1993)\citenamefont {Davies},
  \citenamefont {Foda}, \citenamefont {Jimbo}, \citenamefont {Miwa},\ and\
  \citenamefont {Nakayashiki}}]{Miwa93}%
  \BibitemOpen
  \bibfield  {author} {\bibinfo {author} {\bibfnamefont {B.}~\bibnamefont
  {Davies}}, \bibinfo {author} {\bibfnamefont {O.}~\bibnamefont {Foda}},
  \bibinfo {author} {\bibfnamefont {M.}~\bibnamefont {Jimbo}}, \bibinfo
  {author} {\bibfnamefont {T.}~\bibnamefont {Miwa}},\ and\ \bibinfo {author}
  {\bibfnamefont {A.}~\bibnamefont {Nakayashiki}},\ }\bibfield  {title}
  {\bibinfo {title} {Diagonalization of the ${XXZ}$ {H}amiltonian by vertex
  operators},\ }\href {https://doi.org/10.1007/BF02096750} {\bibfield
  {journal} {\bibinfo  {journal} {Commun. Math. Phys.}\ }\textbf {\bibinfo
  {volume} {151}},\ \bibinfo {pages} {89 } (\bibinfo {year}
  {1993})}\BibitemShut {NoStop}%
\bibitem [{\citenamefont {Lashkevich}(2002)}]{Lash02}%
  \BibitemOpen
  \bibfield  {author} {\bibinfo {author} {\bibfnamefont {M.}~\bibnamefont
  {Lashkevich}},\ }\bibfield  {title} {\bibinfo {title} {Free field
  construction for the eight-vertex model: representation for form factors},\
  }\href {https://doi.org/10.1016/S0550-3213(01)00598-3} {\bibfield  {journal}
  {\bibinfo  {journal} {Nuclear Physics B}\ }\textbf {\bibinfo {volume}
  {621}},\ \bibinfo {pages} {587 } (\bibinfo {year} {2002})}\BibitemShut
  {NoStop}%
\bibitem [{\citenamefont {Dugave}\ \emph {et~al.}(2015)\citenamefont {Dugave},
  \citenamefont {G\"ohmann}, \citenamefont {Kozlowski},\ and\ \citenamefont
  {Suzuki}}]{Dug15}%
  \BibitemOpen
  \bibfield  {author} {\bibinfo {author} {\bibfnamefont {M.}~\bibnamefont
  {Dugave}}, \bibinfo {author} {\bibfnamefont {F.}~\bibnamefont {G\"ohmann}},
  \bibinfo {author} {\bibfnamefont {K.~K.}\ \bibnamefont {Kozlowski}},\ and\
  \bibinfo {author} {\bibfnamefont {J.}~\bibnamefont {Suzuki}},\ }\bibfield
  {title} {\bibinfo {title} {On form-factor expansions for the ${XXZ}$ chain in
  the massive regime},\ }\href
  {http://stacks.iop.org/1742-5468/2015/i=5/a=P05037} {\bibfield  {journal}
  {\bibinfo  {journal} {Journal of Statistical Mechanics: Theory and
  Experiment}\ }\textbf {\bibinfo {volume} {2015}},\ \bibinfo {pages} {P05037}
  (\bibinfo {year} {2015})}\BibitemShut {NoStop}%
\end{thebibliography}
%

\end{document}